\documentclass[11pt,a4paper]{article}
\usepackage{jheppub,xcolor}
\usepackage[utf8]{inputenc}
\graphicspath{{./figs/}}

\usepackage{amssymb}
\usepackage{amsmath}
\usepackage{dcolumn}	
\usepackage{bm}			
\usepackage{bbm} 		
\usepackage{multirow}
\usepackage{slashed}
\usepackage{blkarray}
\usepackage{booktabs}
\usepackage{makecell}
\usepackage{longtable}
\usepackage{placeins}
\usepackage{graphicx}
\usepackage{float}
\usepackage{multirow}
\usepackage{verbatim}
\usepackage{hyperref}
\usepackage{subcaption}
\usepackage{lscape} 
\usepackage{tablefootnote} 
\usepackage[font=small,labelfont=bf]{caption}
\usepackage[compat=1.1.0]{tikz-feynman}
\usetikzlibrary{positioning}

\definecolor{Red}{rgb}{1.,0.,0.}
\definecolor{aquamarine}{rgb}{0.6,1.,0.87}
\newcommand{\lag}{\mathcal{L}}
\newcommand{\mcM}{\mathcal{M}}
\newcommand{\mcO}{\mathcal{O}}

\newcommand{\cpW}{\ensuremath{c_{\varphi W}}}
\newcommand{\cpWt}{\ensuremath{c_{\varphi \widetilde{W}}}}

\title{CP-odd effects at NLO in SMEFT $WH$ and $ZH$ production}

\author[a]{Alejo~N.~Rossia,}
\author[a]{and Eleni~Vryonidou}
\affiliation[a]{Dept. of Physics and Astronomy, University of Manchester, Manchester M13 9PL, UK}

\emailAdd{alejo.rossia@manchester.ac.uk, eleni.vryonidou@manchester.ac.uk}

\date{\today}
\abstract{ CP-violation (CPV) is a rare phenomenon in the Standard Model whilst there is compelling indirect evidence for additional CPV sources in the Universe. The search for CPV effects at the LHC is thus one of the best-motivated precision tests of the Standard Model (SM) and an excellent probe of New Physics. NLO QCD corrections can affect the predictions for those measurements substantially. We study the impact of NLO QCD corrections in $WH$ and $ZH$ production in the Standard Model Effective Field Theory with bosonic CP-odd dimension-6 operators. We analyze the angular distributions at LO of those processes that can be used to probe CPV effects. We then show how NLO QCD effects modify those distributions. We encounter that the corrections have a clear angular dependence and differ between the SM, the dimension-6 squared and their interference, emphasising the need for an exact inclusion of NLO QCD in precision computations. We then perform a phenomenological analysis of $WH$ production at the LHC to study the impact of NLO QCD effects on the projected bounds on the CP-odd Wilson Coefficient  $c_{\varphi\widetilde{W}}$. NLO QCD effects in the signal improve the bounds by $\sim10\%$ but reduce the significance of the interference.
}


\begin{document}
\begin{flushright}
COMETA-2024-020
\end{flushright}
\maketitle
\newpage
\section{Introduction}

The success of the Standard Model in describing the physics explored at the LHC has spurred an intense programme to test its predictions with increasing precision.  The capabilities of the LHC as a precision machine will grow substantially over the coming two decades thanks to the unprecedented amount of data to be collected during Run 3 and the future High-Luminosity LHC (HL-LHC) program. A complete understanding of future experimental results requires theoretical predictions of matching precision.

The LHC precision programme aims to obtain clues about where New Physics (NP) beyond the Standard Model (SM) lies and what it might look like.  One of the most striking predictions of the SM scrutinised by the LHC is the small amount of CP violation (CPV) controlled by the complex phase in the Cabibbo-Kobayashi-Maskawa (CKM) matrix in the EW-Yukawa sector and the $\theta_{\text{QCD}}$ parameter in the QCD sector~\cite{Cabibbo:1963yz,Kobayashi:1973fv}. In particular, $\theta_{\text{QCD}}$ is usually neglected due to being strictly bounded. Thus, CPV in the SM is a rare phenomenon that is accidentally suppressed and requires an interplay between the EW gauge sector and both quark Yukawa matrices. The latter fact is stressed by the CPV order parameter in the SM being the flavour-basis invariant Jarlskog invariant $J$~\cite{Jarlskog:1985ht,Jarlskog:1985cw,Bernabeu:1986fc}.

However, there is evidence of larger CPV effects in the Universe. In particular, the matter-antimatter asymmetry in the Universe can be explained via the baryogenesis process, which requires CPV according to Sakharov's criteria~\cite{Sakharov:1967dj,Canetti:2012zc}. The SM does not provide enough CPV to account for the large observed asymmetry, hence additional CPV sources are required~\cite{Riotto:1998bt,Canetti:2012zc}. Moreover, the evidence for non-vanishing neutrino masses allows for the occurrence of CPV in the lepton sector, contributing to the baryon asymmetry via leptogenesis~\cite{Fukugita:1986hr,Davidson:2008bu}.

Many studies have been performed on CPV effects at low energies, for example in B-mesons and kaon decays. All the CPV effects observed at those scales so far are compatible with the SM predictions and a complex CKM phase near unity, $\delta=1.147(26)$~\cite{ParticleDataGroup:2022pth}. When new sources of CPV are expected to arise from heavy NPs, the high precision of these low-energy probes is counteracted by higher theoretical uncertainties. Hence, the search for CPV effects in higher energy processes must be seen as complementary. The closer the energy of the process to the scale of NP, the less suppressed the new CPV effects will be. As such, high-energy searches for CPV are well motivated and a topic of intense activity by the LHC experiments. Among high-energy probes, the Higgs boson discovery sparked several CPV studies to probe its CP nature and the one of its interactions with vector bosons and fermions, see e.g.~\cite{CMS:2021nnc,CMS:2021sdq,ATLAS:2022akr,CMS:2022dbt,ATLAS:2023cbt,ATLAS:2023mqy}. 
This programme will be intensified and diversified since testing the CPV predictions of the SM is one of the most attractive issues to be tackled at the LHC Run 3 and HL-LHC. 

In the absence of light new particles, the Standard Model Eective Field Theory
(SMEFT) provides a robust framework to parameterize deviations from the SM caused
by heavy NP in a mostly model-independent manner. SMEFT thus constitutes an attractive and natural
framework that can guide the exploration of new CPV phenomena with minimal assumptions on the UV completion of the SM. The development and use of SMEFT are mature and widespread, from the computation of predictions at NLO in QCD, to the interpretation of experimental data and the matching of this EFT to specific NP models, see e.g.~\cite{Isidori:2023pyp} for a recent review.

The SMEFT is defined as a quantum field theory with the same field content and gauge symmetry as the SM, but with a Lagrangian extended by an infinite tower of higher-dimensional operators $\mcO_{i}$,
\begin{equation}
    \lag_{\text{SMEFT}}=\lag_{\text{SM}}+\sum_{i}^{\infty} \frac{c_{i}}{\Lambda^{d_i-4}}\mcO_{i},
\end{equation}
where $d_i$ is the dimension of the operator, $c_i$ are the dimensionless Wilson Coefficients (WCs), and $\Lambda$ is the energy cutoff of the EFT, above which the new particles must be included as dynamical degrees of freedom and the SMEFT loses its validity. Dimension-6 operators provide the leading deviations from the SM and we will focus on them in this work.

At dimension 6, SMEFT introduces $705$ CPV parameters, 6 of which are the WCs of CP-odd bosonic operators and the rest are complex phases in the WCs of fermionic operators~\cite{Degrande:2021zpv,Bonnefoy:2021tbt}. Furthermore, CP-even dimension-6 operators can interfere with the SM Jarlskog invariant~\cite{Bonnefoy:2023bzx}. This produces a rich set of measurement targets and directs the design of specific CP-odd collider observables which could have been neglected without the SMEFT guidance.
Observing CPV effects in any of those processes would signify New Physics beyond doubt.

Several processes have been studied in this spirit, such as  $WW$, $WZ$, $t\bar{t}H$, $WH$, $ZH$, $H\to 4\ell$, VBF $H+jj$, $W\gamma$, $Zjj$, $tj$~\cite{Panico:2017frx,deBeurs:2018pvs,Banerjee:2019twi,Bishara:2020vix,ATLAS:2020nzk,Bhardwaj:2021ujv,Degrande:2021zpv,ElFaham:2024uop}. Higher-order perturbative corrections could alter the angular distributions that these analyses are based upon. Hence, a systematic study of next-to-leading order corrections in CPV observables is essential to match the expected experimental precision. A step in this direction was recently taken  in~\cite{ElFaham:2024uop}, where the authors study the impact of NLO QCD corrections in $WW/WZ$ production modified by the CP-even and CP-odd triple field strength operators, $O_{W}$ and $O_{\widetilde{W}}$, in angular differential measurements. Some of the novel computational tools used here are shared with that work.

In this work, we take another step towards the systematic inclusion of NLO QCD effects in SMEFT CPV studies by extending the \texttt{SMEFTatNLO} UFO with CPV bosonic operators and studying their effect in $WH/ZH$ production at the (HL-)LHC. The rest of this article is organized as follows. In Section~\ref{sec:SMEFTatNLO_ext}, we set our conventions, show which operators we have added to \texttt{SMEFTatNLO} and how it was done. Section~\ref{sec:HeliAmp} is dedicated to the study of CPV effects in $VH$ production. First, we analyse in detail the CP-odd SM interference at LO and then consider how NLO effects alter this picture for both $WH$ and $ZH$ production. We perform a phenomenological analysis of $WH$ at the (HL-)LHC
and obtain expected bounds on CPV effects from it in Section~\ref{sec:CPV_LHC}. Finally, we conclude in Section~\ref{sec:Conclusions}.

\section{CP-odd operator basis}
\label{sec:SMEFTatNLO_ext}

We use the dimension-6 operator basis implemented in \texttt{SMEFTatNLO}~\cite{Degrande:2020evl}, which consists of the Warsaw basis~\cite{Grzadkowski:2010es} restricted by the flavour symmetry U$(2)_q\times$U$(2)_u\times$U$(3)_d\times(\text{U}(1)_\ell\times\text{U}(1)_e)^3$. Additionally, the \texttt{SMEFTatNLO} basis redefines some 2- and 4-fermion operators via rotations, but these differences with the Warsaw basis are irrelevant for this work since we focus on bosonic operators. We list in Table~\ref{tab:Operator_basis} all the EW bosonic operators in our basis, where $\varphi$ is the Higgs doublet and we follow the \texttt{SMEFTatNLO} conventions~\cite{Degrande:2020evl}. We notice that is not necessary to subtract $\frac{v^2}{2}$ from $\varphi^{\dagger}\varphi$ for the CP-odd operators since their dimension-4 contribution is a total derivative without effect on perturbative physics.
\begin{table}[h!]
    \centering
    \begin{tabular}{c c | c c}
         \multicolumn{2}{c|}{CP even} & \multicolumn{2}{ c}{CP odd}\\
        \hline
         $\mcO_i$ & Definition & $\mcO_i$ & Definition \\
         \hline
         \rule{0pt}{1.15em} 
         $\mcO_W$ & $\varepsilon_{IJK} W^{I}_{\mu\nu} W^{J,\nu\rho} W^{K,\mu}_{\phantom{K,}\rho}$ & $\mcO_{\widetilde{W}}$ & $\varepsilon_{IJK} \widetilde{W}^{I}_{\mu\nu} W^{J,\nu\rho} W^{K,\mu}_{\phantom{K,} \rho}$ \\
         $\mcO_{\varphi B}$ & $\left(\varphi^{\dagger}\varphi -\frac{v^2}{2} \right) B_{\mu\nu} B^{\mu\nu} $ & $\mcO_{\varphi \widetilde{B}}$ & $\varphi^{\dagger}\varphi\, \widetilde{B}_{\mu\nu} B^{\mu\nu}$ \\
         $\mcO_{\varphi W}$ & $\left(\varphi^{\dagger}\varphi -\frac{v^2}{2} \right) W_{\mu\nu}^{I} W^{I,\mu\nu} $ & $\mcO_{\varphi \widetilde{W}}$ & $\varphi^{\dagger}\varphi\, \widetilde{W}_{\mu\nu}^{I} W^{I,\mu\nu}$ \\
         $\mcO_{\varphi WB}$ & $\left(\varphi^{\dagger}\tau^{I} \varphi \right) W_{\mu\nu}^{I} B^{\mu\nu} $ & $\mcO_{\varphi \widetilde{WB}}$ & $\varphi^{\dagger}\tau^{I}\varphi\, \widetilde{W}_{\mu\nu}^{I} B^{\mu\nu}$ \\
         $\mcO_{\varphi }$ & $\left(\varphi^{\dagger}\varphi-\frac{v^2}{2}\right)^3$ & - & - \\
         $\mcO_{\varphi d}$ & $\partial_{\mu}\left(\varphi^{\dagger}\varphi\right)\partial^{\mu}\left(\varphi^{\dagger}\varphi\right)$ & - & - \\
         $\mcO_{\varphi D}$ & $\left(\varphi^{\dagger}D^{\mu}\varphi\right)^{\dagger}\left(\varphi^{\dagger}D_{\mu}\varphi\right)$ & - & - \\
         
    \end{tabular}
    \caption{Bosonic dimension-6 SMEFT operators as implemented in \texttt{SMEFTatNLO}, where $\widetilde{V}_{\mu\nu}=\frac{\varepsilon_{\mu\nu\rho\sigma}}{2}V^{\rho\sigma}$ for $V=W,B$.}
    \label{tab:Operator_basis}
\end{table}

We focus on those operators that can induce CP-violating (CPV) effects, i.e.  $\mcO_{\widetilde{W}}$, $\mcO_{\varphi\widetilde{W}}$, $\mcO_{\varphi\widetilde{B}}$ and $\mcO_{\varphi\widetilde{W}B}$. 
How these operators must be renormalised can be inferred from the 1-loop anomalous dimension matrix~\cite{Jenkins:2013zja, Jenkins:2013wua, Alonso:2013hga}. At order $g_3^2$, where $g_3$ is the QCD gauge coupling, the EW bosonic operators do not run under QCD. 
Thus, the inclusion of the EW operators in \texttt{SMEFTatNLO} is straightforward since it does not require new counterterms.\footnote{The two additional operators built out of gluon fields, $\mcO_{\widetilde{G}}$ and $\mcO_{\varphi\widetilde{G}}$, require renormalization when computing at NLO in QCD and we leave their inclusion for future work. }
We built an extended version of \texttt{SMEFTatNLO} by simply adding their corresponding Feynman rules in the original \texttt{SMEFTatNLO}. We have cross-checked these Feynman rules against \texttt{SMEFTsim}~\cite{Brivio:2020onw} after taking into account the differences in the conventions between both models, see Appendix E in~\cite{Brivio:2020onw}.

\section{CP violation in $WH$ and $ZH$ production}
\label{sec:HeliAmp}

\subsection{Interference anatomy}
\label{sec:HeliAmp_Interf_Anato}

The operators $\mcO_{\varphi\widetilde{W}}$, $\mcO_{\varphi\widetilde{B}}$, and $\mcO_{\varphi W\widetilde{B}}$, generate CP-violating 
$VVH$, $VVHH$, and $VVVHH$ vertices, with $V=W,\, B$. These vertices follow the structure:
\begin{align}
    &\frac{c_{\varphi\widetilde{V}}}{\Lambda^2} \varphi^{\dagger} \varphi\, V_{\mu\nu}^{a} \widetilde{V}'^{\rho\sigma,a} \nonumber \\
    =&\frac{c_{\varphi\widetilde{V}}}{\Lambda^2} \left(h^2 + 2 h v + v^2\right) \varepsilon^{\mu\nu\rho\sigma} \partial_{\mu} V_{\nu}^{a} \partial_{\rho} V'{}_{\sigma}^{a} + \text{aTQGC},
\end{align}
where $a$ is the gauge group index and aTQGC collects the terms related to the non-abelian self-interactions which are irrelevant for the rest of this work.
This general expression can be specialized to the photon, $W$ and $Z$ bosons after EWSB as,
\begin{align}
    \left(2 \frac{\cpWt}{\Lambda^2}\right) & \left(h^2 + 2 h v + v^2\right) \varepsilon^{\mu\nu\rho\sigma} \partial_{\mu} W_{\nu}^{+} \partial_{\rho} W_{\sigma}^{-} \nonumber\\
    +\left(c_W^2 \frac{\cpWt}{\Lambda^2} + s_W^2 \frac{c_{\varphi\widetilde{B}}}{\Lambda^2}-\frac{s_{2W}}{2}\frac{c_{\varphi W\widetilde{B}}}{\Lambda^2}\right) & \left(h^2 + 2 h v + v^2\right) \varepsilon^{\mu\nu\rho\sigma} \partial_{\mu} Z_{\nu} \partial_{\rho} Z_{\sigma} \nonumber\\
    -\left(s_{2W} \left(\frac{\cpWt}{\Lambda^2} -\frac{c_{\varphi\widetilde{B}}}{\Lambda^2} \right)+c_{2W} \frac{c_{\varphi W\widetilde{B}}}{\Lambda^2} \right) & \left(h^2 + 2 h v + v^2\right) \varepsilon^{\mu\nu\rho\sigma} \partial_{\mu} Z_{\nu} \partial_{\rho} A_{\sigma} \nonumber\\
    +\left( s_W^2 \frac{\cpWt}{\Lambda^2} + c_W^2 \frac{c_{\varphi\widetilde{B}}}{\Lambda^2} + \frac{s_{2W}}{2}\frac{c_{\varphi W\widetilde{B}}}{\Lambda^2} \right) & \left(h^2 + 2 h v + v^2\right) \varepsilon^{\mu\nu\rho\sigma} \partial_{\mu} A_{\nu} \partial_{\rho} A_{\sigma},
\end{align}
where we have included the contribution of all the relevant dim.-6 operators.

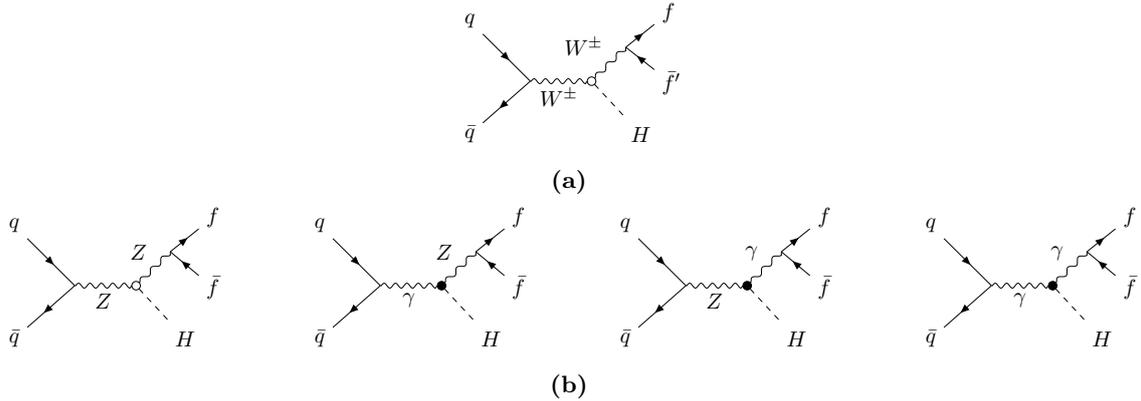
\begin{figure}[h!]
\centering
\begin{subfigure}{\textwidth}
\centering
\scalebox{0.75}{
\begin{tikzpicture}
\begin{feynman}
\vertex (a1) {$q$};
\vertex[below right = of a1] (v1);
\vertex[below=2cm of a1] (a2) {$\bar q$};
\vertex[empty dot, right=1cm of v1] (l1) {};
\vertex[above right=0.6cm and 0.6cm of l1] (l2);
\vertex[right=3cm of a2] (l3){$H$};
\vertex[below right=0.3cm and 0.5cm of l2](f3){$\bar{f}'$};
\vertex[above right=0.3cm and 0.5cm of l2](f4){$f$};
\diagram* {
(a1) -- [fermion, arrow size=1pt] (v1),
(a2) -- [anti fermion, arrow size=1pt] (v1) -- [photon, edge label'=$W^\pm$] (l1),
(l1) -- [photon, edge label=$W^\pm$] (l2),
(l1) -- [scalar] (l3),
(f3) -- [fermion, arrow size=1pt] (l2) -- [fermion, arrow size=1pt] (f4)
};
\end{feynman}
\end{tikzpicture}}
\caption{}
\label{fig:diags_WH_SMEFT}
\end{subfigure}
\hfill
\begin{subfigure}{1.0\textwidth}
\centering
\scalebox{0.75}{
\begin{tikzpicture}
\begin{feynman}
\vertex (a1) {$q$};
\vertex[below right = of a1] (v1);
\vertex[below=2cm of a1] (a2) {$\bar q$};
\vertex[empty dot, right=1cm of v1] (l1) {};
\vertex[above right=0.6cm and 0.6cm of l1] (l2);
\vertex[right=3cm of a2] (l3){$H$};
\vertex[below right=0.3cm and 0.5cm of l2](f3){$\bar{f}$};
\vertex[above right=0.3cm and 0.5cm of l2](f4){$f$};
\diagram* {
(a1) -- [fermion, arrow size=1pt] (v1),
(a2) -- [anti fermion, arrow size=1pt] (v1) -- [photon, edge label'=$Z$] (l1),
(l1) -- [photon, edge label=$Z$] (l2),
(l1) -- [scalar] (l3),
(f3) -- [fermion, arrow size=1pt] (l2) -- [fermion, arrow size=1pt] (f4)
};
\end{feynman}
\end{tikzpicture}}
\hfill
\scalebox{0.75}{
\begin{tikzpicture}
\begin{feynman}
\vertex (a1) {$q$};
\vertex[below right = of a1] (v1);
\vertex[below=2cm of a1] (a2) {$\bar q$};
\vertex[dot, right=1cm of v1] (l1) {};
\vertex[above right=0.6cm and 0.6cm of l1] (l2);
\vertex[right=3cm of a2] (l3){$H$};
\vertex[below right=0.3cm and 0.5cm of l2](f3){$\bar{f}$};
\vertex[above right=0.3cm and 0.5cm of l2](f4){$f$};
\diagram* {
(a1) -- [fermion, arrow size=1pt] (v1),
(a2) -- [anti fermion, arrow size=1pt] (v1) -- [photon, edge label'=$\gamma$] (l1),
(l1) -- [photon, edge label=$Z$] (l2),
(l1) -- [scalar] (l3),
(f3) -- [fermion, arrow size=1pt] (l2) -- [fermion, arrow size=1pt] (f4)
};
\end{feynman}
\end{tikzpicture}}
\hfill
\scalebox{0.75}{
\begin{tikzpicture}
\begin{feynman}
\vertex (a1) {$q$};
\vertex[below right = of a1] (v1);
\vertex[below=2cm of a1] (a2) {$\bar q$};
\vertex[dot, right=1cm of v1] (l1) {};
\vertex[above right=0.6cm and 0.6cm of l1] (l2);
\vertex[right=3cm of a2] (l3){$H$};
\vertex[below right=0.3cm and 0.5cm of l2](f3){$\bar{f}$};
\vertex[above right=0.3cm and 0.5cm of l2](f4){$f$};
\diagram* {
(a1) -- [fermion, arrow size=1pt] (v1),
(a2) -- [anti fermion, arrow size=1pt] (v1) -- [photon, edge label'=$Z$] (l1),
(l1) -- [photon, edge label=$\gamma$] (l2),
(l1) -- [scalar] (l3),
(f3) -- [fermion, arrow size=1pt] (l2) -- [fermion, arrow size=1pt] (f4) };
\end{feynman}
\end{tikzpicture}}
\hfill
\scalebox{0.75}{
\begin{tikzpicture}
\begin{feynman}
\vertex (a1){$q$};
\vertex[below right = of a1] (v1);
\vertex[below=2cm of a1] (a2) {$\bar q$};
\vertex[dot,right=1cm of v1] (l1) {};
\vertex[above right=0.6cm and 0.6cm of l1] (l2);
\vertex[right=3cm of a2] (l3){$H$};
\vertex[below right=0.3cm and 0.5cm of l2](f3){$\bar{f}$};
\vertex[above right=0.3cm and 0.5cm of l2](f4){$f$};
\diagram* {
(a1) -- [fermion, arrow size=1pt] (v1),
(a2) -- [anti fermion, arrow size=1pt] (v1) -- [photon, edge label'=$\gamma$] (l1),
(l1) -- [photon, edge label=$\gamma$] (l2),
(l1) -- [scalar] (l3),
(f3) -- [fermion, arrow size=1pt] (l2) -- [fermion, arrow size=1pt] (f4) };
\end{feynman}
\end{tikzpicture}}
\caption{}
\label{fig:diags_ZH_SMEFT}
\end{subfigure}
    \caption{Leading-order Feynman diagrams of $V(\to f\bar{f}')H$ production at hadron colliders, with $V=W^{\pm},\,Z$ in the Standard Model Effective Field Theory with bosonic dimension-6 operators that modify the $VV'H$ vertex. An empty dot means a vertex generated by the Standard Model and by dimension-6 operators, while a black dot means a vertex generated only by dimension-6 operators. (a): For $W^\pm(\to f\bar{f}') H$ production. (b): For $Z(\to f \bar{f})H$ production (only the first two diagrams contribute to the neutrino case). }
    \label{fig:diags_VH}
\end{figure}

We will consider the production of $V(\to f\bar{f}')H$ at hadron colliders, with $V=W^{\pm},\, Z$ from now on and $f$ an arbitrary SM fermion. The need to include the decay of the vector boson to fermions will become clear later. 
We show the leading-order diagram for $W^{\pm}(\to f\bar{f}')H$ in SMEFT with up to one insertion of bosonic dimension-6 operators that modify the $WWH$ vertex in Fig.~\ref{fig:diags_WH_SMEFT}. The presence of the dimension-6 operators under consideration does not introduce new topologies. Notice that among the operators of our interest only $\mcO_{\varphi\widetilde{W}}$ affects $W^{\pm}(\to f\bar{f}')H$ production. 

In Fig.~\ref{fig:diags_ZH_SMEFT}, we show the corresponding leading-order diagrams for $Z(\to f\bar{f})H$. In this case, the dimension-6 operators introduce three new topologies due to the appearance at tree-level of the vertices $Z\gamma H$ and $\gamma\gamma H$. The contribution of any of these additional diagrams can be enhanced or suppressed by applying suitable cuts on the decay products. In particular, requiring the invariant mass of $f\bar{f}$ to be at least $50$~GeV makes the diagram with two photons negligible. We will assume this cut during the rest of this work. Three of the Warsaw basis operators we consider, $\mcO_{\varphi\widetilde{W}}$, $\mcO_{\varphi\widetilde{B}}$, and $\mcO_{\varphi W\widetilde{B}}$, enter in $Z(\to f\bar{f})H$ production. They all generate similar effects, at most changing the relative contributions of the different diagrams in Fig.~\ref{fig:diags_ZH_SMEFT}. We will focus our study on the effects of $\mcO_{\varphi\widetilde{W}}$ as a representative case and to ease the comparison with $W^{\pm}(\to f\bar{f}')H$ production.

\begin{figure}
    \centering
    \includegraphics[width=0.6\textwidth]{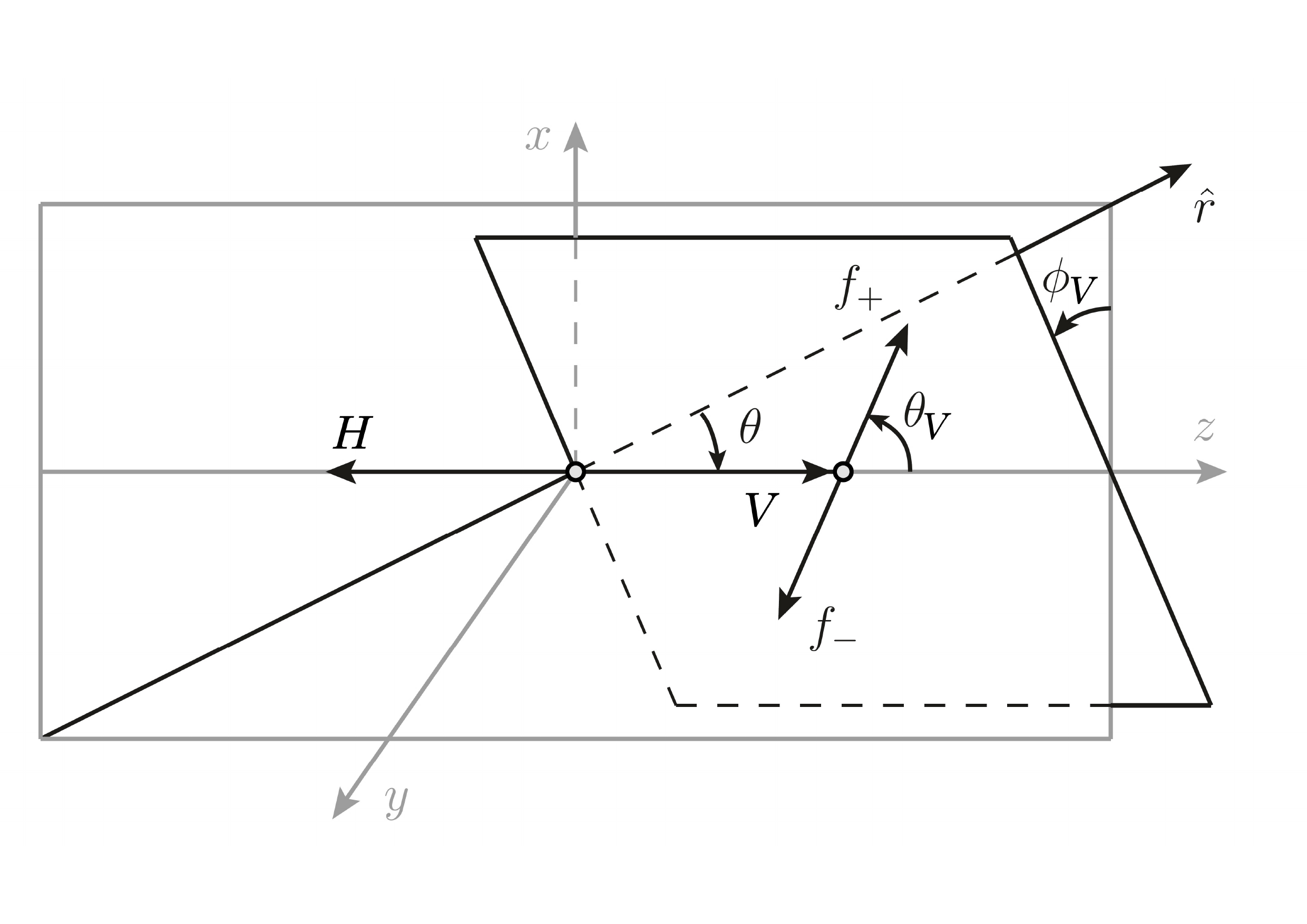}
    \caption{Angles that describe the $pp\to VH\to \bar{f} f H$ collision. The system of coordinates is in the $VH$ rest frame. $f_{+(-)}$ indicates a fermion of positive (negative) helicity. $\hat{r}$ indicates the direction of the boost from the lab frame to the $VH$ rest frame, with respect to which one measures the scattering angle $\theta$.
    $\theta_V$ and $\phi_V$ are the polar and azimuthal decay angle of the positive-helicity fermion produced in the decay of $V$ and are defined in the $V$ rest frame. 
    }
    \label{fig:def_angles}
\end{figure}

Probing unambiguously CP-odd effects in the $VH$ process requires a careful design of specific observables for two reasons. First, the interference between amplitudes generated by CP-odd operators and the SM vanishes in the fully inclusive case. Second, the dim-6 amplitude squared behaves, in general, in a similar way for the CP-odd operator and its CP-even counterpart. Hence, the distinction between them must rely on differential measurements that prevent the interference from vanishing. These differential measurements must split the phase space in portions that are not CP-eigenstates, which discards observables defined by cuts on P-symmetric kinematic variables such as $p_T$, centre-of-mass energy or invariant mass. Angular observables are the preferred remaining option and it is common to define asymmetry coefficients that measure the difference in cross-section between two disjoint angular regions. Not any angular observable is suitable though. In the case of diboson production, the scattering angle is insufficient and one must consider the angles that describe the decay of one of the final bosons. More precisely, the interference of a CP-odd amplitude with the SM is proportional to the sine of the azimuthal angle of the vector boson daughters~\cite{Panico:2017frx}, $\phi_V$, defined in Fig.~\ref{fig:def_angles}.

Thus,
we study the dim-6 interference with the SM including the decay of the vector boson to leptons.  
The choice of leptons over quarks obeys purely to their easier reconstruction at detectors, in particular the possibility of measuring their electric charge that can be used as a reference to define angles.
A convenient definition of the angles that describe the process is the one in Fig.~\ref{fig:def_angles}, as proposed in~\cite{Panico:2017frx,Azatov:2017kzw}.
In the $VH$ rest frame, $\hat{r}$ indicates the direction of the boost from the lab frame, the $z-$axis is defined along the $V$ momentum, the $q\bar{q}'\to VH$ collision occurs in the $x-z$ plane, and $\theta$ is the scattering angle of such process. 
The polar and azimuthal decay angles, $\theta_V$ and $\phi_V$ respectively, describe the decay of $V$ into fermions and are defined in the $V$ rest frame with respect to the fermion of positive helicity.
Notice that $\phi_V$ is also the angle between the plane defined by the $VH$ system and the $\hat{r}$ direction and the decay plane defined by the $V$ momentum and the momentum of its decay products in the $V$ rest frame.

The case of $WH$ production was previously discussed in~\cite{Banerjee:2019twi,Bishara:2020vix}. At high energies, the interference depends on the angles in a simple way\footnote{In our analytical computations, we define the scattering angle $\theta$ with respect to the incoming quark, which differs from the convention in~\cite{Bishara:2020vix}. Thus the scattering angle in both works are related as $\theta \to \theta-\pi$.},
\begin{equation}
    2\text{Re}(\mcM_{\text{SM}}^{*}\mcM_{1/\Lambda^2}) = - \cpWt \frac{ \sqrt{\hat s} \, m_W}{\Lambda^2} (1+\cos(\theta)\cos(\theta_W))\sin(\theta)\sin(\theta_W)\sin(\phi_W)+\mcO(\sqrt{\hat{s}}^0).
\end{equation}
Hence, a double binning on $p_T^H$ and $\phi_W$ gives a CP-odd probe that seizes on the growth with the energy of the interference. This is expected to be a simple and successful strategy at the future FCC-hh~\cite{Bishara:2020vix}. 
It is instructive to notice that the CP-even operator $\mcO_{\varphi W}$ generates an interference term that behaves similarly,
\begin{equation}
    2\text{Re}(\mcM_{\text{SM}}^{*}\mcM_{1/\Lambda^2}) = \cpW \frac{ \sqrt{\hat s} \, m_W}{\Lambda^2} (1+\cos(\theta)\cos(\theta_W))\sin(\theta)\sin(\theta_W)\cos(\phi_W)+\mcO(\sqrt{\hat{s}}^0).
\end{equation}
The only difference with the CP-odd case is in the dependence on the azimuthal decay angle $\phi_W$ and, therefore, being differential on it is the only way to probe genuine CP-odd effects.

\begin{figure}[th!]
    \centering
    \includegraphics[width=0.49\textwidth]{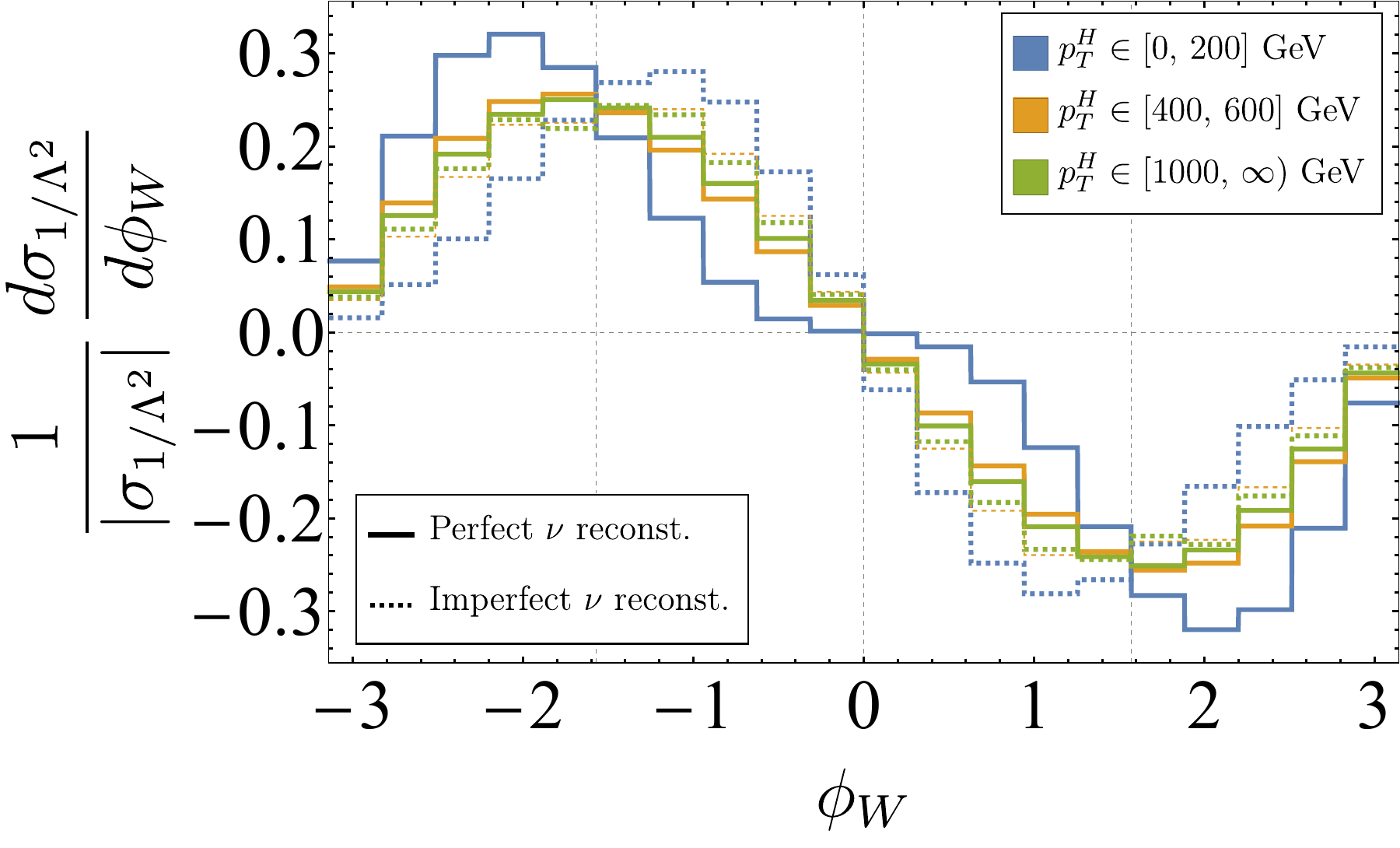}
    \hfill
    \includegraphics[width=0.49\textwidth]{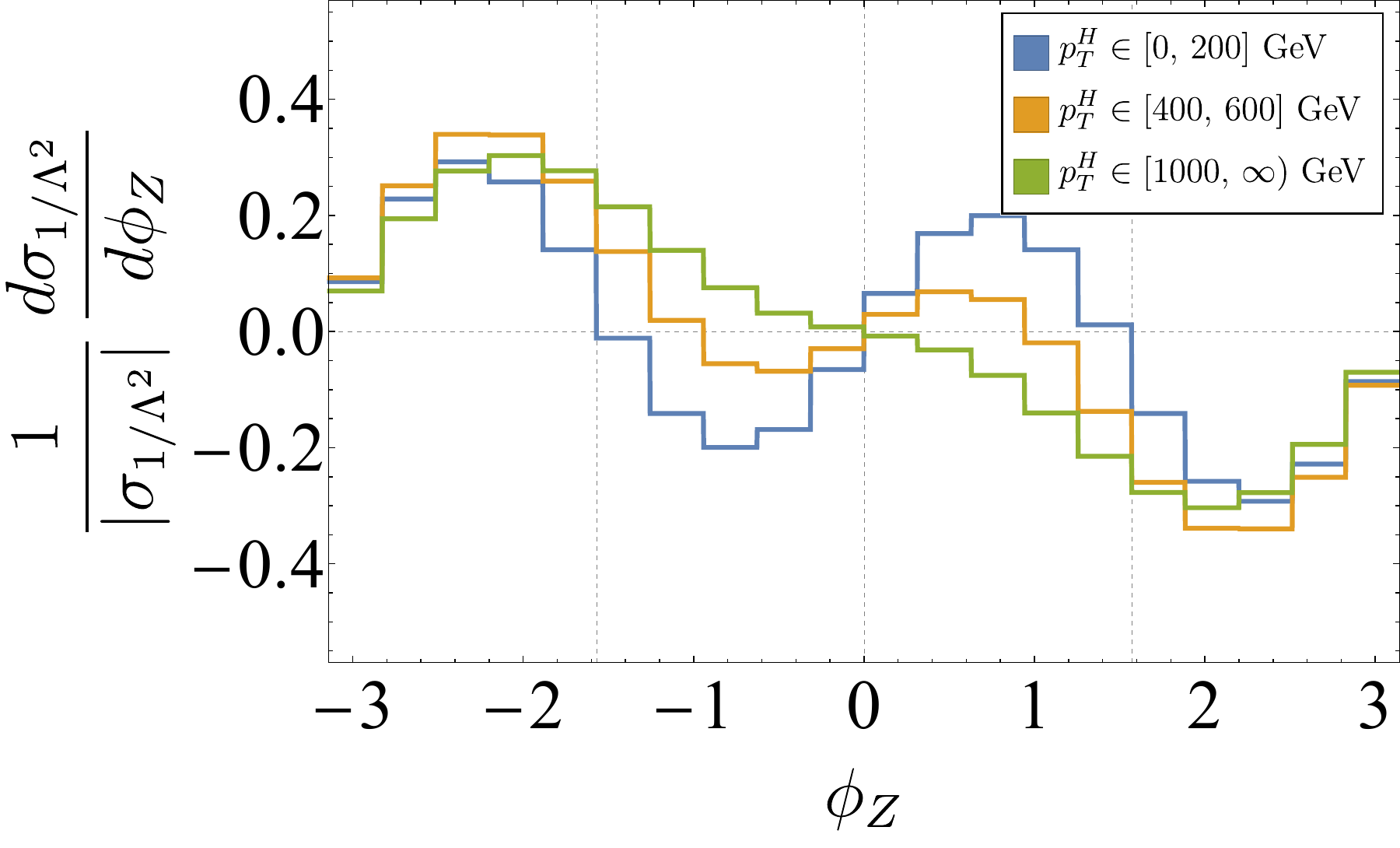}
    \caption{Angular distribution of the interference between the SM and $\mcO_{\varphi\widetilde{W}}$ for $pp\to VH\to \ell \bar{\ell}' H$ at the LHC ($\sqrt{s}=13$~TeV) and LO QCD in different $p_T^H$ bins. The normalization factor is the integral of the absolute value of the interference, $|\sigma_{1/\Lambda^2}|=\int |\frac{d\sigma_{1/\Lambda^2}}{d\phi_V}|d\phi_V$. \textbf{Left:} $W^-H$, where the full lines indicate the ideal case of being able to measure the four-momentum of the neutrino, while the dashed indicates the case of reconstructing it with the criterion explained in the text. \textbf{Right:} $ZH$.
}
    \label{fig:LO_angular_distr_ZH}
\end{figure}

At the lower energies explored at the LHC, the angular dependence shows two distinct modes~\cite{Banerjee:2019twi}.
This is clearer after integrating the interference over the polar decay and scattering angles and keeping its full dependence on $\sqrt{\hat s}$,
\begin{equation}
    \int 2 \text{Re}(\mcM_{\text{SM}}^{*}\mcM_{1/\Lambda^2}) \text{dc}(\theta)\text{dc}(\theta_W) = \frac{\cpWt}{\Lambda^2}\frac{g^4 m_W}{36 \Gamma_W^2}\sqrt{\hat{s}}( f_{1}^{W} \sin(\phi_W) + f_{2}^{W} \frac{m_W}{\sqrt{\hat{s}}} \sin(2\phi_W) ) ,
    \label{eq:interf_WH_cpwtilde_ideal}
\end{equation}
where dc$(\alpha) = \text{d}\cos(\alpha)$~and~$f_{1,2}^{W}$ are functions of $\frac{m_W}{\sqrt{\hat s}}$, $\frac{m_H}{\sqrt{\hat s}}$ and $\Gamma_W$ that tend to a constant for $\hat{s}\to\infty$\footnote{This expression is valid for either charge of the $W$ and its decay products.}. Despite their different energy behaviour, the $\sin(\phi_W)$ mode dominates at all energies, as seen in the left panel of Fig.~\ref{fig:LO_angular_distr_ZH}, since $|f_1^{W}/f_{2}^{W}| = \frac{9\pi^2}{64}\left(1-\frac{m_H^2-m_W^2}{\hat{s}}\right)\geqslant\frac{9\pi^2}{64}\left( 2\frac{m_W}{m_H+m_W}\right)> 1$ and $\sqrt{\hat{s}}\geqslant m_H + m_W$. 
The interference vanishes after integrating over $\phi_W$, regardless of the treatment of the other kinematical variables, thus making $\phi_W$ the only CP-sensitive observable in $WH$ production.

The experimentally accessible angular distribution for $WH$ differs from the one shown in Eq.~\eqref{eq:interf_WH_cpwtilde_ideal} due to the impossibility of measuring the neutrino 4-momenta at hadron colliders. 
One can reconstruct the neutrino momentum by requiring the $W$ boson to be as close to on-shell as possible. 
When the lepton transverse mass is smaller than the $W$ mass, $m_{T\ell\nu}<m_W$, there are two solutions for the neutrino momentum that ensure an on-shell $W$ boson, producing an ambiguity that reduces to $\phi_W\to \pi-\phi_W$ at high energies~\cite{Franceschini:2017xkh, Bishara:2020vix}. 
If $m_{T\ell\nu} \geqslant m_W$, the reconstructed $W$ is made as on-shell as possible by imposing $\eta_\nu = \eta_\ell$. The high-energy ambiguity does not affect the energy-leading piece in Eq.~\eqref{eq:interf_WH_cpwtilde_ideal} but it makes the subleading piece inaccessible since $\sin(2(\pi-x))=-\sin(2x)$. For comparison, in the case of the CP-even interference, the energy-leading piece vanishes under this ambiguity but the subleading one remains invariant. 

In the case of the ambiguous solution, one must pick one solution when performing a collider analysis. We choose the solution that maximises the invariant mass of the reconstructed $WH$ system, thus increasing the sensitivity of the analysis to the high centre-of-mass region. 
This reconstruction procedure distorts the differential angular distribution as shown in the left panel of Fig.~\ref{fig:LO_angular_distr_ZH} with dashed lines, whilst the full lines are obtained assuming one can measure the neutrino four-momentum. The difference is starker at the lower energies, where the $\sin(2\phi_W)$ mode, only captured by the perfect reconstruction, is most relevant. 
From now on, unless said otherwise, all the $WH$ distributions shown were obtained with this reconstruction procedure and its effect on NLO results can be found in Appendix~\ref{app:Angular_amb}.

Now, let us consider $q\bar{q}\to ZH\to \ell^{+}\ell^{-} H$ in the limit of massless fermions. 
Since the $Z$ boson couples to both left- and right-handed fermions, the azimuthal decay angle defined for a fermion of fixed helicity is not experimentally accessible. 
Hence, we choose to define the angle against the fermion of positive charge and comment on the effect of this choice below.
After integrating over the polar angles of the $Z$ and its decay products, the SM-$\mcO_{\varphi\widetilde{W}}$ interference acquires the same structure as in the $WH$ case,
\begin{equation}
    \int 2 \text{Re}(\mcM_{\text{SM}}^{*}\mcM_{1/\Lambda^2})\, \text{dc}(\theta)\text{dc}(\theta_Z) =\frac{\cpWt}{\Lambda^2}\frac{g^4 m_Z}{144 \,\Gamma_Z^2 c_{W}^2}\sqrt{\hat{s}}( f_{1}^{Z} \sin(\phi_Z) + f_{2}^{Z} \frac{m_Z}{\sqrt{\hat{s}}} \sin(2\phi_Z) ),
    \label{eq:interf_Zh_cpwtilde}
\end{equation}
where $f_{1,2}^{Z}$ are functions of $s_W$, $\frac{m_Z}{\sqrt{\hat{s}}}$, $\frac{m_H}{\sqrt{\hat{s}}}$ and $\Gamma_Z$ that behave like a constant in the high-energy limit. In fact,
\begin{align}
    f_1^{Z} \xrightarrow[\hat{s}\to\infty]{} & - 9 \pi^2 (1-2 |Q_q| s_{W}^2)(1-4 s_{W}^2),\\
    f_2^{Z} \xrightarrow[\hat{s}\to\infty]{} & 64 (1-2 |Q_q| s_{W}^2)(1-4s_{W}^2+8 s_{W}^4),
\end{align} 
where $|Q_q|$ is the charge of the initial quark. Each of these angular functions corresponds to one of the angular moments associated with CP-odd operators in~\cite{Banerjee:2019twi}.
As in the $WH$ case, the interference can be observed only with a measurement differential in the azimuthal decay angle and the angular distribution is different between the leading and sub-leading terms, with $\sin(2 \phi_Z)$ at lower energies and $\sin(\phi_Z)$ in the high-energy limit. Since $f_1^{Z}/f_{2}^{Z}\simeq 0.15 f_1^{W}/f_{2}^{W}$, there is an energy regime in which the $\sin(2\phi_Z)$ dominates, in contrast with $WH$. The transition between modes occurs at $p_T^H\gtrsim 1$~TeV, as seen in the right panel of Fig.~\ref{fig:LO_angular_distr_ZH}.

The general expression in Eq.~\ref{eq:interf_Zh_cpwtilde} is valid both when the angle $\phi_Z$ is defined for a fermion of fixed charge and when it is done with a fermion of fixed helicity. 
The function $f_2^{Z}$ is not affected by the angle definition since it is associated with the angular mode that is invariant under $\phi_Z\to\phi_Z+\pi$. On the other hand, $f_1^{Z}$ changes such that,
\begin{equation}
    f_1^{Z}|_{\text{ideal}} \xrightarrow[\hat{s}\to\infty]{} -9\pi^2 (1- 2 |Q_q| s_{W}^2)(1-4 s_{W}^2+8 s_{W}^4),
\end{equation}
where $|_{\text{ideal}}$ indicates that this is valid when the angle is defined with a fermion of negative helicity. 
Since $f_1^{Z}/f_1^{Z}|_{\text{ideal}}\simeq 1/7$ with almost no dependence on energy, defining the azimuthal angle for fixed charge harms the sensitivity to CP-odd operators. At the same time, the energy-dependent angular behaviour dictates the need for different angular binnings in different energy regions.

Another consequence of how we define the azimuthal decay angle can be seen by computing the interference with general $Z$ couplings. If the angle is defined for fixed helicity, one finds that,
\begin{align}
    f_{1,2}^{Z}|_{\text{ideal}} \propto  \left(\left(g_{Z,L}^{\ell}\right)^2+\left(g_{Z,R}^{\ell}\right)^2\right).
\end{align}
where $g_{Z,L(R)}^{\psi}$ is the coupling of the $Z$ to the left(right)-handed fermion $\psi$. However, if the angle is defined against a fermion of fixed charge, both functions behave differently,
\begin{align}
    f_{1}^{Z}\propto&\left(g_{Z,L}^{q}-g_{Z,R}^{q}\right)\left(\left(g_{Z,L}^{\ell}\right)^2-\left(g_{Z,R}^{\ell}\right)^2\right),\\
f_{2}^{Z}\propto&\left(\left(g_{Z,L}^{\ell}\right)^2+\left(g_{Z,R}^{\ell}\right)^2\right),
\end{align}
hence the energy-growing $\sin(\phi_Z)$ mode is only accessible for a chirally-coupled $Z$ boson. Thus, the inability to measure the final fermion helicity reduces our sensitivity to CP-odd effects but makes the $\sin(\phi_{Z})$ mode a hallmark of the chiral couplings of the $Z$ boson.

Finally, we note that the interference between the $\mcO_{\varphi\widetilde{W}}$ diagrams with a virtual photon and the SM survives the integration over the decay angles and vanishes only after averaging over the direction of the incoming quark, i.e. over $\theta$ and $\theta+\pi$. Thus, there is no possibility of defining a CP-odd observable at hadron colliders being inclusive in the decay of the $Z$ boson. 

\subsection{NLO corrections at fixed order}
\label{sec:NLO_FO}

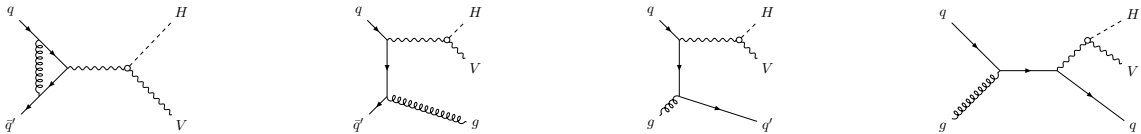
\begin{figure}[h!]
    \centering
\scalebox{0.5}{
\begin{tikzpicture}
\begin{feynman}
\vertex (a1) {$q$};
\vertex[below right = 0.75cm and 0.75cm of a1] (v1);
\vertex[below=1.5cm of v1] (v2); 
\vertex[below=3cm of a1] (a2) {$\bar q'$};
\vertex[below right=0.75cm and 0.75cm of v1] (v3);
\vertex[empty dot, right=1.5cm of v3] (v4) {};
\vertex[right=4.5cm of a1] (f1){$H$};
\vertex[right=4.5cm of a2] (f2){$V$};
\diagram* {
(a1) -- [fermion, arrow size=1pt] (v1),
(v1) -- [fermion, arrow size=1pt] (v3),
(a2) -- [anti fermion, arrow size=1pt] (v2) -- [anti fermion, arrow size=1pt] (v3),
(v1) -- [gluon] (v2),
(v3) -- [photon] (v4),
(v4) -- [photon] (f2),
(v4) -- [scalar] (f1)
};
\end{feynman}
\end{tikzpicture}}
\hfill
\scalebox{0.5}{
\begin{tikzpicture}
\begin{feynman}
\vertex (a1) {$q$};
\vertex[below=3cm of a1] (a2) {$\bar{q}'$};
\vertex[below right=0.75cm and 0.75cm of a1] (v1);
\vertex[below=1.5cm of v1] (v2);
\vertex[right=1.5cm of v1, empty dot] (v3) {};
\vertex[above right=0.75cm and 0.75cm of v3] (f1) {$H$};
\vertex[below=1.5cm of f1] (f2) {$V$};
\vertex[below=1.5cm of f2] (f3) {$g$};
\diagram* {
(a1) -- [fermion, arrow size=1pt] (v1) -- [fermion, arrow size=1pt] (v2) -- [fermion, arrow size=1pt] (a2),
(v1) -- [photon] (v3),
(v3) -- [photon] (f2),
(v2) -- [gluon] (f3),
(v3) -- [scalar] (f1) };
\end{feynman}
\end{tikzpicture}}
\hfill
\scalebox{0.5}{
\begin{tikzpicture}
\begin{feynman}
\vertex (a1) {$q$};
\vertex[below=3cm of a1] (a2) {$g$};
\vertex[below right=0.75cm and 0.75cm of a1] (v1);
\vertex[below=1.5cm of v1] (v2);
\vertex[right=1.5cm of v1, empty dot] (v3) {};
\vertex[above right=0.75cm and 0.75cm of v3] (f1) {$H$};
\vertex[below=1.5cm of f1] (f2) {$V$};
\vertex[below=1.5cm of f2] (f3) {$q'$};
\diagram* {
(a1) -- [fermion, arrow size=1pt] (v1) -- [fermion, arrow size=1pt] (v2) -- [gluon] (a2),
(v1) -- [photon] (v3),
(v3) -- [photon] (f2),
(v2) -- [fermion, arrow size=1pt] (f3),
(v3) -- [scalar] (f1) };
\end{feynman}
\end{tikzpicture}}
\hfill
\scalebox{0.5}{
\begin{tikzpicture}
\begin{feynman}
\vertex (a1) {$q$};
\vertex[below=3cm of a1] (a2) {$g$};
\vertex[below right = 1.5cm and 1.5cm of a1] (v3);
\vertex[ right=1.5cm of v3] (v4);
\vertex[above right = 0.75cm and 0.75cm of v4,empty dot] (v5) {};
\vertex[right=5cm of a1] (f1){$H$};
\vertex[below=1.5cm of f1] (f2){$V$};
\vertex[right=5cm of a2] (f3){$q$};
\diagram* {
(a1) -- [fermion, arrow size=1pt] (v3) -- [fermion, arrow size=1pt] (v4) -- [fermion, arrow size=1pt] (f3),
(a2) -- [gluon] (v3),
(v4) -- [photon] (v5),
(v5) -- [photon] (f2),
(v5) -- [scalar] (f1)};
\end{feynman}
\end{tikzpicture}}
    \caption{Feynman diagrams of $q\bar{q}'\to VH$ at NLO in QCD. The vertex marked with an empty dot can be a SM vertex or an insertion of a bosonic dimension-6 operator.}
    \label{fig:feyn_Wh_1loop}
\end{figure}

NLO QCD corrections enter these processes in two ways: via virtual QCD corrections to the $q\bar{q} V$ vertex and via real emission. The latter also implies the opening of new production channels such as $q g\to V H j$. We show in Figure~\ref{fig:feyn_Wh_1loop} all the possible NLO topologies in the SM and for one insertion of the bosonic operators that modify the $VV'H$ vertex. The inclusion of such kind of dimension-6 operators does not generate new topologies. When the final vector boson is a $W$ boson, the virtual vector boson must also be a $W$. If the final vector boson is a $Z$ and one considers the $VVH$ vertex at $\mcO(1/\Lambda^2)$, the intermediate vector boson can be a photon. 
 
\begin{figure}
    \centering
    \includegraphics[width=0.475\textwidth]{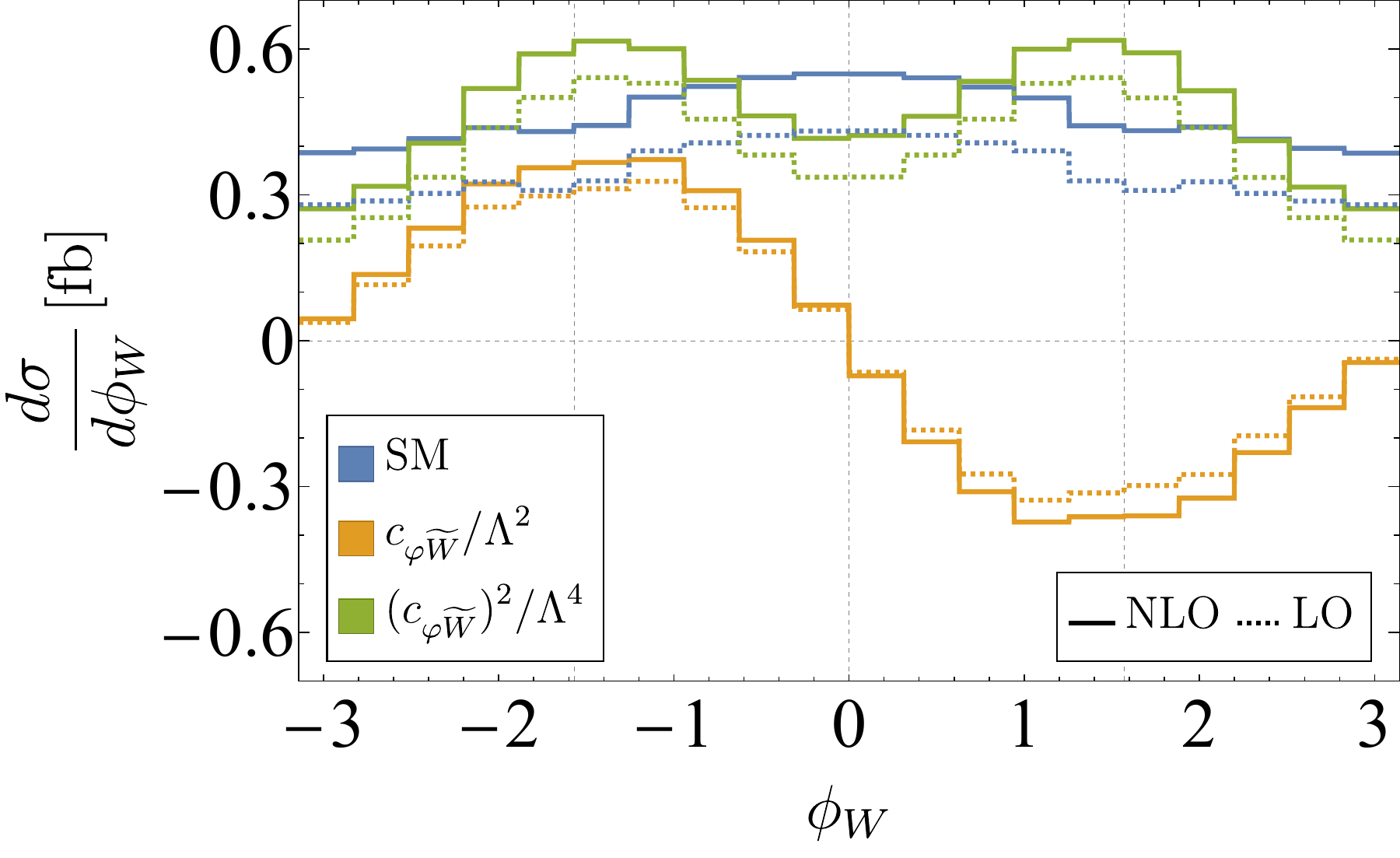}
    \hfill
    \includegraphics[width=0.475\textwidth]{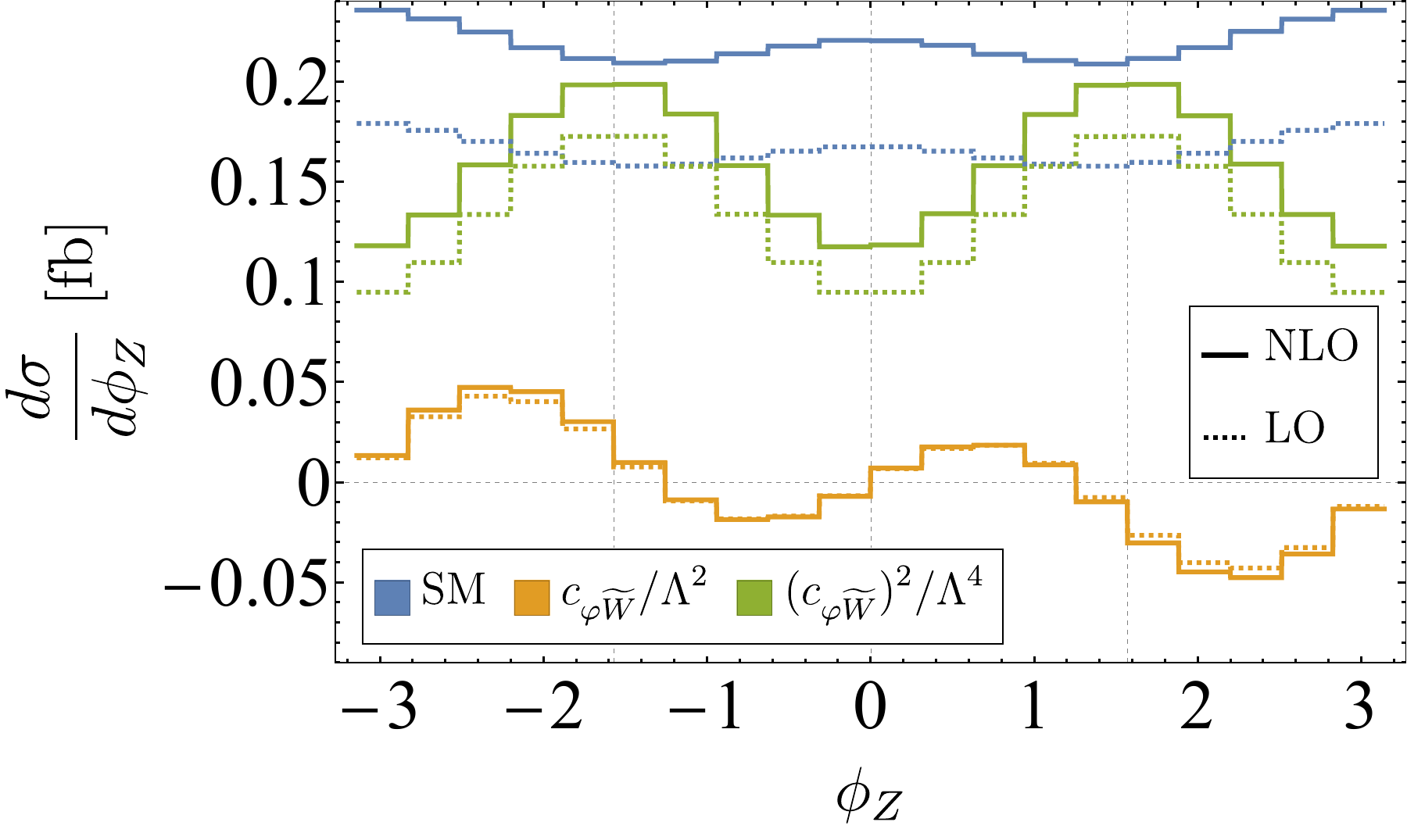}
    \caption{Angular distribution for $pp\to VH\to \ell \ell' H$ at the LHC in the $p_T^H\in [200,400]$~GeV bin when including the effect of $\mcO_{\varphi\widetilde{W}}$. We show separately the SM (blue), SM-EFT interference ($\cpWt/\Lambda^2$, orange), and EFT squared ($(\cpWt)^2/\Lambda^4$, green) pieces, at LO (dashed) and NLO (full) in QCD. \textbf{Left:} $W^{-} H\to e^{-} \bar{\nu}_{e} H$. 
    \textbf{Right:} $ZH\to e^{+}e^{-} H$.
    }
    \label{fig:ZH_cpwtilde_ang_distr_all}
\end{figure}

The presence of NLO corrections could distort the angular distributions of the SM and the ones generated by the CP-odd operators and thus impact the sensitivity to CP-odd effects.
In Fig.~\ref{fig:ZH_cpwtilde_ang_distr_all}, we show the differential distribution with respect to the azimuthal decay angle $\phi_V$ for the processes $pp \to VH \to \ell \ell' H$ in presence of the CP-odd $\mcO_{\varphi\widetilde{W}}$ operator and in the bin $p_{T}^{H}\in[200,\,400]$~GeV. We show separately the SM contribution, the purely EFT $\mcO(\Lambda^{-4})$ contribution, and their interference at LO and NLO in QCD and for $V=W^-,\,\, Z$.
The NLO corrections do not alter significantly the shape of the angular distributions but change their overall normalisation, which impacts the relative importance of the different contributions. $W^{-} H$ shows bigger NLO effects than $ZH$ in the interference term, while the corrections to the SM and EFT squared pieces are similar for both processes. 
In both processes, the interference integrates to zero and has a well-defined shape. It approximates $\sin(\phi_W)$ for $WH$ and $\sin(2\phi_Z)$ in the case of $ZH$. 
For the latter, notice that the selected $p_{T}^{H}$ bin is dominated by the low $\hat s$ regime.
For completeness, we also show the angular distributions generated by the CP-even operator $\mcO_{\varphi W}$ in $W^{-} H$ production in Appendix~\ref{app:app_cp_even_wh}.

\begin{figure}[h!]
    \centering
    \includegraphics[width=0.475\textwidth]{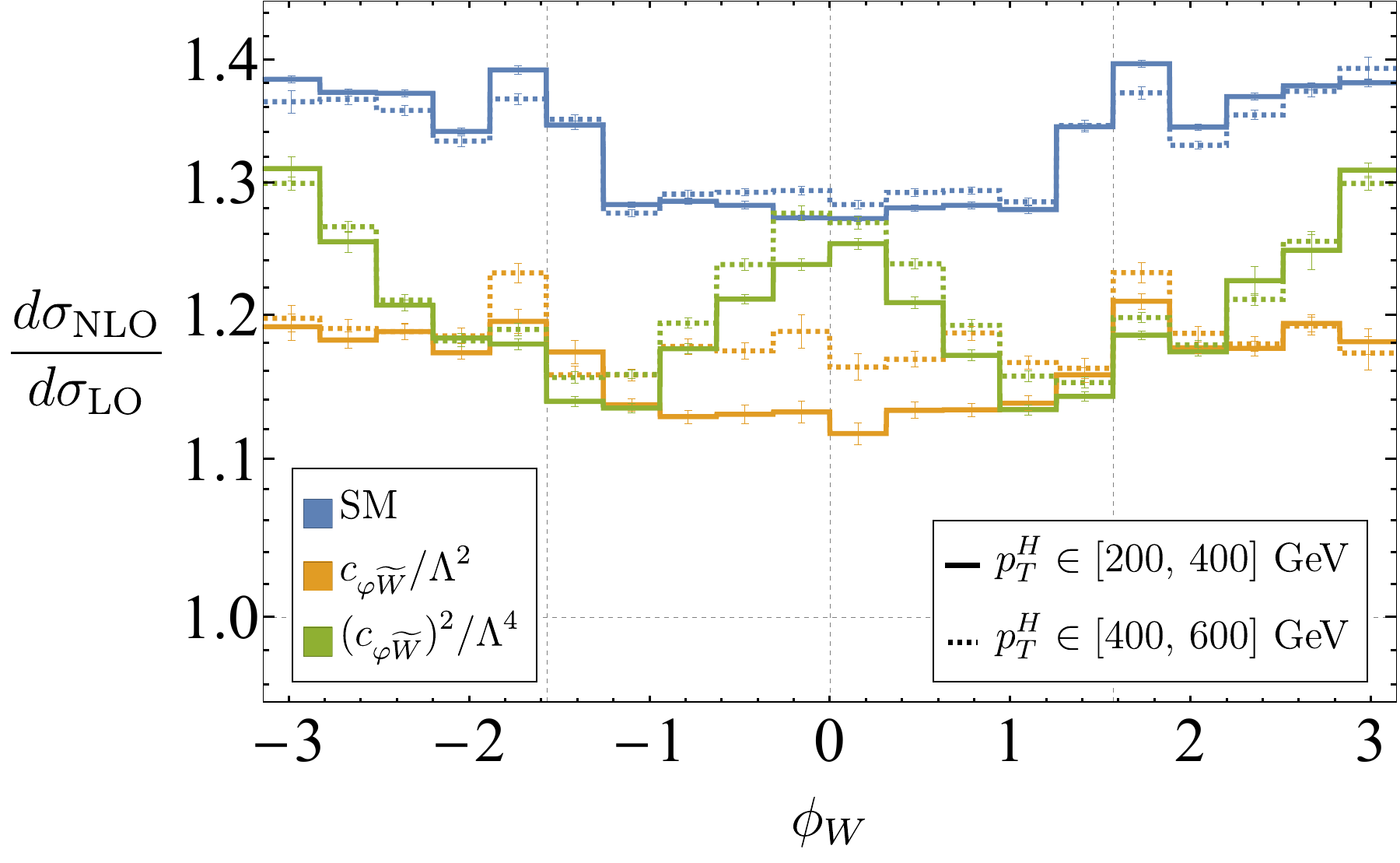}
    \hfill
    \includegraphics[width=0.475\textwidth]{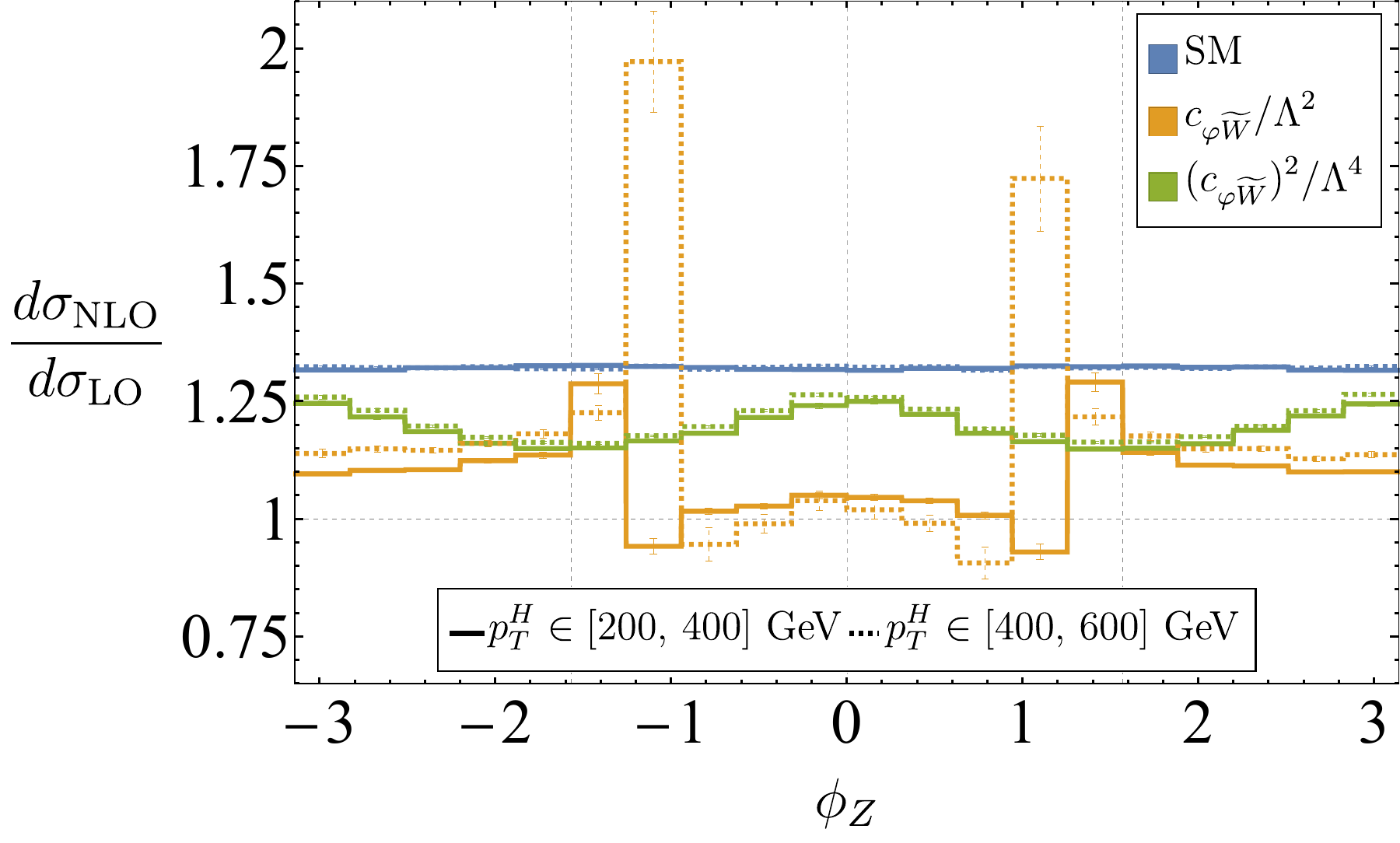}
    \caption{NLO/LO ratio of the angular differential cross section for the SM, interference and EFT squared pieces of the operator $\mcO_{\varphi \widetilde{W}}$. The full (dashed) lines correspond to the $p_T^H\in [200,400]$~GeV ($[400,600]$~GeV) bin. The error bars show the Monte Carlo uncertainty and the vertical dashed lines are located at $\phi_{W/Z}=0,\,\pm\frac{\pi}{2}$. \textbf{Left:} $WH$, with the azimuthal decay angle $\phi_W$. \textbf{Right:} $ZH$, with the azimuthal decay angle $\phi_Z$. 
    }
    \label{fig:ang_dep_kF}
\end{figure}

Another way of studying these corrections is by computing the NLO/LO $k$-factors for each piece of the cross-section and in different energy regimes, which we show in Fig.~\ref{fig:ang_dep_kF}. The SM $k$-factor shows a small dependence on angle and energy and oscillates around $\sim 1.33$ for both $WH$ and $ZH$ production. The interference shows different features between the processes. In $WH$, its angular dependence is similar to the one of the SM while it changes more with energy. The interference in $ZH$ production shows a stronger dependence on the angle and the energy.
The SM-$\mcO_{\varphi\widetilde{W}}$ interference in $ZH$ shows a $k$-factor below 1 for $|\phi_Z|\lesssim \frac{\pi}{2}$ and a high $k$-factor for $\phi_Z \simeq \frac{\pi}{2}$. 
This occurs because the LO and NLO contributions change sign at different points and hence the NLO correction suppresses the interference in certain regions.
The $k$-factor for the EFT squared pieces depends strongly on the angle in a similar way for both $WH$ and $ZH$ production.

The non-trivial dependence of the NLO corrections on the angular variables, both in $WH$ and $ZH$ stresses the importance of including full NLO effects since they might not be captured accurately by inclusive $k$-factors. Furthermore, the difference in the $k$-factors between the interference and EFT squared pieces remarks how inaccurate a SM $k$-factor approach could be. 

\begin{figure}[h!]
    \centering
    \includegraphics[width=0.475\textwidth]{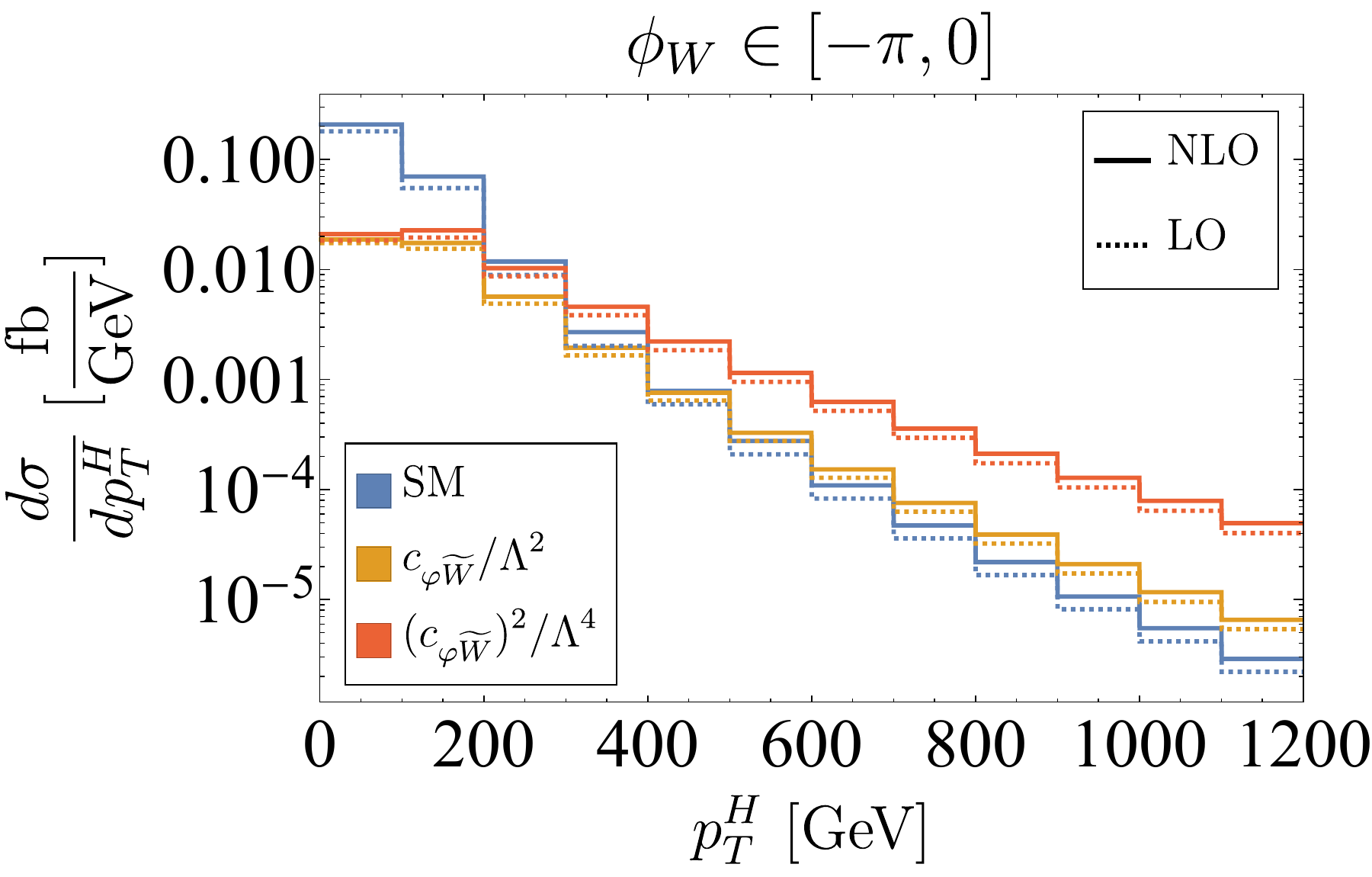}
    \hfill
    \includegraphics[width=0.475\textwidth]{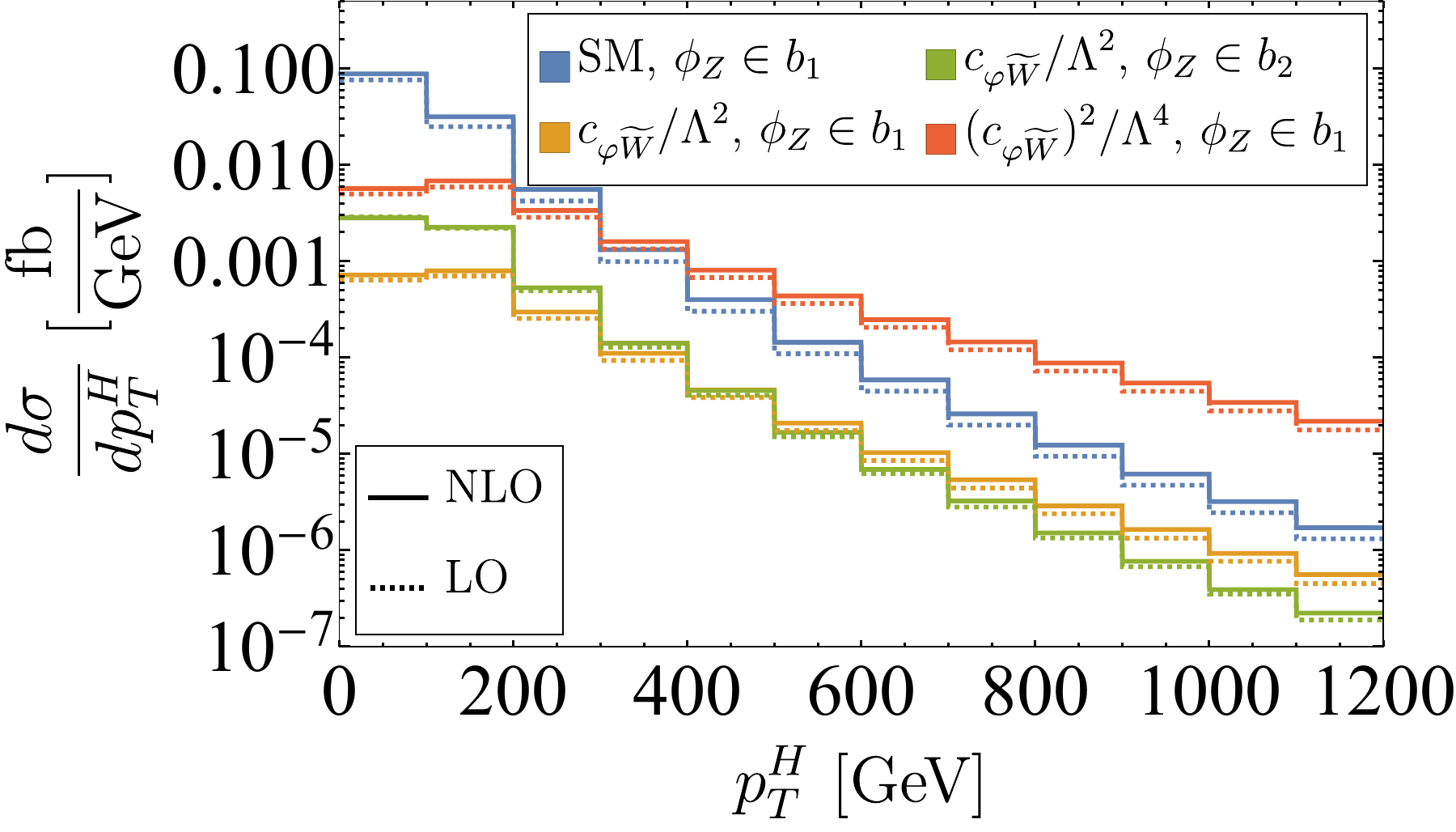}
    \caption{$p_T^H$ distribution integrating over regions where the interference is positive. We show separately the SM (blue), interference (orange or green) and EFT squared (red) contributions. \textbf{Left:} $WH$, with the azimuthal decay angle $\phi_W\in[-\pi,\,0]$. \textbf{Right:} $ZH$, with the azimuthal decay angle $\phi_Z$. In this case, for the interference, we show the result of two possible binning choices, where $b_1=[-\pi,0]$ and $b_2=[ -\pi, -\frac{\pi}{2} ] \cup [0,\frac{\pi}{2}]$. For the SM and EFT squared contributions, both choices produce identical results.}
    \label{fig:pth_distr_SM_int}
\end{figure}

\begin{figure}[h!]
    \centering
    \includegraphics[width=0.6\textwidth]{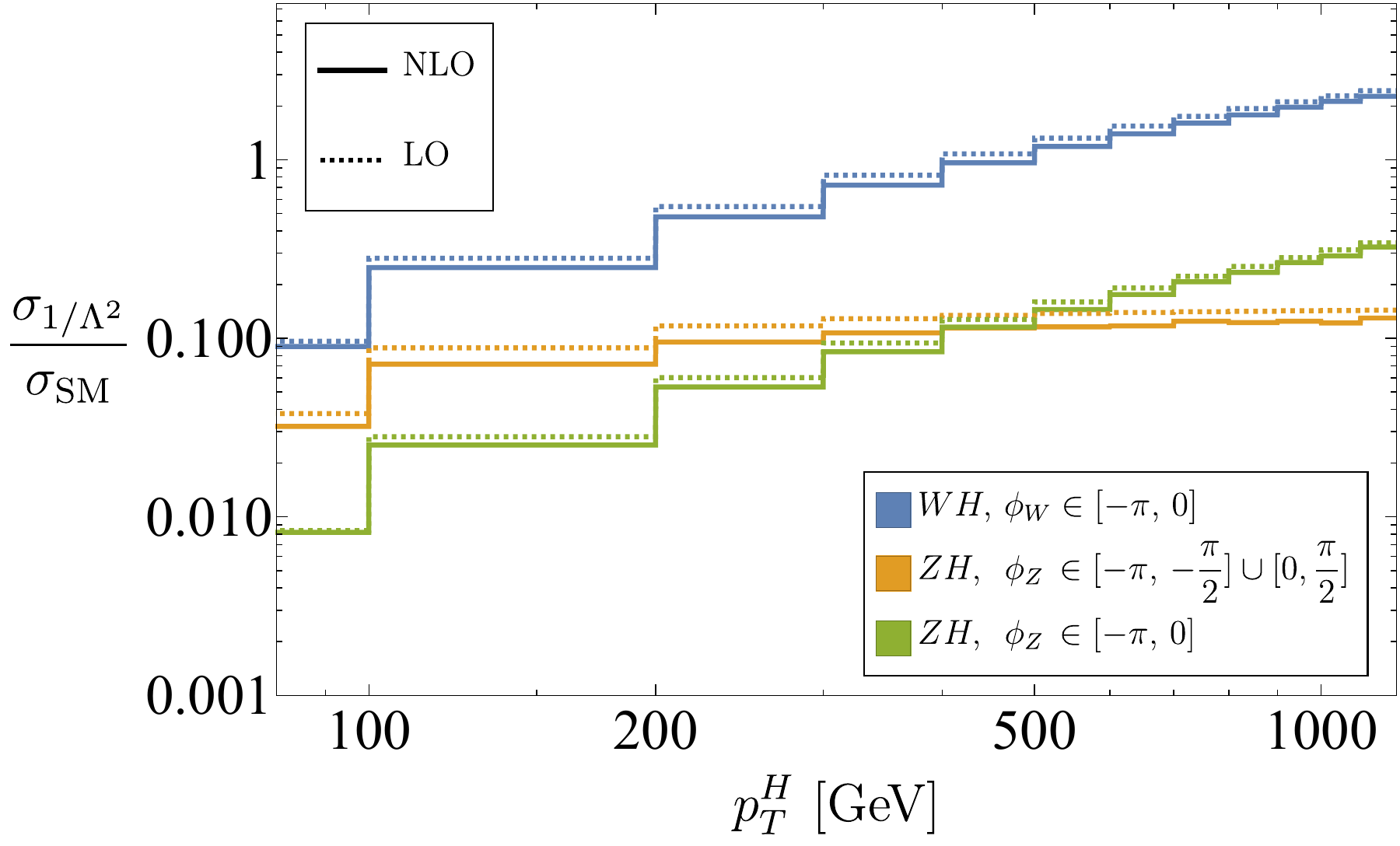}
    \caption{Ratio of interference over SM as a function of $p_T^H$ for $WH$ (blue) and $ZH$ (orange and green) at LO (dashed) and NLO (full line) in angular bins where the interference is positive. For $ZH$, there are two such bins: $\phi_Z\mod{ \pi } \in [0,\frac{\pi}{2}]$ (orange) and $\phi_Z\in[-\pi,\,0]$ (green).}
    \label{fig:pth_distr_ratio_int_SM}
\end{figure}

The differential distributions also confirm the need for angular binning to observe the energy growth in the interference term and therefore we study the $p_T^H$ distribution in different angular bins. We show in Fig.~\ref{fig:pth_distr_SM_int} the $p_T^H$ distribution of the SM, interference, and EFT squared pieces for $WH$ and $ZH$ in the angular bins where the interference is positive. For $WH$, this bin is simply defined as $\phi_W<0$. 
As it can be seen on Fig.~\ref{fig:pth_distr_ratio_int_SM}, the interference grows linearly with energy with respect to the SM amplitude squared. Even at low energies, $p_T^H \lesssim 300$~GeV, the interference is sizeable, which justifies why this process is so promising for CPV studies at the LHC. 
NLO QCD corrections just increase the cross-section in all cases without changing the energy behaviour. In addition, NLO QCD suppresses the ratio interference over SM as shown in Fig.~\ref{fig:pth_distr_ratio_int_SM}.

On the other hand, for $ZH$ two possible angular bins yield positive interference. One is the bin $\phi_Z \in [-\pi,\,\frac{\pi}{2}]\cup[0,\frac{\pi}{2}]$ and the other means integrating $\phi_Z$ over $[-\pi,0]$. As discussed in Section~\ref{sec:HeliAmp_Interf_Anato}, each option is justified by the dominant angular distribution at either low or high energies and produces equivalent results for the SM and EFT squared pieces of the cross-section.
The bin $\phi_Z \in [-\pi,\,\frac{\pi}{2}]\cup[0,\frac{\pi}{2}]$ renders a bigger interference when $p_T^H \lesssim 500$~GeV. Above that energy, the greater interference is given by the bin $\phi_Z\in[-\pi,0]$. Moreover, the latter shows a distinctive energy behaviour since it picks the energy-growing piece of the interference. This is shown clearly in Fig.~\ref{fig:pth_distr_ratio_int_SM}, where we plot the interference over SM ratio as a function of $p_T^H$ for each bin choice and at different QCD orders. While the ratio corresponding to $\phi_Z\in[-\pi,0]$ grows linearly with energy, the one for $\phi_Z \in [-\pi,\,\frac{\pi}{2}]\cup[0,\frac{\pi}{2}]$ initially grows and then plateaus since both interference and SM amplitudes have the same behaviour at high energies. 
NLO QCD effects do not change these behaviours but reduce the relative size of the interference, as expected from the previous discussion. 
 
The interference for $ZH$ is much smaller than the SM contribution in all cases, differently from the $WH$ case. The different energy behaviours for different bin choices justify the use of a 4-bins strategy for $ZH$, with each bin of width $\pi/2$, however, this would be inherently limited by statistics. Therefore, $ZH$ is not a promising channel to study CPV effects at the LHC and could only be efficiently exploited with sophisticated analysis techniques and/or at future colliders.

\begin{table}[h!]
    \centering
\scalebox{0.6}{
    \begin{tabular}{c|c|c|c|c|c|c|c|}
     \multicolumn{2}{c|}{$\sigma$ [fb]}    &  \multicolumn{2}{c|}{SM} & \multicolumn{2}{c|}{$\mcO\left(\Lambda^{-2}\right)$} & \multicolumn{2}{c|}{$\mcO\left(\Lambda^{-4}\right)$} \\
     \cline{1-8}
     $p_T^H\,[\text{GeV}] \in$ & $\phi_W $    & LO & NLO & LO & NLO & LO & NLO \\
     \hline
     \multirow{2}{*}{$[0,\,200]$} 
     & $\phi_W<0$ & $23.525(7)$ & $27.85(6)$ & $3.055(3)$ & $3.46(2)$ & $3.7887(8)$ & $4.379(8)$ \\
     \cline{2-8}
     & $\phi_W>0$ & $23.510(7)$ & $27.83(6)$ & $-3.054(3)$ & $-3.45(2)$ & $3.7881(8)$ & $4.36(1)$  \\
     \hline
     \hline
     \multirow{2}{*}{$[200,\,400]$} 
     & $\phi_W<0$ & $1.0958(1)$ & $1.4529(9)$ & $6.544(2)\times10^{-1}$ & $7.73(1)\times10^{-1}$ & $1.2510(2)$ & $1.489(1)$  \\
     \cline{2-8}
     & $\phi_W>0$ & $1.0960(1)$ & $1.452(1)$ & $-6.548(2)\times10^{-1}$ & $-7.73(1)\times10^{-1}$ & $1.2509(2)$ & $1.489(2)$ \\
     \hline
     \hline
     \multirow{2}{*}{$[400,\,\infty)$} 
     & $\phi_W<0$ & $9.575(1)\times 10^{-2}$ & $1.268(1)\times 10^{-1}$ & $1.1920(3)\times10^{-1}$ & $1.434(2)\times10^{-1}$ & $4.0654(6)\times10^{-1}$ & $4.908(5)\times10^{-1}$ \\
     \cline{2-8}
     & $\phi_W>0$ & $9.577(1)\times 10^{-2}$ & $1.270(1)\times 10^{-1}$ & $-1.1919(3)\times10^{-1}$ & $-1.435(2)\times10^{-1}$ & $4.0667(6)\times10^{-1}$ &    $4.914(5)\times10^{-1}$
    \end{tabular}
    }
    \caption{Cross section (in fb) of $pp\to e^{-}\, \bar{\nu}_e\, h$ at $\sqrt{s}=13$~TeV in the SM and when affected by $\mcO_{\varphi \widetilde{W}}$. }
    \label{tab:pp_WH_xs_allorders}
\end{table}

We finalise this section by showing the cross-section of $VH$ at different QCD and EFT orders and in various energy and angular bins in Table~\ref{tab:pp_WH_xs_allorders} and Table~\ref{tab:pp_ZH_xs_allorders}. In both cases, one can verify the growth with energy of the interference with respect to the SM by taking the ratio between the columns $\mcO(\Lambda^{-2})$ and SM. One can also observe the $k$-factor $<1$ for the interference in $ZH$ production discussed before by comparing the LO and NLO columns in the bins $\phi_Z\in[-\frac{\pi}{2},0],\,\,[0,\frac{\pi}{2}]$ and $p_T^H\in[0,200],\,\,[200,\,400]$~GeV.

\begin{table}[]
    \centering
    \scalebox{0.6}{
    \begin{tabular}{c|c|c|c|c|c|c|c|}
     \multicolumn{2}{c|}{$\sigma$ [fb]}    &  \multicolumn{2}{c|}{SM} & \multicolumn{2}{c|}{$\mcO\left(\Lambda^{-2}\right)$} & \multicolumn{2}{c|}{$\mcO\left(\Lambda^{-4}\right)$} \\
     \cline{1-8}
      $p_T^H\,[\text{GeV}] \in$ & $\phi_Z\in$ & LO & NLO & LO & NLO & LO & NLO \\
     \hline
     \multirow{4}{*}{$[0,\,200]$}   
     & $[-\pi,-\frac{\pi}{2}]$ & $5.1887(8)$ & $6.131(8)$ & $  \phantom{-9} 3.211(1) \times10^{-1}$ & $\phantom{-9} 3.30(3)\times10^{-1}$    & $5.435(2) \times10^{-1}$ & $ 6.2(1) \times10^{-1}$ \\
     \cline{2-8}
     & $[-\frac{\pi}{2},0]$ & $4.9144(8)$    & $5.825(7)  $ & $ -1.8720(9)\times 10^{-1}$ & $\,\,\, -1.79(1)\times 10^{-1}$ & $5.437(2) \times 10^{-1}$ & $ 6.3(1) \times 10^{-1}$ \\
     \cline{2-8}
     & $[0,\frac{\pi}{2}]$  & $4.9126(8) $    & $5.824(9)  $ & $ \phantom{-}1.8729(9)\times 10^{-1}$ & $\phantom{-9} 1.78(1)\times 10^{-1}$   & $5.434(2) \times 10^{-1}$ & $ 6.3(1) \times 10^{-1}$ \\
     \cline{2-8}
     & $[\frac{\pi}{2},\pi]$ & $5.1906(8) $   & $6.131(7)  $ & $\,\,\, -3.212(1) \times 10^{-1}$ & $-3.293(9)\times 10^{-1}$ & $5.441(2) \times 10^{-1}$ & $ 6.3(1) \times 10^{-1}$ \\
     \hline
     \hline
    \multirow{4}{*}{$[200,\,400]$}  & $[-\pi,-\frac{\pi}{2}]$ & $2.6647(2)\times 10^{-1}$ & $3.517(2) \times 10^{-1}$ & $\phantom{-} 4.8513(6)\times 10^{-2}$ & $\phantom{-9} 5.40(1)\times 10^{-2}$ & $2.0988(3) \times 10^{-1}$ & $ 2.485(5) \times 10^{-1}$ \\
     \cline{2-8}
                                  & $[-\frac{\pi}{2},0]$ & $2.5467(2)\times 10^{-1}$ & $3.365(2) \times 10^{-1}$ & $ -1.3737(6)\times 10^{-2}$ & $-1.322(9)\times 10^{-2}$ & $2.0990(3) \times 10^{-1}$ & $ 2.485(4) \times 10^{-1}$ \\
     \cline{2-8}
                                  & $[0,\frac{\pi}{2}]$ & $2.5470(2)\times 10^{-1}$ & $3.363(2) \times 10^{-1}$ & $\phantom{-} 1.3738(6)\times 10^{-2}$ & $\phantom{-} 1.318(9)\times 10^{-2}$ & $2.0984(3) \times 10^{-1}$ & $ 2.487(4) \times 10^{-1}$  \\
     \cline{2-8}
                                  & $[\frac{\pi}{2},\pi]$ & $2.6647(2)\times 10^{-1}$ & $3.518(2) \times 10^{-1}$ & $ -4.8490(6)\times 10^{-2}$ & $\,\,\,-5.40(1)\times 10^{-2}$ & $2.0989(3) \times 10^{-1}$ & $ 2.486(3) \times 10^{-1}$  \\
     \hline
     \hline
\multirow{4}{*}{$[400,\,\infty)$}   & $[-\pi,-\frac{\pi}{2}]$ & $2.5148(2)\times 10^{-2}$ & $ 3.319(2) \times 10^{-2}$ & $ \phantom{-17} 7.112(1)\times 10^{-3}$ & $\phantom{-} 8.26(2)\times 10^{-3}$ & $7.8189(8) \times 10^{-2}$ & $ 9.394(7) \times 10^{-2}$ \\
     \cline{2-8}
                                  & $[-\frac{\pi}{2},0]$ & $2.4458(2)\times 10^{-2}$ & $3.226(2) \times 10^{-2}$ & $ \phantom{-117} 3.55(1)\times 10^{-4}$ & $\phantom{-6} 7.1(2)\times 10^{-4}$ & $7.8184(8) \times 10^{-2}$ & $ 9.398(7) \times 10^{-2}$ \\
     \cline{2-8}
                                  & $[0,\frac{\pi}{2}]$ & $2.4463(2)\times 10^{-2}$ & $3.226(2) \times 10^{-2}$ & $\quad\,\,\, -3.55(1)\times 10^{-4}$ & $\,\,\, -6.9(2)\times 10^{-4}$ & $7.8186(8) \times 10^{-2}$ & $ 9.395(8) \times 10^{-2}$  \\
     \cline{2-8}
                                  & $[\frac{\pi}{2},\pi]$ & $2.5152(2)\times 10^{-2}$ & $3.319(2) \times 10^{-2}$ & $ -7.1137(11) \times 10^{-3}$ & $ -8.21(2) \times 10^{-3}$ & $7.8191(8) \times 10^{-2}$ & $ 9.401(7) \times 10^{-2}$  \\
    \end{tabular}
    }
    \caption{Cross section (in fb) of $pp\to e^{+}\, e^{-}\, H$ at $\sqrt{s}=13$~TeV in the SM and when affected by $\mcO_{\varphi \widetilde{W}}$. 
    }
    \label{tab:pp_ZH_xs_allorders}
\end{table}

\section{CP violation at the LHC with $WH$}
\label{sec:CPV_LHC}

\subsection{Analysis strategy}
\label{sec:CPV_LHC_Strat}

As indicated in the previous section and noted in previous studies~\cite{Banerjee:2019twi,Barrue:2023ysk}, $WH$ is the most promising $VH$ channel for CP-violation studies at the LHC.
One must use the whole energy range since 
the sensitivity to CP-odd operators is expected to come mostly from the angular binning, but a second binning in energy is required to fully exploit the features of the interference~\cite{Bishara:2020vix,Barrue:2023ysk}. 
The most sensitive channel at the LHC is $WH\to\ell\nu b \bar{b}$, for which studies show that the resolved-regime contribution to the sensitivity to dimension-6 SMEFT operators at the LHC is not negligible. Hence, we
combine the scale-invariant b-tagging algorithm and (HL-)LHC analysis developed in~\cite{Bishara:2022vsc} with the CP-odd sensitive angular binning proposed for FCC-hh in~\cite{Bishara:2020vix}, which previous studies indicate should offer nearly-optimal sensitivity~\cite{Barrue:2023ysk,Englert:2024ufr}.
The $H\to b\bar{b}$ decay channel presents further difficulties due to a high background level generated mainly by $t\bar{t}$ and $W b \bar{b}$ production. We take the background simulations from~\cite{Bishara:2022vsc}, which were performed at NLO QCD in the case of $Wb\bar{b}$ and at LO with an additional hard jet for $t\bar{t}$, and are compatible with published ATLAS results~\cite{Bishara:2022vsc}. 
This work combines for the first time a double-binning analysis of $WH$ at the LHC with a scale-invariant b-tagging that covers the whole energy range and, most importantly, full NLO QCD corrections in the signal. 

The collider events are first separated between the boosted and resolved categories by the b-tagging algorithm according to the presence or not of a boosted Higgs candidate. In the resolved category, the event must contain a pair of resolved b-jets that constitute the Higgs candidate. In the boosted category, the event must have one boosted jet identified via the mass-drop-tagging (MDT) procedure~\cite{Butterworth:2008iy} and with 2 b-tags in order to be a Higgs candidate. In either category, the Higgs candidate must have an invariant mass within the $[90,120]$~GeV window. Our implementation is as described in~\cite{Bishara:2022vsc} and was calibrated against published ATLAS results~\cite{ATLAS:2020fcp,ATLAS:2020jwz}.

We apply further selection cuts to enhance the $S/B$ ratio. In both categories, we require the events to have one charged lepton and no untagged jets within their corresponding acceptance regions, defined in App.~\ref{app:app_detec_analysis} and based on ATLAS LHC Run 2 analyses~\cite{ATLAS:2020fcp,ATLAS:2020jwz}. In the boosted category, two other selection cuts are of high relevance: $|\Delta y (W,H_{\text{cand}})| \leqslant 1.4$ and $|\eta^{H_{\text{cand}}}|\leqslant 2.0$, while in the resolved category we instead apply a cut $\Delta R_{bb}\leqslant 2.0$. We provide details of all the used cuts in App.~\ref{app:app_detec_analysis}. 
All the events that pass the selection cuts are split into different bins according to $p_T^H$ and $\phi_W$. We summarize the binning strategy in Table~\ref{tab:binning}. Only the $p_T^H$ binning depends on whether the event is classified as boosted or resolved.

\begin{table}
\centering
\scalebox{0.8}{
\begin{tabular}{c|c|c}
Category & $p_{T}^H$ bins & $\phi_W / \pi$ bins \\ 
\hline
\multirow{2}{*}{Resolved} & \multirow{2}{*}{$\lbrace 0, \,175, \,250, \,\infty  \rbrace$} & \multirow{4}{*}{ 
$ \begin{cases} 
[-1,0],\,[0,1] \\
[-1,-\frac{1}{2}],\, [-\frac{1}{2},0],\,[0,\frac{1}{2}],\,[\frac{1}{2},1]  \end{cases}
$ } \\
 & & \\
\cline{1-2}
\multirow{2}{*}{Boosted}  & \multirow{2}{*}{$\lbrace 0,\, 175,\, 250,\, 300,\,\infty \rbrace$} & \\
 & &
\end{tabular}
}
\caption{Bin edges for the $p_T^H$ and angular binning used in the $WH$ analysis. Only the $p_T^H$ binning depends on the event category, boosted or resolved. }
\label{tab:binning}
\end{table}

\subsection{Differential distributions}

\begin{figure}
    \centering
    \includegraphics[width=0.475\textwidth]{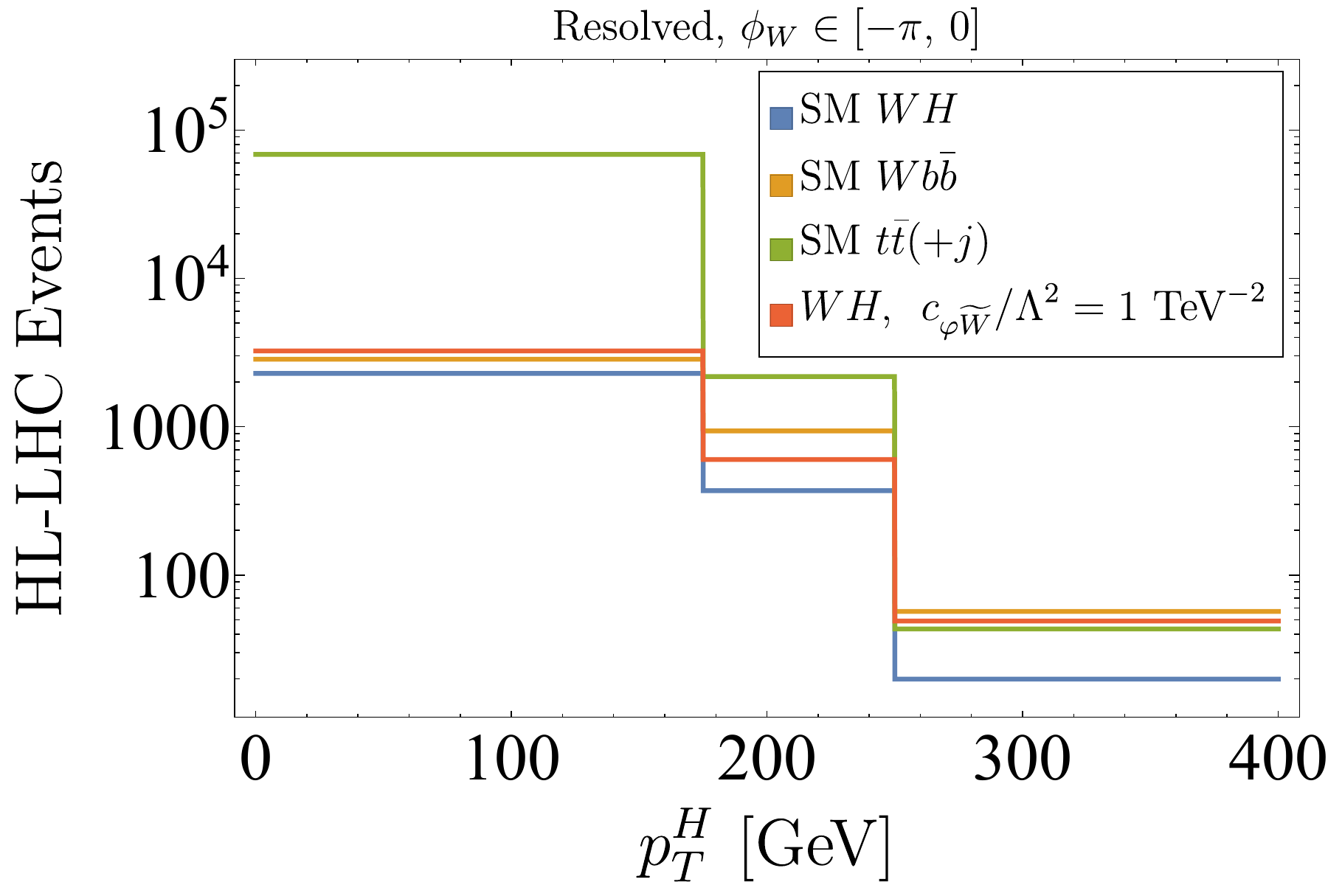}\hfill
    \includegraphics[width=0.475\textwidth]{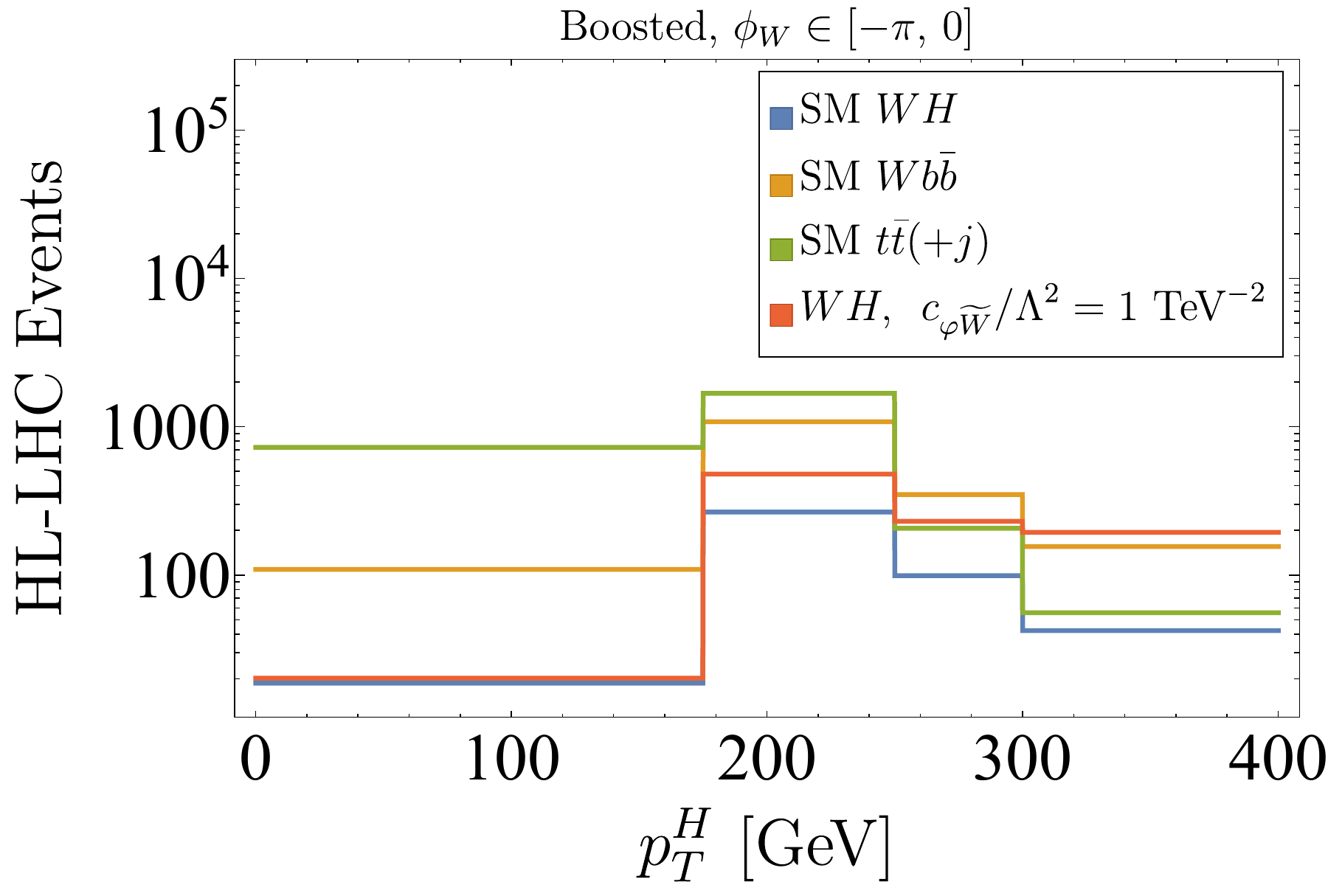}\hfill
    \caption{$p_{T}^{H}$ distributions at the HL-LHC ($3~\text{ab}^{-1}$) after full analysis with events that have a negative $\phi_W$ angle. We show the $WH$, $Wb\bar{b}$ and $t\bar{t} j$ processes. For $WH$, we show the SM case and when $c_{\varphi\widetilde{W}} = 1.0$~TeV$^{-2}$. $WH$ and $Wb\bar{b}$ are computed at NLO QCD. On the left, the result in the resolved regime. On the right, the one corresponding to the boosted regime.}
    \label{fig:pth_distr_SM_sig_bkgd}
\end{figure}

We first consider the number of HL-LHC events in different bins of $p_T^H$ for the signal and background processes, which is shown in Fig.~\ref{fig:pth_distr_SM_sig_bkgd}, where we consider only those events with $\phi_W\in[-\pi,0]$ to make the CP-odd interference effects visible. We observe a clear growth with energy of the relative $WH$ contribution when $c_{\varphi \widetilde{W}} =  1.0$~TeV$^{-2}$. 

\begin{figure}
    \centering
    \includegraphics[width=0.475\textwidth]{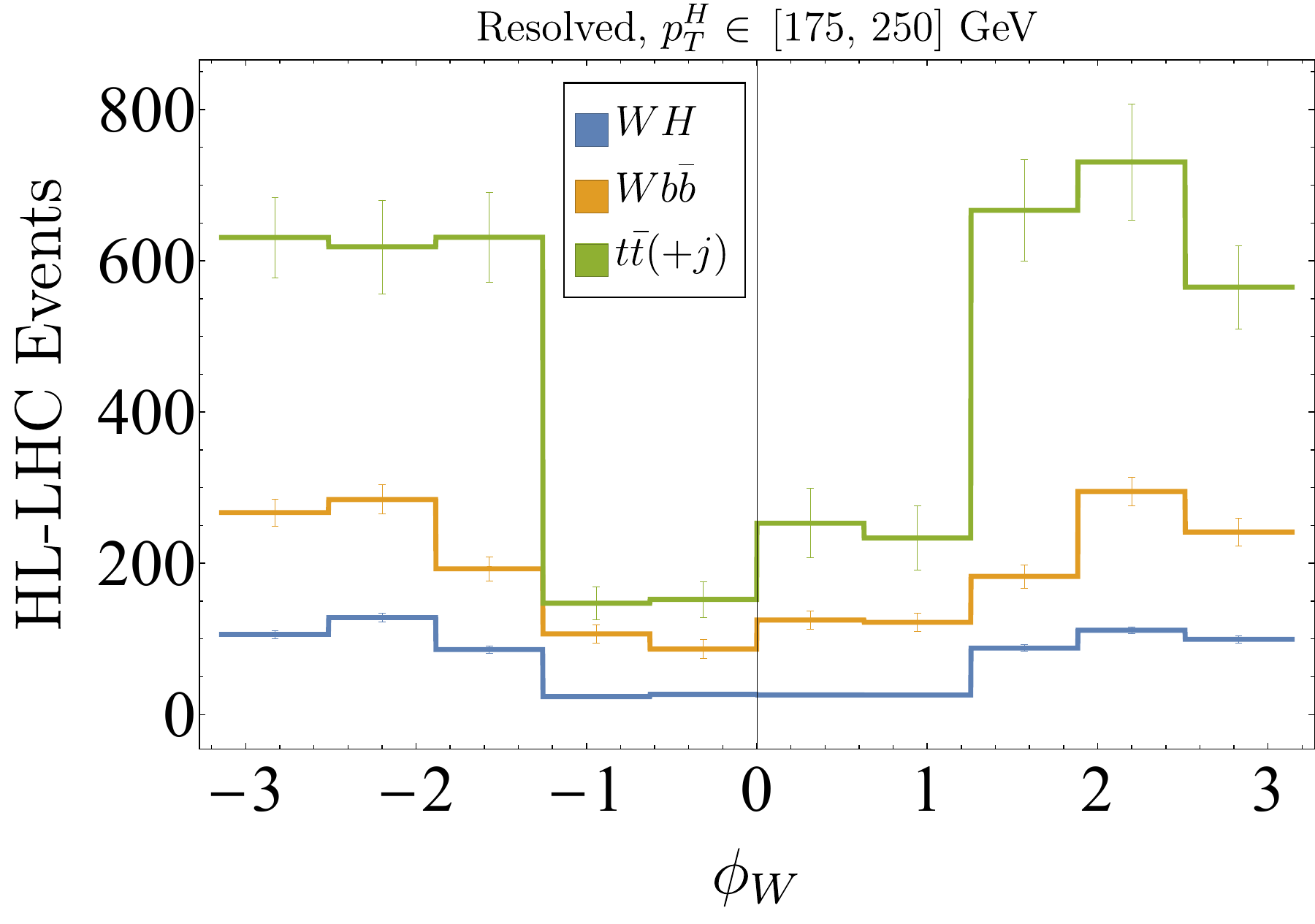}\hfill
    \includegraphics[width=0.475\textwidth]{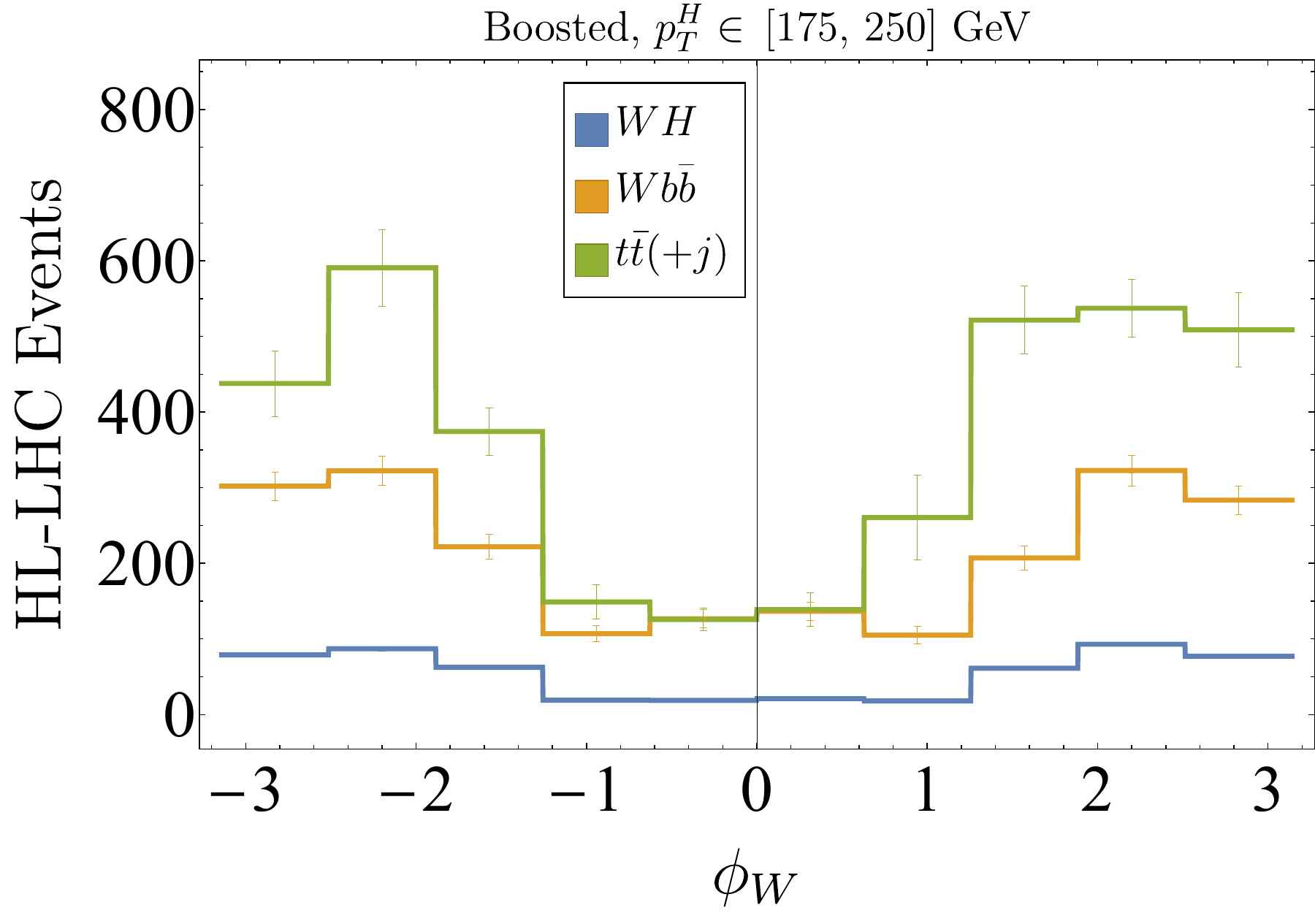}\hfill
    \caption{Angular distributions at the HL-LHC after full analysis in the SM. We show the $WH$, $Wb\bar{b}$ and $t\bar{t} j$ processes. The first two are simulated at NLO QCD. On the left, the result in the resolved regime. On the right, the one corresponding to the boosted regime. The error bars show the Monte Carlo uncertainty.}
    \label{fig:Ang_distr_SM_sig_bkgd}
\end{figure}

In Fig.~\ref{fig:Ang_distr_SM_sig_bkgd}, we show the angular distribution of events after the full analysis for the SM signal and background events. Within statistical errors, these distributions are CP even and show a preference for large values of $|\phi_W|$.
Selecting events with $|\phi_W|<\pi/2$ could increase the $S/B$ ratio but would damage the sensitivity to CP-odd BSM effects since most of the alteration caused by $\mcO_{\varphi\widetilde{W}}$ is generated in the region that is cut away.

\begin{figure}
    \centering
    \includegraphics[width=0.475\textwidth]{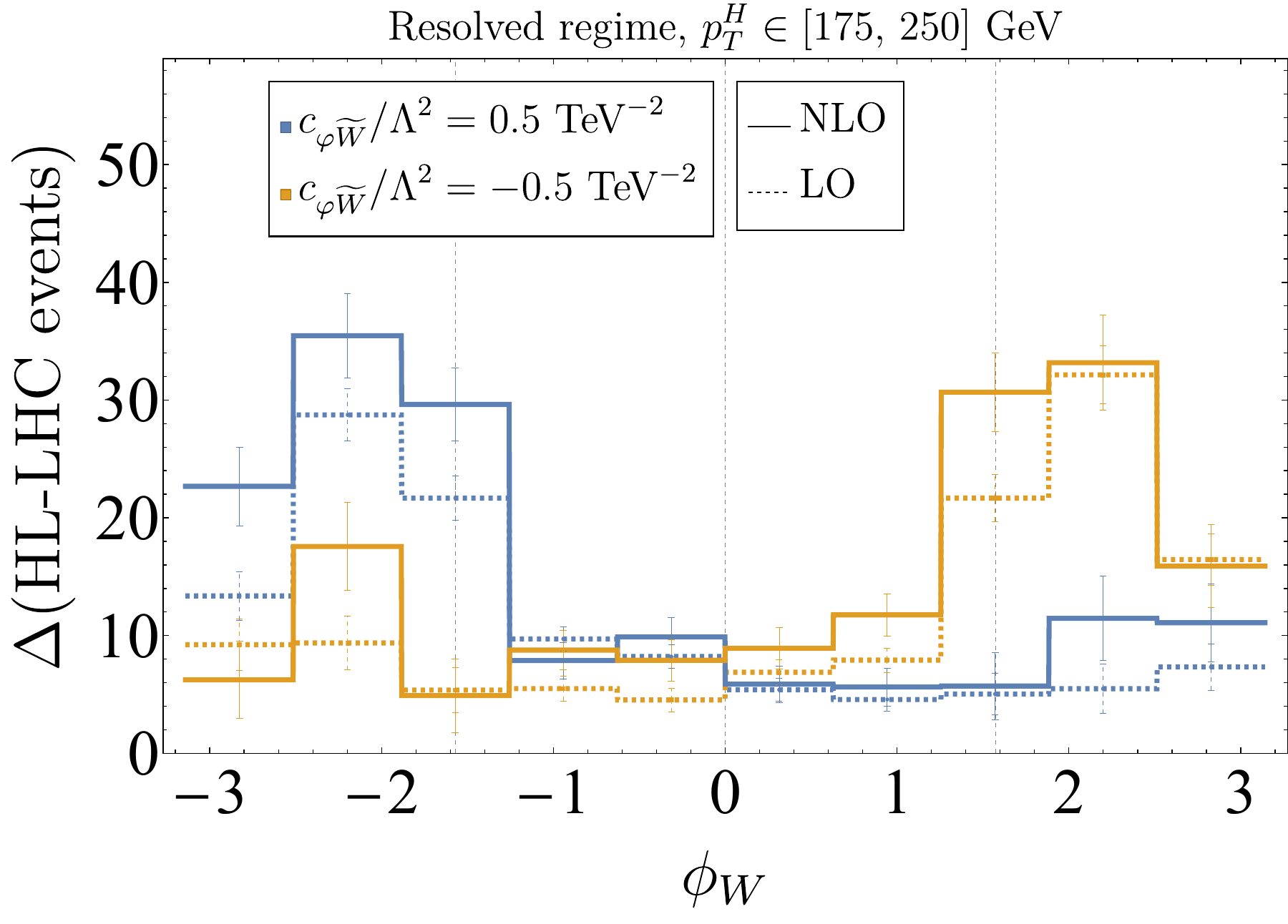}\hfill
    \includegraphics[width=0.475\textwidth]{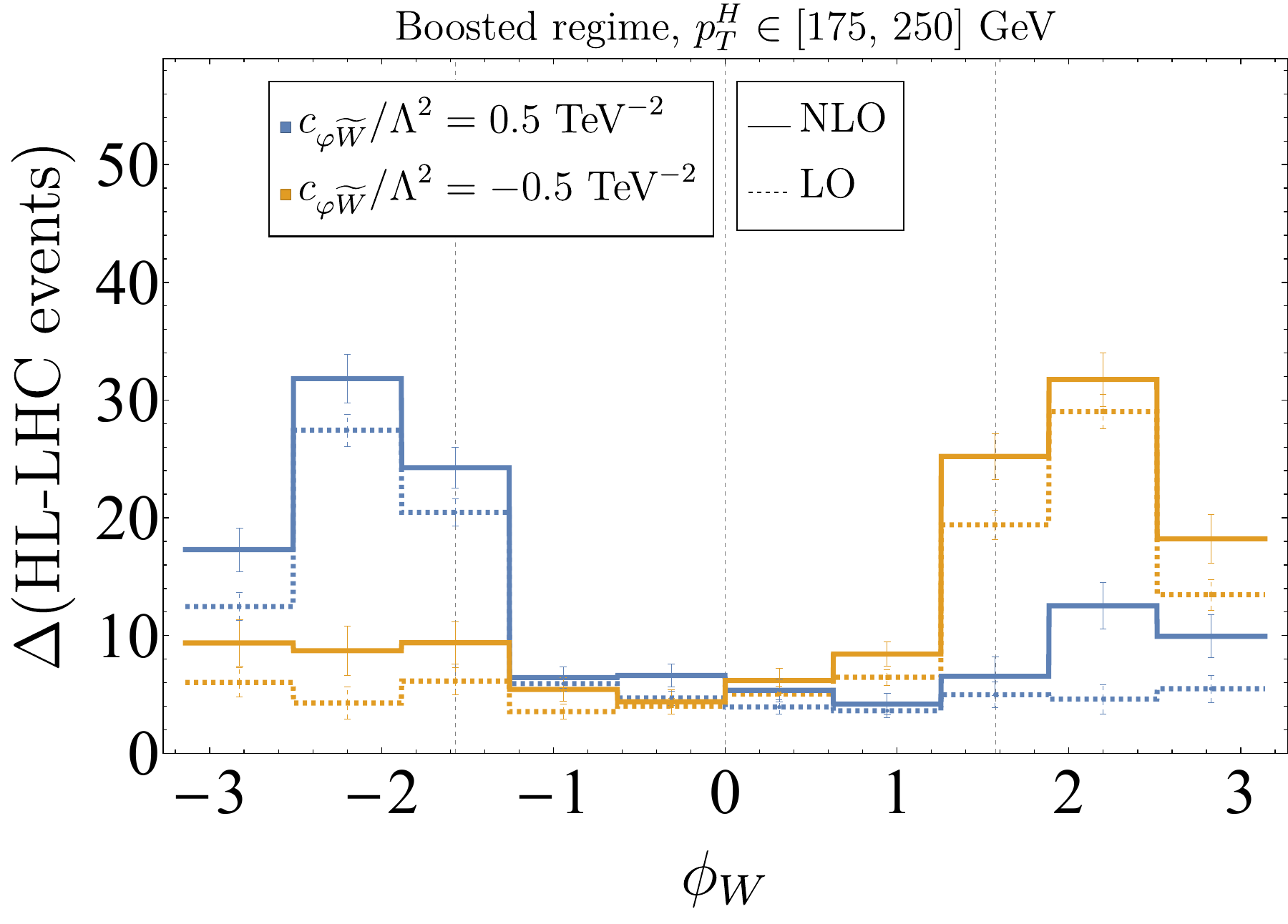}\hfill
    \caption{
    Difference in the number of events from $pp\to W H\to \ell \nu b\bar{b}$ at the HL-LHC with respect to the SM prediction, defined in Eq.~\eqref{eq:diff_evts_LHC}, as a function of $\phi_W$ from the full simulated analysis. We show two different values of $c_{\varphi \widetilde{W}}$, $\pm 0.5$~TeV$^{-2}$ and two different computation orders in QCD: NLO (full) and LO (dashed). 
    The resolved (boosted) regime is shown on the left (right) panel and in both cases, we show the bin $p_T^H\in[175,250]$~GeV. The vertical dashed lines are located at $\phi_W=\pm\frac{\pi}{2},\,0$.
    }
    \label{fig:ang_distr_cpwtilde_HLLHC_Res_Boos}
\end{figure}

\begin{figure}[h!]
    \centering
    \includegraphics[width=0.6\textwidth]{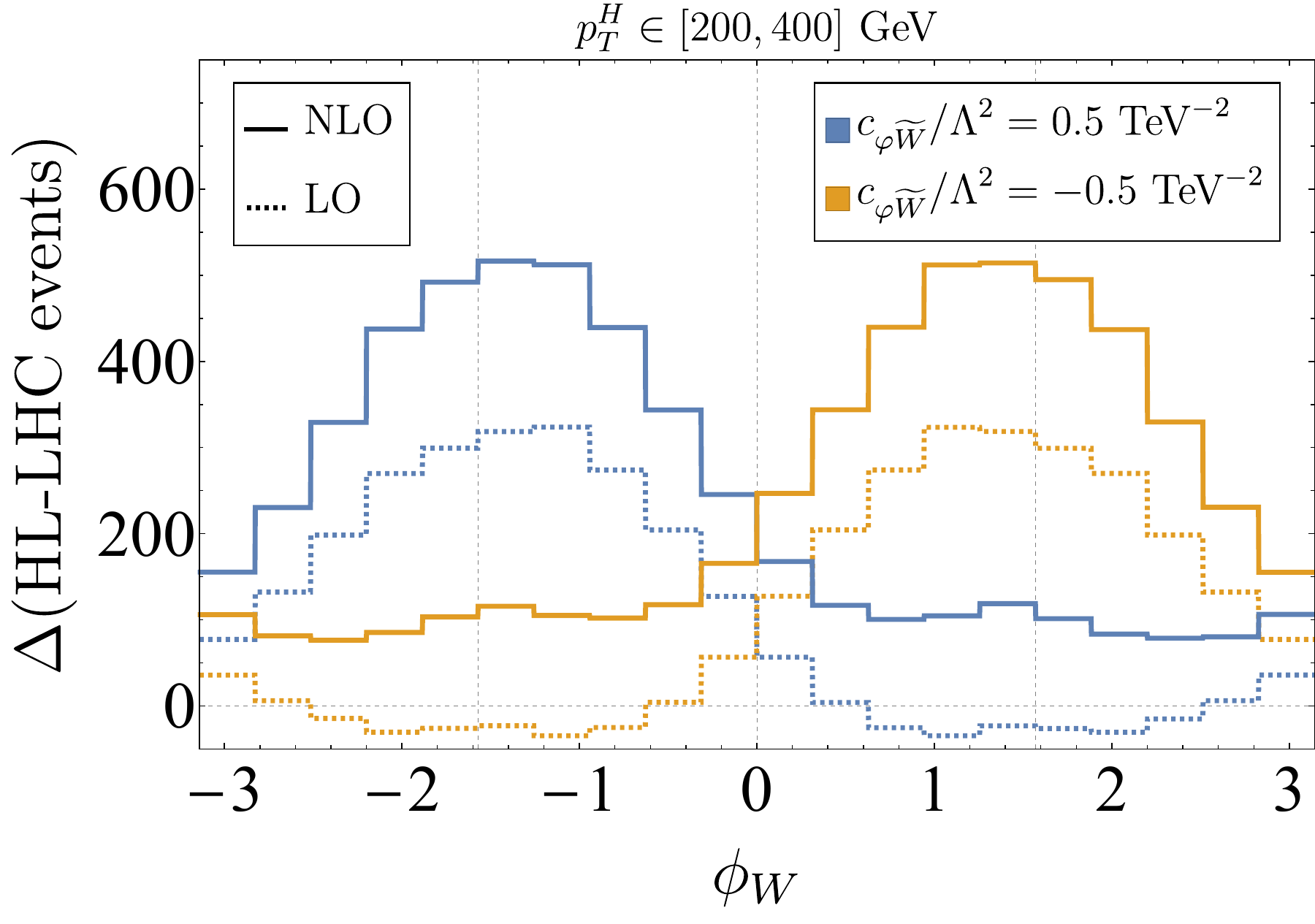}
    \caption{
    Difference in the number of events from $pp\to W^{-}H\to \ell^- \bar{\nu}\, b\bar{b}$ at the HL-LHC with respect to the SM prediction, defined in Eq.~\eqref{eq:diff_evts_LHC}, as a function of $\phi_W$ from a fixed-order computation. We show two different values of $c_{\varphi \widetilde{W}}$, $\pm0.5$~TeV$^{-2}$ and two different computation orders in QCD: NLO (full) and LO (dashed). We show the result in the region $p_{T}^{H}\in[200,400]$~GeV.
    }
  \label{fig:FO_reconstPhiWCompLONLO}
\end{figure}

Then, we consider the effects of turning on the WC $\cpWt$ in Fig.~\ref{fig:ang_distr_cpwtilde_HLLHC_Res_Boos}, where we show the difference in the number of events at the HL-LHC, $N|_{\text{HL-LHC}}$, with respect to the SM prediction, i.e.,
\begin{equation}
\left(N - N_{\text{SM}}\right)|_{\text{HL-LHC}} =\lag\cdot\left(\sigma -\sigma_{\text{SM}}\right) = \lag\cdot \left( \frac{\cpWt}{\Lambda^2} \, \sigma_{1/\Lambda^2} + \frac{\cpWt^2}{\Lambda^4} \, \sigma_{1/\Lambda^4} \right),
\label{eq:diff_evts_LHC}
\end{equation}
as a function of the angle $\phi_W$, where $\lag=3$~ab$^{-1}$ is the HL-LHC luminosity.
We plot two values of the WC, $\cpWt/\Lambda^2 =\pm0.5$~TeV$^{-2}$, which are representative of the bounds to be discussed in the next section, at a mid-energy bin, $p_T^H\in[175,\,250]$~GeV, for both the boosted and resolved regimes. The effect of the interference is visible for these values in the form of an asymmetry between the regions with negative and positive $\phi_W$. 
NLO effects increase the number of events in all cases and reduce the asymmetry between negative and positive $\phi_W$, as expected from the fixed-order analysis in section~\ref{sec:NLO_FO}. We show in Fig.~\ref{fig:FO_reconstPhiWCompLONLO} a plot analogue to Fig.~\ref{fig:ang_distr_cpwtilde_HLLHC_Res_Boos} obtained at fixed order without cuts and 
for the case of a negatively charged final lepton ($\ell^{-}=e^{-},\,\mu^{-}$). This allows us to see how the LO interference is big enough to generate a negative difference of events in a certain angular region. 
The differences in the shape between Fig.~\ref{fig:ang_distr_cpwtilde_HLLHC_Res_Boos} and~\ref{fig:FO_reconstPhiWCompLONLO} are due to the detector simulation, acceptance and selection cuts considered in the first one.

\subsection{Projected sensitivity at the LHC}
\label{sec:sensitivity_LHC}
\subsubsection{One-operator analysis}
\label{sec:sensitivity_LHC_cpwt_only}

We present the $95\%$~C.L. projected bounds on $\cpWt$ at the LHC Run 3 and HL-LHC in Table~\ref{tab:all_bounds} for different systematic uncertainties: $1\%$, $5\%$ and $10\%$. Of these three scenarios, $10\%$ is similar to current systematic uncertainties in this channel, $5\%$ is a realistic expectation for HL-LHC and $1\%$ is considered an optimistic limit case. We show the results for 2 different angular binnings as defined in Table~\ref{tab:binning}, for a signal process simulated at either LO or NLO in QCD, and when considering the cross-section up to linear ($\mcO\left(\Lambda^{-2}\right)$) or quadratic ($\mcO\left(\Lambda^{-4}\right)$) order in $\cpWt$. At the LHC Run 3, this analysis is limited by statistical uncertainties since a higher number of bins and/or the variation of the systematics cause small effects. Accordingly, the results benefit greatly from the higher statistics at the HL-LHC and the bounds improve by a factor of $1.5-2$ depending on the systematics level. A total removal of the background processes would improve the bounds by a factor of at least $2$ at the HL-LHC.

These bounds are dominated by the piece of the cross-section that is quadratic in the WC, as can be seen from comparing the columns $\mcO\left(\Lambda^{-2}\right)$
and $\mcO\left(\Lambda^{-4}\right)$ in Table~\ref{tab:all_bounds}.
The bounds at the linear level are worse by a factor between $2$ and $3.7$ in all cases.
In the ideal case of negligible backgrounds, the interference terms would provide $\sim 80\%$ of the total bound, indicating the importance of reducing the background. We notice that the bounds are expected to be symmetric around $0$ even when including the quadratic dependence. This is because the interference terms are odd, whilst 
the SM and quadratic terms are even, thus not allowing any odd terms in the $\chi^2$ after summing over all the bins. Any asymmetry in the bounds is, then, due to numerical imprecision.
The possible impact of CP-odd dimension-8 operators is hard to estimate since they might generate different angular distributions.

NLO QCD effects in the signal have beneficial effects both for LHC Run 3 and HL-LHC. The bounds on $\cpWt$ improve by $\sim10\%$ at NLO regardless of luminosity and systematic uncertainties. However, due to the smaller QCD corrections for the interference discussed before, the NLO bounds are more quadratics-dominated than the LO ones, as can be seen in Table~\ref{tab:all_bounds}. There, for equal systematics and number of angular bins, the degrading factor from quadratic bounds to linear is always bigger for the case of an NLO signal.

The HL-LHC projected bounds perceive a positive impact from an increase in the number of angular bins, in particular for higher systematics. 
The linear bounds in Table~\ref{tab:all_bounds} show a clearer gain from a higher number of angular bins since the angular binning provides more information about the signal interference. 
We checked that adding more angular bins does not significantly improve the sensitivity of the analysis beyond reducing the impact of high systematic uncertainties.

\begin{table}[]
    \centering
    \resizebox{\textwidth}{!}{
    \begin{tabular}{c|c|cc|cc|cc|cc}
       \multicolumn{2}{c|}{QCD order} & \multicolumn{4}{c|}{LO} & \multicolumn{4}{c}{NLO} \\
       \hline
       \multicolumn{2}{c|}{ $\# \phi_{W}$ bins}  & \multicolumn{2}{c|}{$2$} & \multicolumn{2}{c|}{$4$} & \multicolumn{2}{c|}{$2$} & \multicolumn{2}{c}{$4$} \\
       \hline
       \multicolumn{2}{c|}{EFT order} & $\mcO\left(\Lambda^{-2}\right)$ & $\mcO\left(\Lambda^{-4}\right)$ & $\mcO\left(\Lambda^{-2}\right)$ & $\mcO\left(\Lambda^{-4}\right)$ & $\mcO\left(\Lambda^{-2}\right)$ & $\mcO\left(\Lambda^{-4}\right)$ & $\mcO\left(\Lambda^{-2}\right)$ & $\mcO\left(\Lambda^{-4}\right)$ \\
       \hline
       \multirow{3}{*}{$\lag=300$~fb$^{-1}$} & $1\%$ Syst.  &$[-2.24,2.24]$&$[-0.64,0.64]$&$[-2.18,2.18]$&$[-0.63,0.63]$&$[-2.12,2.12]$&$[-0.56,0.56]$&$[-2.06,2.06]$&$[-0.56,0.56]$ \\
                                             & $5\%$ Syst.  &$[-2.64,2.64]$&$[-0.67,0.67]$&$[-2.48,2.48]$&$[-0.65,0.65]$&$[-2.44,2.44]$&$[-0.60,0.59]$&$[-2.30,2.30]$&$[-0.58,0.58]$ \\
                                             & $10\%$ Syst. &$[-3.21,3.21]$&$[-0.72,0.72]$&$[-2.90,2.90]$&$[-0.69,0.69]$&$[-2.91,2.91]$&$[-0.64,0.64]$&$[-2.64,2.64]$&$[-0.61,0.61]$ \\
                                               \hline
       \multirow{3}{*}{$\lag = 3$~ab$^{-1}$} & $1\%$ Syst.  & $[-0.77,0.77]$&$[-0.35, 0.35]$&$[-0.74,0.74]$&$[-0.35, 0.35]$&$[-0.72,0.72]$&$[-0.32,0.31]$& $[-0.69,0.69]$&$[-0.31,0.31]$ \\
                                             & $5\%$ Syst.  & $[-1.24,1.24]$&$[-0.43, 0.43]$&$[-1.08,1.08]$&$[-0.40, 0.40]$&$[-1.11,1.11]$&$[-0.38,0.38]$& $[-0.96,0.96]$&$[-0.35,0.35]$ \\
                                             & $10\%$ Syst. & $[-1.92,1.92]$ & $[-0.54, 0.53]$& $[-1.53,1.53]$&$[-0.47, 0.47]$& $[-1.68,1.68]$&$[-0.47,0.47]$& $[-1.32,1.32]$&$[-0.41,0.41]$ \\        
    \end{tabular}
    }
    \caption{Projected $95\%$ C.L. bounds on $\cpWt$, with $\Lambda=1$~TeV, from $WH$ production at the LHC Run 3 ($\lag = 300$~fb$^{-1}$) and the HL-LHC ($\lag = 3$~ab$^{-1}$). 
    We show the result with the signal process simulated at LO or NLO in QCD, with the use of 2 or 4 angular bins in each $p_T$ bin and when considering the signal cross-section linear ($\mcO\left(\Lambda^{-2}\right)$) or quadratic ($\mcO\left(\Lambda^{-4}\right)$) dependence on $\cpWt$. 
    }
    \label{tab:all_bounds}
\end{table}

\begin{figure}[h!]
    \centering
    \includegraphics[width=0.75\textwidth]{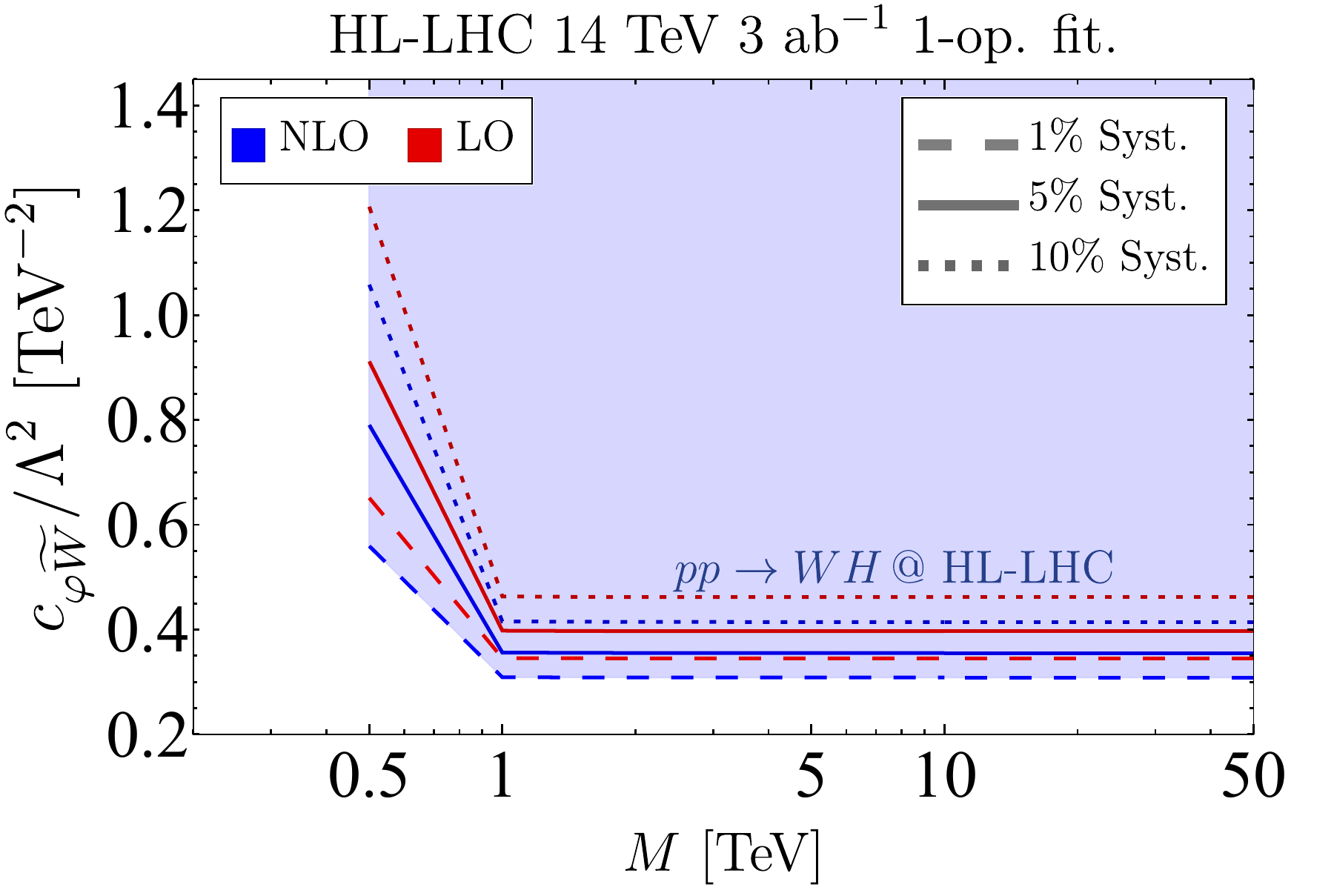}
    \caption{ Projected
    $95\%$ C.L. on $c_{\varphi\widetilde{W}}$ from $pp\to WH$ at the HL-LHC as a function of the maximal invariant mass of the $WH$ system, $M$, for different systematic uncertainties and choices of QCD order for the signal. 
    We use 4 angular bins in all cases.
}
  \label{fig:cpwt_bound_vsM}
\end{figure}

The EFT validity can be studied by adding a cut on the maximal invariant mass of the reconstructed $WH$ system. Assuming that the EFT description is valid up to energy $M$, one removes all the signal and background events with $m_{WH}>M$ and recomputes the bounds. The result of this procedure for the HL-LHC projected bounds with 4 angular bins is shown in Fig.~\ref{fig:cpwt_bound_vsM}, from where one can see that the bounds presented before are valid for any EFT cutoff $\gtrsim 1$~TeV.
This validity region coincides with the analogous analysis without the angular binning performed in previous studies~\cite{Bishara:2022vsc}.
The leading contribution to the $\chi^2$ comes from the highest-energy bin in the boosted category, $p_{T}^{H}\in[300,\,\infty)$~GeV, in agreement with the bounds being dominated by the quadratic contributions that grow with energy like $\hat{s}$.

The analysis presented here but without the angular binning was extended to FCC-hh, where the bounds on CP-even operators are improved by a factor of $\sim 10$ with respect to HL-LHC~\cite{Bishara:2022vsc}. We expect a similar improvement for the bounds on $\cpWt$, which would make $WH(\to b\bar{b})$ at least as good as $WH(\to \gamma\gamma )$ to probe this operator at the FCC-hh since the latter is expected to give a bound of $|\cpWt|\lesssim 8\times10^{-2}$ TeV$^{-2}$~\cite{Bishara:2020vix}.

The impact of $\cpWt$ on $WH$ production at the LHC was previously studied in~\cite{Banerjee:2019twi}, where they estimate a bound of $|\cpWt|\lesssim 0.5$~TeV$^{-2}$ 
at the HL-LHC from a boosted decision tree and angular moments-based analysis, in agreement with our results. A more systematic study was performed in~\cite{Barrue:2023ysk}, where they find a  LHC Run 3 ($\lag = 300$~fb$^{-1}$) bound of $|\cpWt|<0.144$~TeV$^{-2}$ when considering the quadratic contributions. Those results should be compared with our estimated LO bound with $1\%$ syst. and $4$ angular bins, $|\cpWt|<0.64$~TeV$^{-2}$, which is $\sim 4$ times looser. This difference can be attributed to several factors such as different strategies to simulate detector effects and our usage of NLO-corrected backgrounds.
In a broader context, $\cpWt$ has been probed in other Higgs processes such as Higgs decays to EW vector boson~\cite{Bernlochner:2018opw}, WBF-induced $H\gamma$ production~\cite{Biekotter:2018rhp}, and VBF-induced $H\to\gamma\gamma$~\cite{ATLAS:2022tan}, with the latter giving the leading experimental bound $\frac{\cpWt}{\Lambda^2}\in[-0.53,1.07]$~TeV$^{-2}$.
Thus, $WH$ seems a competitive process to study the CP-violating effects induced by $\cpWt$ at the (HL-)LHC but the combination with other channels could significantly improve the collider sensitivity to this WC. Even in that case, the collider bound is not expected to reach the levels of the indirect bound from electron EDMs, $\frac{\cpWt}{\Lambda^2}\lesssim2\cdot 10^{-5}$~TeV$^{-2}$~\cite{Panico:2018hal}, which however relies on the absence of other CP-odd contributions.

\subsubsection{Interplay between CP-odd and CP-even operators}
\label{sec:sens_LHC_2D_analysis}

The double angular binning in principle offers the possibility of distinguishing the effects of the CP-even and CP-odd operators $\mcO_{\varphi W}$ and $\mcO_{\varphi \widetilde{W}}$. We evaluate this by extending our analysis with the dependence of the signal on $\cpW$ at NLO QCD and up to quadratic order in the WC. 
We show in Table~\ref{tab:2d_bounds} the projected HL-LHC $95\%$~C.L. bounds on $\cpWt$ and $\cpW$ from a 2-dimensional Gaussian $\chi^2$ for different systematic assumptions and after either profiling or setting to zero the other WC. Those results were obtained with the 4 angular bins presented before.
The individual bounds on $\cpW$ and $\cpWt$ show a similar sensitivity to both WCs, with tighter bounds on the CP-even one for low systematics and a smaller sensitivity to the systematic uncertainties for the CP-odd WC.

\begin{table}[h!]
    \centering
    \begin{tabular}{c|c|c|c}
      WC & Syst. & Profiled fit & Individual fit  \\
      \hline
      \multirow{3}{*}{$\frac{\cpW}{\Lambda^2}\,[\text{TeV}^{-2}]$} & $1\%$ & $[-0.24,0.12]$  & $[-0.16,0.12]$ \\
      & $5\%$ & $[-0.52,0.17]$ & $[-0.52,0.17]$ \\
      & $10\%$ & $[-0.63,0.23]$ & $[-0.63,0.23]$ \\
      \hline
      \multirow{3}{*}{$\frac{\cpWt}{\Lambda^2}\,[\text{TeV}^{-2}]$} & $1\%$ & $[-0.38,0.38]$ & $[-0.31,0.31]$ \\
      & $5\%$ & $[-0.42,0.42]$ & $[-0.35,0.35]$ \\
      & $10\%$ & $[-0.47,0.47]$ & $[-0.41,0.41]$ \\
    \end{tabular}
    \caption{Projected $95\%$ C.L. bounds on $\cpW$ and $\cpWt$ from the analysis of $pp\to WH \to \ell\nu b\bar{b}$ at the HL-LHC. We simulated the signal at NLO in QCD and used 4 angular bins. }
    \label{tab:2d_bounds}
\end{table}

The individual bound on $\cpW$ receives important contributions from both the interference and quadratic pieces as evidenced by its asymmetry around $0$. Furthermore, the positive bound for $1\%$ and $5\%$ systematics is dominated by the interference contribution, while both the interference and squared pieces influence the negative bound. The best bound from each $p_T^H$ bin is given by the resolved $[175,\,250]$~GeV bin, with the three highest energy bins in the boosted category giving very similar positive bounds to the one mentioned before but looser negative bounds. 

\begin{figure}[h!]
    \centering
    \includegraphics[width=0.6\linewidth]{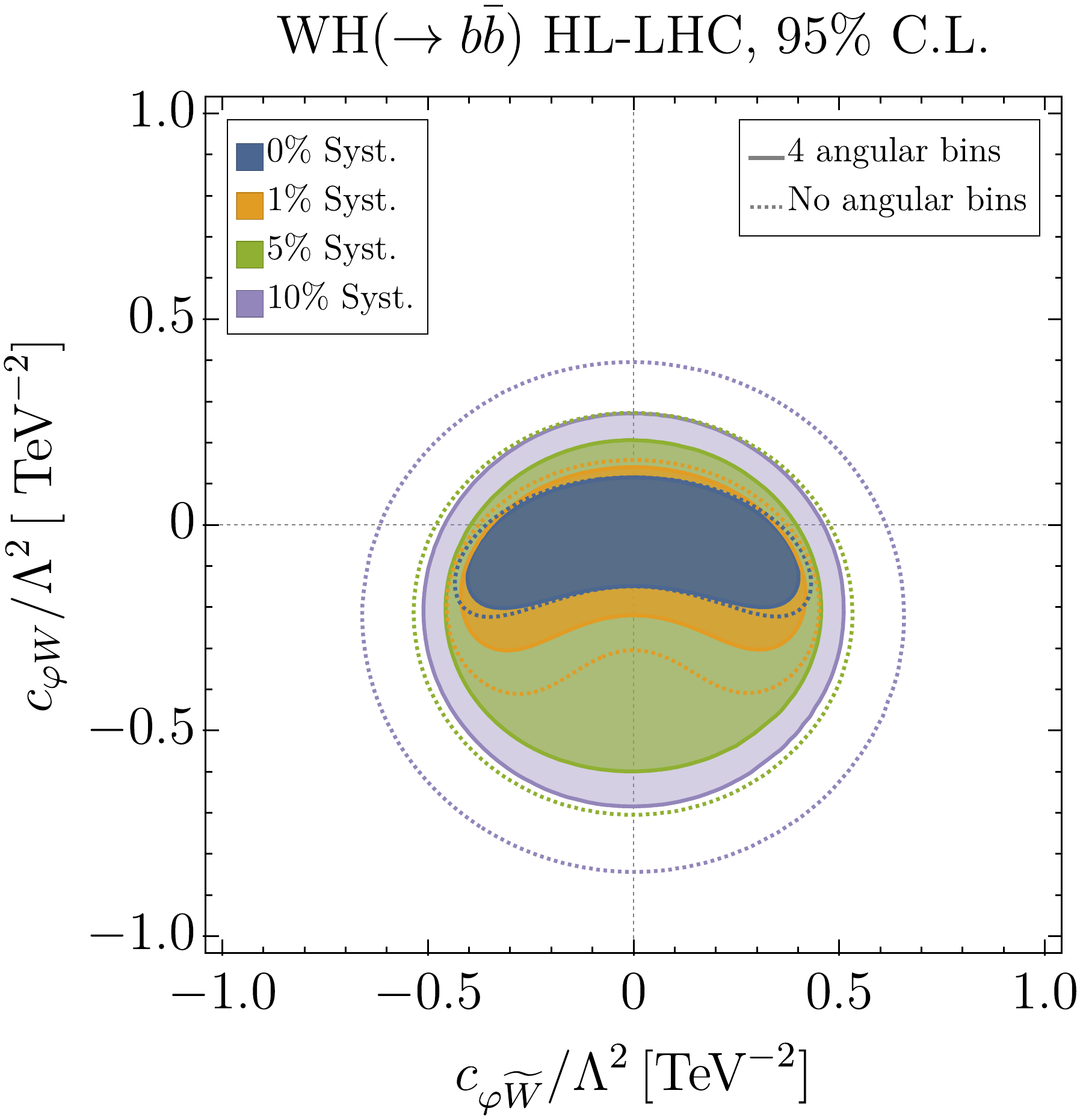}
    \caption{Projected $95\%$ C. L. on the $\cpWt - \cpW$ plane from an analysis of $WH$ production at the HL-LHC with double binning on $p_T^H$ and $\phi_W$. We show the result of using 4 angular bins (full line and shaded region) or no angular binning (dashed line) under different assumptions of the systematic uncertainty.
    }
    \label{fig:2d_bound_cpw_cpwt}
\end{figure}

The small difference between the individual and profiled-fit bounds for $\cpW$ is mainly caused by a small $\mcO_{\varphi W}$-$\mcO_{\varphi \widetilde{W}}$ interference, see App.~\ref{app:app_xs} for details, and helped by the inherent symmetry of the bounds on $\cpWt$. On the other hand, there is a $\sim 20\%$ difference between individual and profiled bounds for $\cpWt$ that originates in the asymmetry of the bounds on $\cpW$. This can be seen by plotting the $95\%$ C.L. allowed region in the $\cpWt-\cpW$ plane, which we show in Fig.~\ref{fig:2d_bound_cpw_cpwt} with full lines and for different systematic uncertainties.

To elucidate the impact of the angular binning on these results, we show with dashed lines in Fig.~\ref{fig:2d_bound_cpw_cpwt} the result of binning only on $p_T^H$. Using 4 angular bins improves the sensitivity to both WCs for any non-zero value of systematic uncertainties. 
This improvement is similar for both the CP-odd and CP-even operators, as expected in a measurement with uncertainties dominated by the systematics. The results for the extreme case of $0\%$ systematics show a sensitivity improvement only along the $\cpWt$ direction since the angular binning only provides additional information about the CP-odd effects.

It is interesting to notice that the sensitivity to $\cpW$ and $\cpWt$ from $WH$ production is expected to profit differently from future higher-energy hadron colliders. While the growth with energy in the CP-odd interference uncovered by the angular binning will lead to a clear gain in the bounds on $\cpWt$, the bounds on $\cpW$ will not improve as much and might become quadratics-dominated. The higher efficiency of the angular binning in probing CP-odd effects at higher energies will further decorrelate the effects of these two operators~\cite{Bishara:2020vix}.

\section{Conclusions}
\label{sec:Conclusions}

In this article, we have studied NLO QCD corrections to the effects caused by CP-odd dimension-6 bosonic SMEFT operators in $WH$ and $ZH$ production at the (HL-)LHC. We considered the operators $\mcO_{\varphi\widetilde{W}}$, $\mcO_{\varphi\widetilde{B}}$, and $\mcO_{\varphi W\widetilde{B}}$. We implemented the mentioned operators together with $\mcO_{\widetilde{W}}$ in an extension of the \texttt{SMEFTatNLO} UFO~\cite{Degrande:2020evl}.

CPV effects in these processes are observed in the angular distributions of the decay products of the vector bosons. We focused on the leptonic decays of the $W$ and $Z$ and revisited the leading-order interference between the SM amplitudes and the ones generated by the CP-odd operators in $WH$ and $ZH$ production. We analysed their angular and energy behaviour, establishing which angular binning strategies would increase our sensitivity to the interference in each case. We found that two angular modes are always present, with distinctive energy-growth patterns, but their relative importance is different in $WH$ and $ZH$. In particular, the one that grows with energy, $\sin(\phi_{W/Z})$, dominates at all energies for $WH$. 

Then, we performed a detailed fixed-order analysis of the NLO QCD effects on these processes. The NLO QCD corrections showed a non-trivial dependence on the angles for both processes, highlighting the need to include them fully in precision computations. These perturbative corrections tend to increase the cross-section in all cases. However, that increase is relatively bigger for the SM and $\mcO(\Lambda^{-4})$ contributions, thus reducing the relative importance of the interference and making the measurement of CPV effects more difficult.

Between the two processes, $WH$ is the more relevant from a phenomenological point of view since it has a bigger cross-section. Hence, it is the main target for experimental searches at the (HL-)LHC. On the other hand, 
$ZH$ shows richer features as its angular distribution shows different modulations depending on the energy. At lower energies, the $ZH$ interference shows a clear $\sin(2\phi_Z)$ distribution that morphs into a $\sin(\phi_Z)$ at higher energies. Its NLO QCD corrections show the same behaviour but with a different energy dependence, leading to $k$-factors smaller than 1 in some angular and energy regions.

We then performed a phenomenological analysis of $WH\to \ell\nu b\bar{b}$ at the (HL-)LHC to study the impact of the NLO QCD corrections in the projected bounds on $c_{\varphi \widetilde{W}}$. We combined a scale-invariant b-tagging strategy, an ATLAS-based analysis used in previous work and a second angular binning originally proposed for $WH$ at future colliders. We showed that this combination makes $WH$ one of the best channels to study the effects of $c_{\varphi \widetilde{W}}$ at the (HL-)LHC.

The NLO QCD effects, included for the first time in this work, improved the bounds by $\sim 10\%$ but made them more reliant on the quadratic piece of the cross-section. Hence, although NLO QCD corrections allow us to probe this operator better, it also hinders our ability to distinguish CP-odd from CP-even effects. We also showed how a higher number of angular bins has a positive but small effect on the bounds.

We then extended the analysis by considering the effects of the CP-even WC $\cpW$ on the same process. We show that the sensitivity to $\cpW$ is slightly better than to $\cpWt$ but comes from lower energy bins since its interference with the SM does not grow with energy. The bound on $\cpW$ is partially led by its interference with the SM, which causes asymmetrical bounds and a worsening of the bounds on $\cpWt$ of $\sim 20\%$ when profiling over $\cpW$. We showed how the angular binning helps to decorrelate the effects of $\cpW$ and $\cpWt$, although this would be noticeable at the HL-LHC only for unrealistically low systematics.

The precision programme for CPV effects at the (HL-)LHC has a long road ahead, hence this work can be extended in several directions. First, a combined analysis of all diboson channels, or at least $VH$, at the (HL-)LHC would be enlightening to evaluate the correlations among all the different dimension-6 operators that affect these processes. The use of double binning in energy and angle could help decorrelate most of their effects.
NLO EW corrections is another frontier that must be explored since these are non-negligible for $VH$ production in the SM. Furthermore, they might alter the angular distributions induced by purely bosonic operators.
Finally the effect of dimension-8 operators in $VH$ production has been studied mostly via energy and $p_T$ distributions, while their effect on angular distributions is unknown and could be of great interest.

\section*{Acknowledgments}
We thank H. El Faham, C. Severi, and M. Thomas for useful discussions. We thank DESY Theory Group for
allowing us to use their computational resources for parts of this project. This work is supported by the European Research Council (ERC) under the European Union’s Horizon 2020 research and innovation programme (Grant
agreement No. 949451) and a Royal Society University Research Fellowship through grant URF/R1/201553.
The authors acknowledge support from the COMETA COST Action CA22130.

\appendix

\section{Angular ambiguities at NLO}
\label{app:Angular_amb}
In this appendix, we show how the ambiguities in defining the azimuthal decay angle $\phi_V$ discussed in Section~\ref{sec:HeliAmp_Interf_Anato} affect the differential cross-sections at NLO.

\subsection{Neutrino reconstruction in $WH$}
\label{app:Angular_amb_WH}

\begin{figure}
    \centering
    \includegraphics[width=0.6\textwidth]{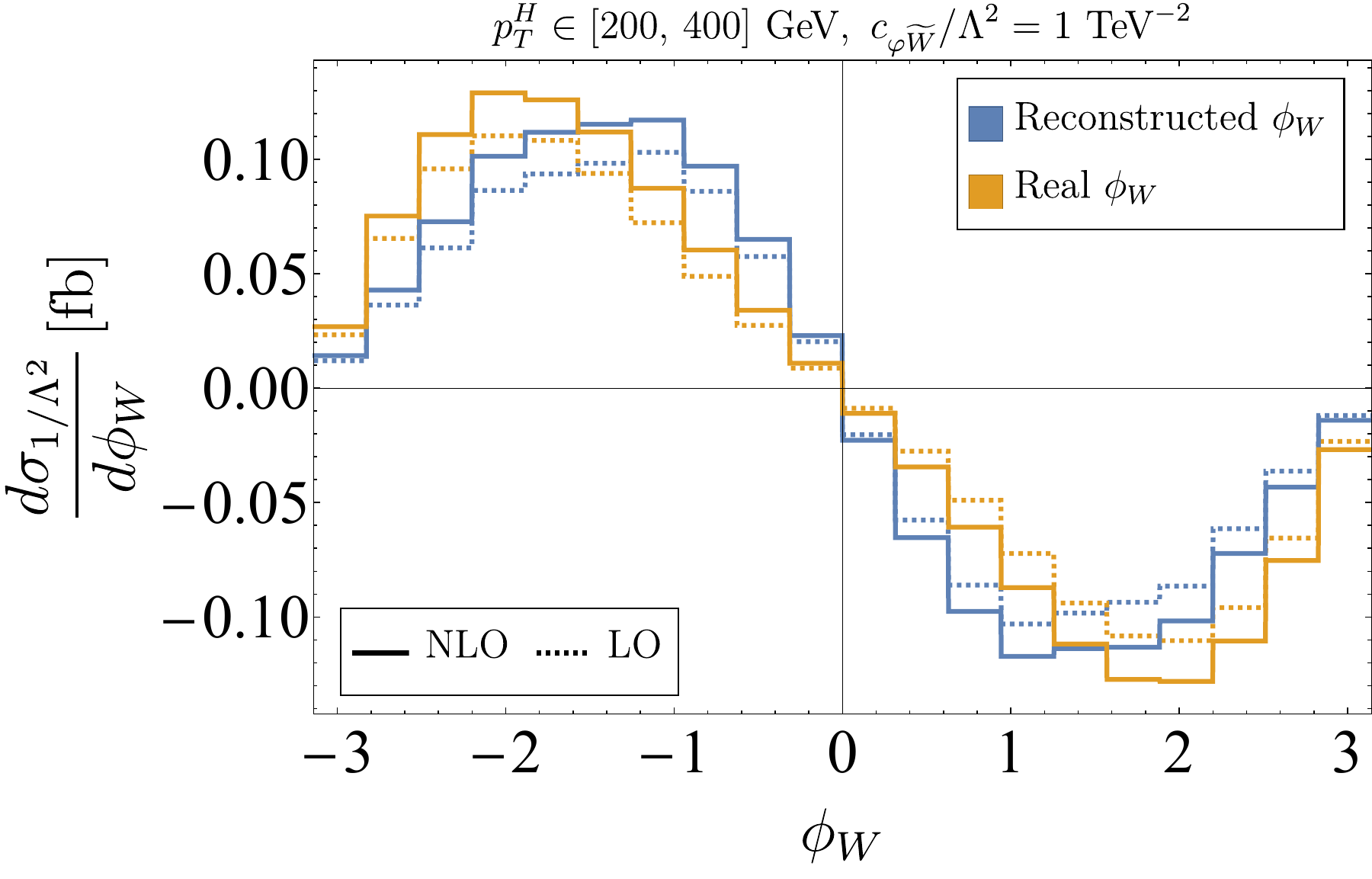}
    \caption{ Angular distribution of the SM-$\mcO_{\varphi \widetilde{W}}$ interference for $pp\to WH\to e^{-} \bar{\nu}_e H$ at the LHC in the region $p_T^H\in [200,400]$~GeV.
    The NLO and LO results are shown by full and dashed lines respectively. The blue lines correspond to the experimentally accessible distribution where the neutrino four-momentum is reconstructed following the procedure explained in the main text. The orange line shows the distribution in the case of having access to the full neutrino momentum.
    }
    \label{fig:FO_compPhiW}
\end{figure}

We show in Fig.~\ref{fig:FO_compPhiW} the differential cross-section against $\phi_W$ for the interference between SM and $\mcO_{\varphi\widetilde{W}}$ at LO and NLO in QCD in the bin $p_T^H\in[200,\,400]$~GeV. We show in different colours the effect of reconstructing $\phi_W$ from the full knowledge of the final neutrino or after reconstructing the neutrino following the procedure explained in Section~\ref{sec:HeliAmp_Interf_Anato}. Both LO and NLO differential distributions are similarly affected by the reconstruction procedure.

\subsection{Helicity ambiguity in $ZH$}
\label{app:Angular_amb_ZH}

\begin{figure}
    \centering
    \includegraphics[width=0.75\textwidth]{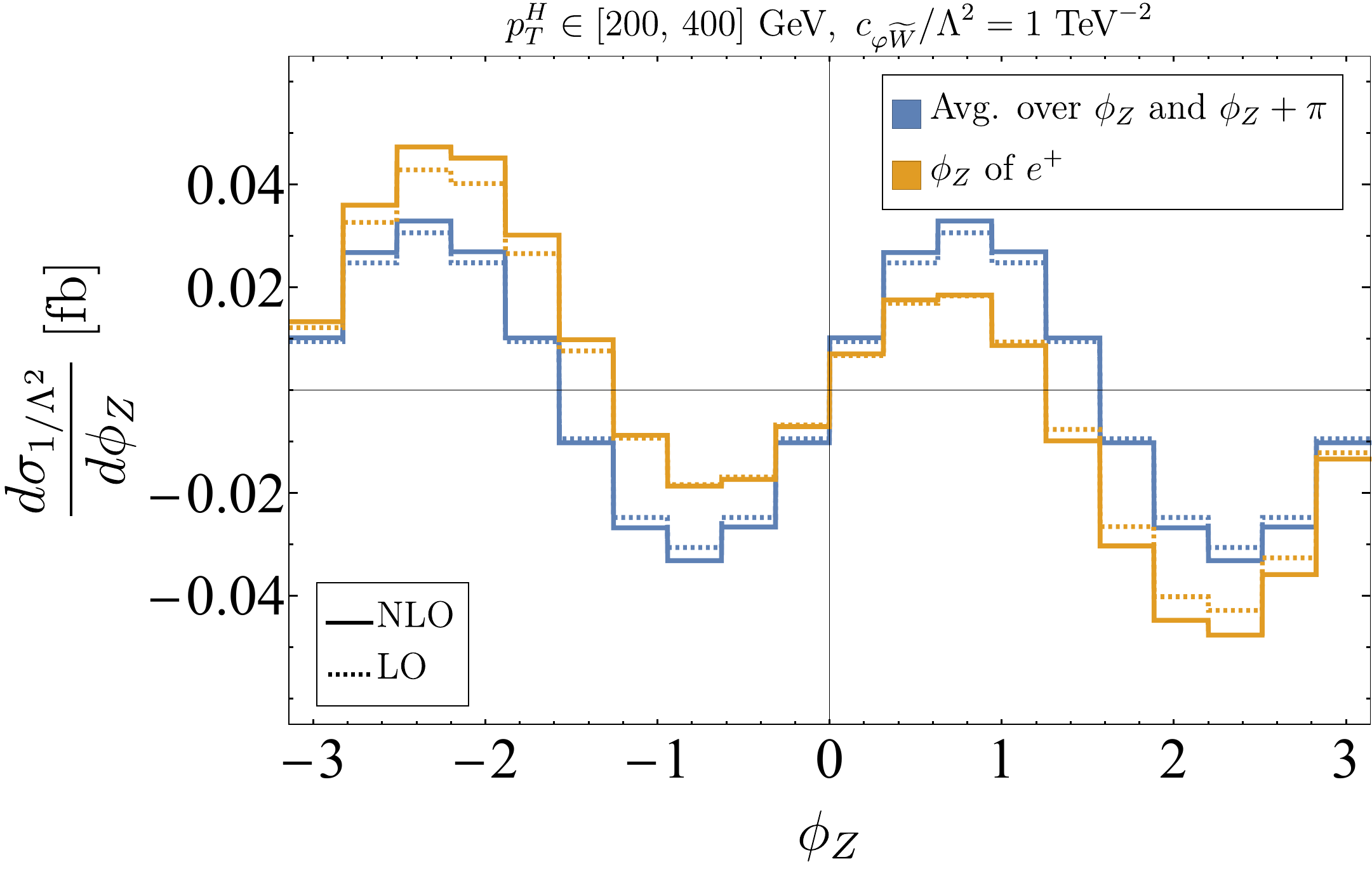}
    \caption{Angular distribution of the SM-$\mcO_{\varphi \widetilde{W}}$ interference for $pp\to ZH\to e^{+} e^{-} H$ at the LHC in the region $p_T^H\in [200,400]$~GeV. We show the distribution when the angle is defined w.r.t. to $e^{+}$ and after averaging between $\phi_Z$ and $\phi_Z + \pi$ due to the helicity ambiguity.
    }
    \label{fig:ZH_cpwtilde_ang_distr}
\end{figure}

As explained in the main text, defining the azimuthal angle $\phi_Z$ with respect to a fermion of fixed charge does not erase any of the 2 angular modes present in the interference. If one insists on defining the angle for fixed helicity, the experimentally accessible distribution would change since one has to account for the ignorance of the helicity of the final fermion. This can be done by averaging the distribution for a fixed charge over $\phi_Z$ and $\phi_Z+\pi$. The effect of such averaging in the distribution can be seen in Fig.~\ref{fig:ZH_cpwtilde_ang_distr}, where we show the differential cross-section with respect to $\phi_Z$ in the region $p_T^H\in [200,400]$~GeV. The averaging over both solutions forces the distribution to be $\sin(2\phi_Z)$, eliminating the $\sin(\phi_Z)$ mode and thus the energy-growing piece of the interference.

\section{Angular distributions and NLO effects for $\mcO_{\varphi W}$ in $WH$}
\label{app:app_cp_even_wh}

\begin{figure}
    \centering
    \includegraphics[width=0.6\textwidth]{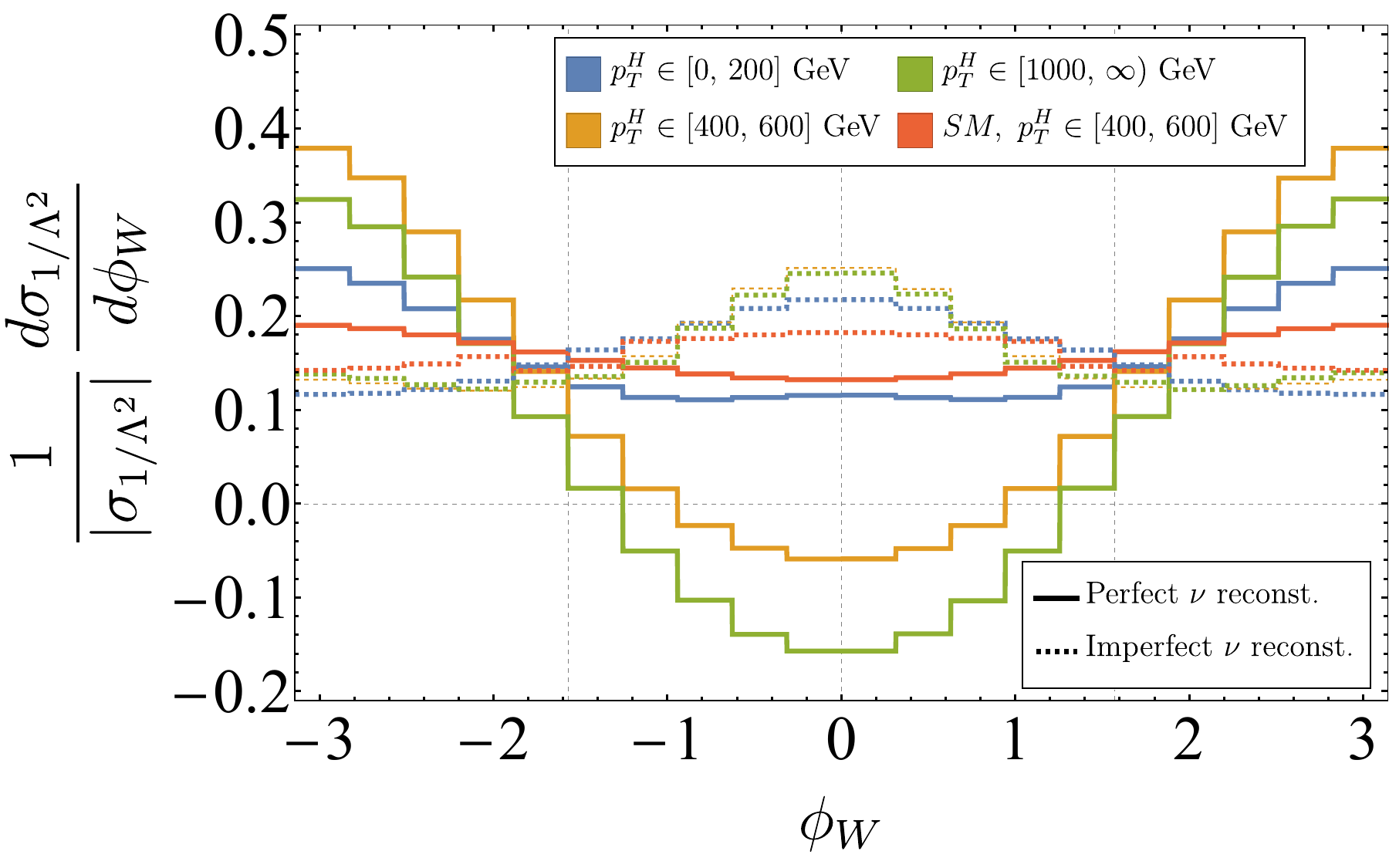}
    \caption{Angular distribution of the interference between the SM and $\mcO_{\varphi W}$ for $pp\to W^{-}H\to e^{-} \bar{\nu}_e H$ at the LHC ($\sqrt{s}=13$~TeV) and LO QCD in different $p_T^H$ bins. The normalization factor is the integral of the absolute value of the interference, $|\sigma_{1/\Lambda^2}|=\int |\frac{d\sigma_{1/\Lambda^2}}{d\phi_V}|d\phi_V$. For reference, we also show the normalized differential distribution of the SM in the middle $p_T^H$ bin. Full lines are obtained assuming a perfect reconstruction of the final neutrino, while dashed lines account for the ambiguity caused by its imperfect reconstruction. The vertical dashed lines are located at $\phi_{W}=0,\,\pm\frac{\pi}{2}$.
}
    \label{fig:LO_angular_distr_cpeven_WH}
\end{figure}

In this appendix, we show the angular differential distributions generated by the CP-even operator $\mcO_{\varphi W}$ in the process $pp\to W^{-} H\to e^{-}\bar{\nu}_e H$. We begin by showing in Fig.~\ref{fig:LO_angular_distr_cpeven_WH} the differential distribution at LO QCD generated by the interference with the SM with and without a perfect neutrino reconstruction in different $p_T^H$ bins. For reference, we add the differential distribution of the SM. The comparison of this Figure against the left panel of Fig.~\ref{fig:LO_angular_distr_ZH} shows how the CP-even and CP-odd operators generate very different distributions and how the former mimics the behaviour of the SM. Additionally, the ambiguity caused by the imperfect neutrino reconstruction affects significantly the SM and CP-even distributions.

\begin{figure}
    \centering
    \includegraphics[width=0.6\textwidth]{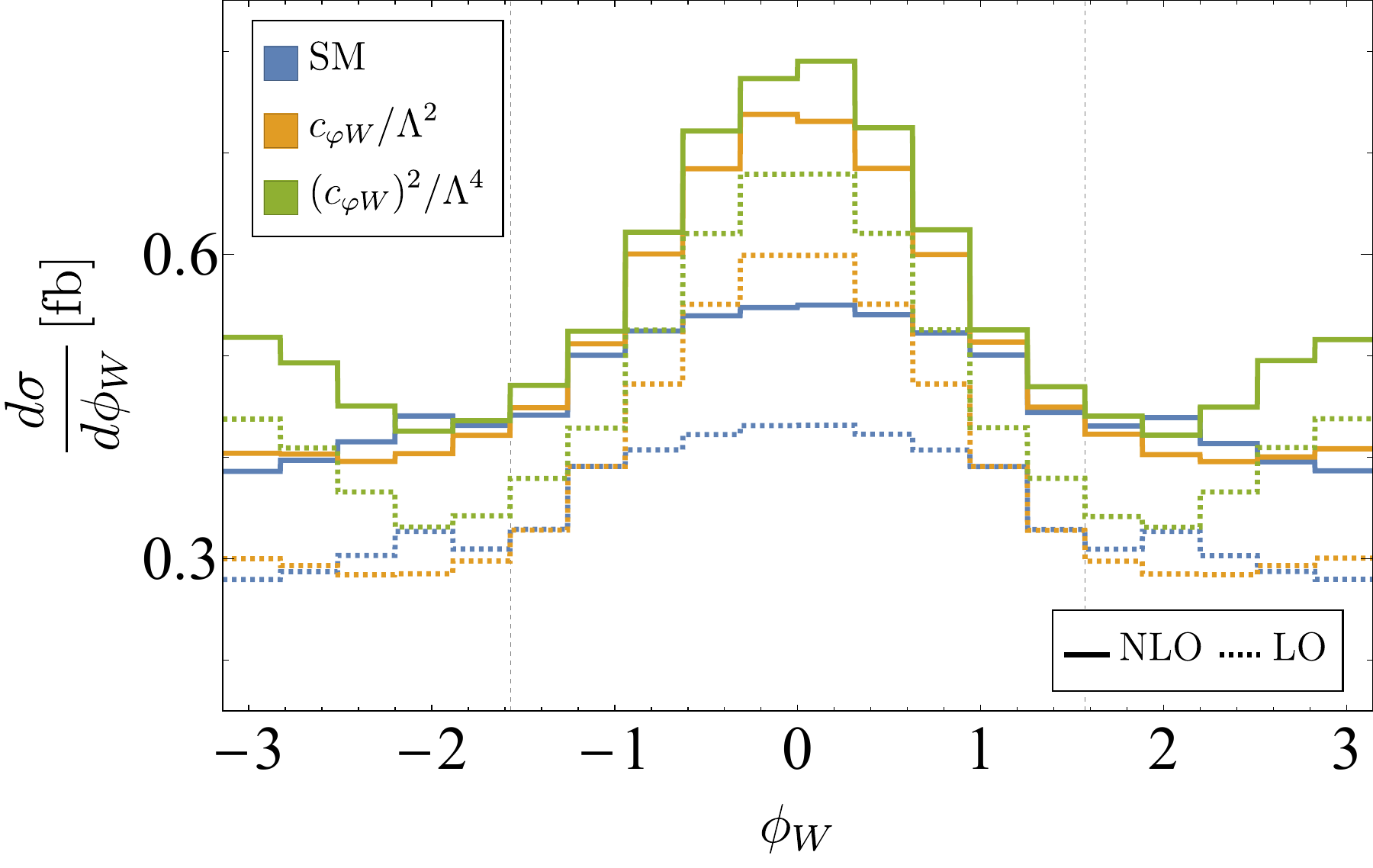}
    \caption{Angular distribution for $pp\to W^{-}H\to e^{-} \bar{\nu}_e H$ at the LHC in the $p_T^H\in [200,400]$~GeV bin when including the effect of $\mcO_{\varphi W}$. We show separately the SM (blue), SM-EFT interference ($\cpW/\Lambda^2$, orange), and EFT squared ($(\cpW)^2/\Lambda^4$, green) pieces, at LO (dashed) and NLO (full) in QCD. The vertical dashed lines are located at $\phi_{W}=0,\,\pm\frac{\pi}{2}$.
    }
    \label{fig:WH_cpw_ang_distr_all}
\end{figure}

Fig.~\ref{fig:WH_cpw_ang_distr_all} shows the differential angular distribution for $pp\to W^{-}H\to e^{-} \bar{\nu}_e H$ at the LHC ($\sqrt{s}=13$~TeV) in the SM, the EFT squared piece with $\mcO_{\varphi W}$ and their interference. We show each contribution at LO and NLO in QCD in the bin $p_T^H\in [200,\,400]$~GeV. The EFT-induced contributions show a distribution more peaked towards $0$ and the EFT-squared piece also presents secondary maxima for $|\phi_W|=\pi$. In comparison against the left panel of Fig.~\ref{fig:ZH_cpwtilde_ang_distr_all}, the CP-even and CP-odd operators generate very different distributions, in particular when they interfere with the SM. Even their squared amplitude contributions show different behaviours.

\begin{figure}[h!]
    \centering
    \includegraphics[width=0.6\textwidth]{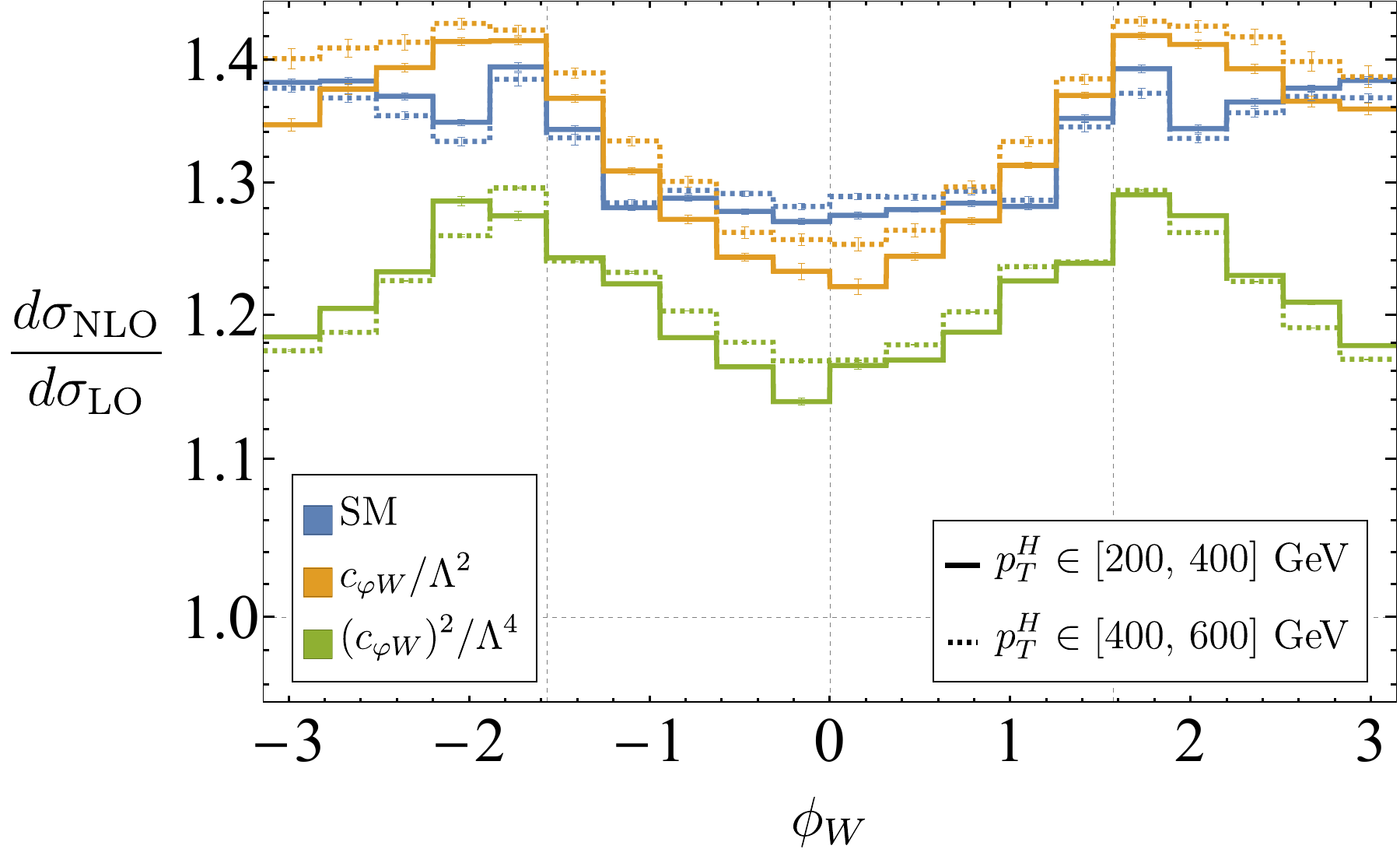}
    \caption{NLO/LO ratio of the angular differential cross section for the SM, interference and EFT squared pieces of the operator $\mcO_{\varphi W}$ in $pp\to W^{-}H\to e^{-} \bar{\nu}_e H$ at the LHC. The full (dashed) lines correspond to the $p_T^H\in [200,400]$~GeV ($[400,600]$~GeV) region. The error bars show the Monte Carlo uncertainty and the vertical dashed lines are located at $\phi_{W}=0,\,\pm\frac{\pi}{2}$.}
    \label{fig:ang_dep_kF_cpW}
\end{figure}

Finally, we show the angular differential NLO/LO $k$-factor for the SM, its interference with $\mcO_{\varphi W}$ and the squared amplitude of $\mcO_{\varphi W}$ in Fig.~\ref{fig:ang_dep_kF_cpW}. We show this for two different $p_T^H$ bins. The interference shows a smaller $k$-factor than the SM and EFT squared pieces with a small dependence on the $p_T^H$ bin. This figure can be compared against the left panel of Fig.~\ref{fig:ang_dep_kF}.

\section{Simulation and analysis details}
\label{app:app_Simu_details}

\subsection{Montecarlo simulation}
\label{app:app_MC_simu}

The Monte Carlo event generation of the signal process in Section~\ref{sec:CPV_LHC} was performed with \texttt{MadGraph5\_aMC@NLO}~v.3.4.1~\cite{Alwall:2014hca}. For consistency with the background processes samples, we used the \texttt{NNPDF23} parton distribution functions~\cite{Ball:2012cx} and the parton shower and Higgs decay were simulated with \texttt{Pythia8.24}~\cite{Sjostrand:2014zea}. The samples were generated for a centre-of-mass energy of $13$~TeV and we neglect any possible gain that might come from the increase in the centre-of-mass energy to $14$~TeV foreseen for the LHC at Run 3 and HL-LHC. 

\begin{table}[t]
\begin{centering}
\resizebox{\textwidth}{!}{
\begin{tabular}{c|ccccccc}
$p_T^H$ bin [GeV]  &$[0,\,230]$&$[230,\,330]$&$[330,\,500]$&$[500,\,700]$&$[700,\,1000]$&$[1000,\,1500]$&$[1500,\,\infty]$ \\
\hline
EW $k$-factor & $0.99$ & $0.95$ & $0.90$ & $0.83$ & $0.75$ & $0.63$ & $0.57$ 
\end{tabular}}
\par\end{centering}
\caption[caption]{NLO EW $k$-factor for $W H (\rightarrow b\bar b)$ per bin of the transverse momentum of the Higgs boson $p_T^H$. The $k$-factor is extracted from ref.~\cite{Frederix:2018nkq}.
}
\label{tab:k_Factor_EW}
\end{table}

\begin{table}[t]
\begin{centering}
\centering\renewcommand*{\arraystretch}{1.5}
\begin{tabular}{c | m{2.5cm}}
 & \centering{$W\rightarrow\ell \nu$} \tabularnewline
\hline
$p_{T,\min}^{\ell}$ {[}GeV{]} & \centering{$25$} \tabularnewline
$|\eta_{max}^{\ell}|$ & \centering{$2.8$} \tabularnewline
$p_{T}^{H}$ & $\{0,\,200,\,400,\,700,\,\infty\}$ \tabularnewline
\end{tabular}
\par\end{centering}
\caption{Parton level generation cuts for the signal process at $13\,\mathrm{TeV}$.
}
\label{tab:gen_cuts}
\end{table}

\begin{table}[t!]
    \centering
    \begin{tabular}{c|c|c|c|c|c|c}
     & Electrons & Muons & Light Jets & $b$-Jets \\
     \hline
     $p_{T}$ [GeV] & $>27$ & $>25$ & $>30\, (20)$ & $>20$ \\
     $|\eta|$ & $<2.5$ & $<2.7$ & $<4.5\, (2.5)$ & $<2.5$ \\
    \end{tabular}
    \caption{Acceptance regions used in our analysis for charged leptons and jets. For light jets, the minimum $p_T$ outside (between) the parenthesis corresponds to the $\eta$ interval outside (between) parenthesis. The values follow \cite{ATLAS:2020fcp,ATLAS:2020jwz}.}
    \label{tab:acceptance_regions}
\end{table}

\begin{table}[h!]
    \centering
    \begin{tabular}{c|c|c}
      Selection cuts   & Boosted category & Resolved category  \\
      \hline
      $p_{T,\text{min}}^{b}$ [GeV] & - & 20 \\
      $p_{T,\text{ min}}^{b,\text{ leading}}$ [GeV] & - & 45 \\
      $\eta_{\text{max}}^{b}$ & - & 2.5 \\
      $\eta_{\text{max}}^{H_{\text{cand}}}$ & 2.0 & - \\
      $\Delta\text{R}^{\text{max}}_{bb}$ & - & 2.0 \\
      $E_{T,\text{min}}^{\text{miss}}$ [GeV] & $\begin{cases}
          50\text{ if } \ell=e\\
          90\text{ if } \ell=\mu
      \end{cases}$ & $\begin{cases}
          30\text{ if } \ell=e\\
          90\text{ if } \ell=\mu
      \end{cases}$ \\
     $|\Delta y\left( W,\,H_{\text{cand}}\right)|_{\text{max}}$ & 1.4 & - \\
     $m_{H_{\text{cand}}}$ [GeV] & \multicolumn{2}{c}{$[90,120]$}
\end{tabular}
    \caption{Selection cuts selected following~\cite{ATLAS:2020fcp,ATLAS:2020jwz}.}
    \label{tab:selection_cuts}
\end{table}

The signal process, $WH \to\ell\nu b\bar{b}$, was simulated at LO and NLO in QCD. For the results presented in section~\ref{sec:sensitivity_LHC} and App.~\ref{app:app_xs}, we accounted for NLO EW corrections via a $p_T^H$ dependent $k$-factor detailed in Table~\ref{tab:k_Factor_EW}. The efficiency of the simulated events to pass the selection cuts was increased via generation-level cuts and binning on $p_T^H$. We detail these cuts and the bin limits in Table~\ref{tab:gen_cuts}. Similar techniques were used for the background processes, for details we refer the reader to Appendix B in~\cite{Bishara:2022vsc}.

\subsection{Detector simulation and analysis}
\label{app:app_detec_analysis}

The detector simulation was performed with an in-house code that performed the jet clustering and substructure analysis, b-tagging and applied acceptance and selection cuts. This code was previously used in~\cite{Bishara:2022vsc,Rossia:2023hen,Englert:2024ufr}. Those events that contain at least one mass-drop-tagged jet are considered as ``boosted'' and the rest are ``resolved'' events. The b-tagging of the jet constituents is performed by angular proximity. The parameters of the clustering and b-tagging algorithms were tuned to reproduce published ATLAS differential distributions~\cite{ATLAS:2020fcp,ATLAS:2020jwz} and we refer the reader to Appendix B of~\cite{Bishara:2022vsc} for the details. The acceptance cuts are based on the ATLAS analysis of the 1-lepton category of $VH$ production~\cite{ATLAS:2020fcp,ATLAS:2020jwz}, agree with the ones used in~\cite{Bishara:2022vsc} and we summarize them in Table~\ref{tab:acceptance_regions}. We require one charged lepton in the acceptance region, which corresponds to the ``tight'' region defined by ATLAS. We veto events with more than 2 $b$-tagged or 0 untagged jets in the resolved category or any $b$-tagged or untagged jet besides a doubly $b$-tagged boosted jet in the boosted category.

The selection cuts are summarized in Table~\ref{tab:selection_cuts}. The $E_{T,\text{ min}}^{\text{miss}}$ cut depends on the flavour of the charged lepton. If it is an electron, we require $E_{T,\text{ min}}^{\text{miss}}= 30 (50)$~GeV in the resolved (boosted) category, while for a muon, the cut is implemented as min$\lbrace p_{T,\text{ min}}^{\mu,E_{T}^{\text{miss}} } \rbrace\geqslant 90$~GeV to replicate the ATLAS analysis~\cite{ATLAS:2020fcp,ATLAS:2020jwz}. The invariant mass windows for the Higgs candidate is set to $[90,120]$~GeV. The two most efficient cuts in reducing the background are the Higgs invariant mass and the jet veto~\cite{Bishara:2022vsc}. Our choice of selection cuts is nearly optimal since replacing it with a Machine Learning-based strategy improves its sensitivity to the leading SMEFT effects by $\lesssim 7\%$~\cite{Englert:2024ufr}.

\section{Signal and background number of events}
\label{app:app_xs}

\begin{table}[t!]
		\centering
\resizebox{\textwidth}{!}{
			\begin{tabular}{|@{\hspace{.35em}}c|c|c@{\hspace{.25em}}|c@{\hspace{.5em}}|}
				\hline
				\multicolumn{4}{|c|}{Resolved category, HL-LHC } \tabularnewline
				\hline
				\multirow{2}{*}{
				\hspace{-2.5em}
				\begin{tabular}{c}
				$p_{T}^{H}$ bin\\
				$[$GeV$]$
				\end{tabular}
				\hspace{-2.5em}}  &\multirow{2}{*}{$\phi_W$ bin} & \multicolumn{2}{c|}{Number of expected signal events}\tabularnewline
				\cline{3-4} & & \rule{0pt}{1.15em}LO & NLO \tabularnewline
				\hline
				\multirow{4}{*}[0pt]{$[0-175]$} & $[-\pi,-\frac{\pi}{2}]$ & 
    $\begin{aligned} \rule[-1.0em]{0pt}{2.5em} 1455 \,& + (119\pm15) \,\cpWt + 465 \left(\cpWt\right)^2\\
                                               + 1336 \,&\cpW + 628 \left(\cpW\right)^2+ (10\pm 3)\,\cpW \cpWt
				\end{aligned}$ & 
    $\begin{aligned} \rule[-1.0em]{0pt}{2.5em} 1800 \,& + (110\pm16) \,\cpWt + 550 \left(\cpWt\right)^2 \\
                                               + 1650 \,&\cpW + 770 \left(\cpW\right)^2+ (4\pm 8)\,\cpW \cpWt
				\end{aligned}$\tabularnewline
				\cline{2-4}
				& $[-\frac{\pi}{2},0]$ & 
$\begin{aligned} \rule[-1.0em]{0pt}{2.5em} 577 \,& + (83\pm10) \,\cpWt + 244  \left(\cpWt\right)^2\\
                                         + 566 \,&\cpW + 304 \left(\cpW\right)^2+ (17\pm 3)\,\cpW \cpWt

    \end{aligned}$ & 
$\begin{aligned} \rule[-1.0em]{0pt}{2.5em} 720 \,& + (76\pm 10) \,\cpWt + 275 \left(\cpWt\right)^2\\
                                                + 730  \,&\cpW + 392 \left(\cpW\right)^2+(39\pm 5)\,\cpW \cpWt
				\end{aligned} $\tabularnewline
				\cline{2-4}
				& $[0,\frac{\pi}{2}]$ & 
$\begin{aligned} \rule[-1.0em]{0pt}{2.5em} 571 \,& + (-65\pm10) \,\cpWt + 236 \left(\cpWt\right)^2\\
                                         + 566 \,&\cpW + 305 \left(\cpW\right)^2+ (-18\pm 2)\,\cpW \cpWt
				\end{aligned}$ & 
$\begin{aligned} \rule[-1.0em]{0pt}{2.5em} 711 \,& + (-70\pm 11) \,\cpWt + 314 \left(\cpWt\right)^2\\
                                        + 730 \,&\cpW + 400 \left(\cpW\right)^2+(-29\pm 7.4)\,\cpW \cpWt
				\end{aligned}$\tabularnewline
                \cline{2-4}
				& $[\frac{\pi}{2},\pi]$ & 
$\begin{aligned} \rule[-1.0em]{0pt}{2.5em} 1453 \,& + ( -121\pm15) \,\cpWt + 465 \left(\cpWt\right)^2\\
                                        + 1298 \,&\cpW + 601 \left(\cpW\right)^2+ (-5\pm 4)\,\cpW \cpWt
				\end{aligned}$ & 
$ \begin{aligned} \rule[-1.0em]{0pt}{2.5em} 1790 \,& + (-110\pm16) \,\cpWt + (540\pm17) \left(\cpWt\right)^2\\
                                           + 1670 \,&\cpW + 790 \left(\cpW\right)^2+(-4\pm 8)\,\cpW \cpWt
				\end{aligned} $\tabularnewline
                \hline
		\multirow{4}{*}[0pt]{$[175-250]$} & $[-\pi,-\pi/2]$ & 
$\begin{aligned} \rule[-1.0em]{0pt}{2.5em} 237 \,& + ( 34 \pm 3) \,\cpWt + 154 \left( \cpWt \right)^2\\
                                        + 235 \,&\cpW + 191 \left(\cpW\right)^2+ (4.8\pm 1.5)\,\cpW \cpWt
				\end{aligned}$ & $ 
\begin{aligned} \rule[-1.0em]{0pt}{2.5em} 293 \,& + ( 40 \pm 4) \,\cpWt + 202 \left(\cpWt\right)^2\\
                                        + 291 \,&\cpW + 234 \left(\cpW\right)^2 + (2\pm 3)\,\cpW \cpWt
				\end{aligned} $\tabularnewline
				\cline{2-4}
				& $[-\pi/2,0]$ & 
$\begin{aligned} \rule[-1.0em]{0pt}{2.5em} 69 \,& + ( 16.3 \pm 1.8 ) \,\cpWt + 75 \left( \cpWt \right)^2\\
                                        + 78 \,&\cpW + 78 \left(\cpW\right)^2+ (5.1\pm 0.5)\,\cpW \cpWt
				\end{aligned}$ & 
$\begin{aligned} \rule[-1.0em]{0pt}{2.5em} 87 \,& + (13\pm2) \,\cpWt + 92 \left(\cpWt\right)^2\\
                                        + 98 \,&\cpW + 97 \left(\cpW\right)^2+(9\pm 1)\,\cpW \cpWt
				\end{aligned}$  \tabularnewline
				\cline{2-4}
				& $[0,\pi/2]$ & 
$\begin{aligned} \rule[-1.0em]{0pt}{2.5em} 68 \,& + ( -14.1 \pm 1.8 ) \,\cpWt + 70 \left( \cpWt \right)^2\\
                                        + 80.8 \,&\cpW + 82.1 \left(\cpW\right)^2+ (-6.1\pm 1.3)\,\cpW \cpWt
				\end{aligned}$ & 
$ \begin{aligned} \rule[-1.0em]{0pt}{2.5em} 88 \,& + (-15\pm4) \,\cpWt + 89 \left(\cpWt\right)^2\\
                                        + 101 \,&\cpW + 101 \left(\cpW\right)^2+(-11.5 \pm 1.8)\,\cpW \cpWt
				\end{aligned} $\tabularnewline
                \cline{2-4}
				& $[\pi/2,\pi]$ & 
$\begin{aligned} \rule[-1.0em]{0pt}{2.5em} 238 \,& + (-41\pm3) \,\cpWt + 150 \left(\cpWt\right)^2\\
                                         + 232 \,&\cpW + 189 \left(\cpW\right)^2+ (-4\pm 2)\,\cpW \cpWt
				\end{aligned}$ & 
$ \begin{aligned} \rule[-1.0em]{0pt}{2.5em} 296 \,& + (-35\pm4) \,\cpWt + 186 \left(\cpWt\right)^2\\
                                           + 293 \,&\cpW + 237 \left(\cpW\right)^2+(-10\pm 6)\,\cpW \cpWt
				\end{aligned} $\tabularnewline
                \hline 
		\multirow{4}{*}{$[250-\infty]$} & $[-\pi,-\pi/2]$ & 
    $\begin{aligned} \rule[-1.0em]{0pt}{2.5em} 13 \,& + (3.4 \pm 0.6) \,\cpWt + 16.1 \left( \cpWt \right)^2\\
                                           + 13.5 \,&\cpW + 19.1 \left(\cpW\right)^2+ (-0.3\pm 0.3)\,\cpW \cpWt
				\end{aligned}$ & 
    $\begin{aligned} \rule[-1.0em]{0pt}{2.5em} 16 \,& + (4.6\pm0.7) \,\cpWt + 20.0 \left(\cpWt\right)^2\\
                                           + 17.0 \,&\cpW + 24.4 \left(\cpW\right)^2+(-0.6\pm 0.7)\,\cpW \cpWt
				\end{aligned}$\tabularnewline
				\cline{2-4}
				& $[-\pi/2,0]$ & 
$\begin{aligned} \rule[-1.0em]{0pt}{2.5em} 4 \,& + (1.5\pm0.3) \,\cpWt + (7.8\pm0.5) \left( \cpWt \right)^2\\
                                               + (4.5\pm0.3) \,&\cpW + (7.6\pm0.4) \left(\cpW\right)^2+ (0.3\pm 0.1)\,\cpW \cpWt
				\end{aligned}$ & 
$\begin{aligned} \rule[-1.0em]{0pt}{2.5em} 4 \,& + (1.6\pm0.4) \,\cpWt + (10.0\pm0.6) \left(\cpWt\right)^2\\
                                           + (5.7\pm0.4) \,&\cpW + 10.3 \left(\cpW\right)^2+(1.4\pm 0.2)\,\cpW \cpWt
				\end{aligned} $\tabularnewline
				\cline{2-4}
				& $[0,\pi/2]$ & 
$\begin{aligned} \rule[-1.0em]{0pt}{2.5em} 3 \,& + (-1.4\pm0.4) \,\cpWt + (8.2\pm0.5) \left( \cpWt \right)^2\\
                                               + (4.5\pm0.3) \,&\cpW + (7.6\pm 0.4) \left(\cpW\right)^2+ (-1\pm 0.5)\,\cpW \cpWt
				\end{aligned}$ & 
$ \begin{aligned} \rule[-1.0em]{0pt}{2.5em} 4 \,& + (-2.2\pm0.4) \,\cpWt + (9.9\pm0.6) \left(\cpWt\right)^2\\
                                        + (5.6\pm0.3) \,&\cpW + 10.1 \left(\cpW\right)^2+(-0.5\pm 0.4)\,\cpW \cpWt
				\end{aligned} $\tabularnewline
                \cline{2-4}
				& $[\pi/2,\pi]$ & 
$\begin{aligned} \rule[-1.0em]{0pt}{2.5em} 13 \,& + (-3.0\pm0.6) \,\cpWt + (14.9\pm0.8) \left( \cpWt \right)^2\\
                                         + 14.1 \,&\cpW + 19.6 \left(\cpW\right)^2+ (0.1\pm 0.3)\,\cpW \cpWt
				\end{aligned}$ & 
$ \begin{aligned} \rule[-1.0em]{0pt}{2.5em} 16 \,& + (-3.9\pm0.7) \,\cpWt + 21.3 \left(\cpWt\right)^2\\
                                           + 17.2 \,&\cpW + 24.3 \left(\cpW\right)^2+(0.8\pm 0.5)\,\cpW \cpWt
				\end{aligned} $
                \tabularnewline
                \hline 
			\end{tabular}
}
				\caption{ Number of expected signal events from $pp\to WH\to \ell\nu b\bar{b}$ at the HL-LHC in the resolved category after applying the analysis described in App.~\ref{app:app_detec_analysis} and assuming $\Lambda=1$~TeV. We quote the Monte Carlo uncertainties when they are higher than $5\%$.
	}
	\label{tab:App_sigma_Wh_HLLHC_res_LO_NLO}
	\end{table}

\begin{table}[h!]

  \centering
\resizebox{\textwidth}{!}{
			\begin{tabular}{|@{\hspace{.35em}}c|c|c@{\hspace{.25em}}|c@{\hspace{.5em}}|}
				\hline
				\multicolumn{4}{|c|}{Boosted category, HL-LHC } \tabularnewline
				\hline
				\multirow{2}{*}{
				\hspace{-2.5em}
				\begin{tabular}{c}
				$p_{T}^{H}$ bin\\
				$[$GeV$]$
				\end{tabular}
				\hspace{-2.5em}}  &\multirow{2}{*}{$\phi_W$ bin} & \multicolumn{2}{c|}{Number of expected signal events}\tabularnewline
				\cline{3-4} & & \rule{0pt}{1.15em}LO & NLO \tabularnewline
				\hline
		\multirow{4}{*}[0pt]{$[0-175]$} & $[-\pi,-\frac{\pi}{2}]$ & 
    $\begin{aligned} \rule[-1.0em]{0pt}{2.5em} 9 \,& + (1.3\pm0.7) \,\cpWt + (7.8\pm0.9) \left(\cpWt\right)^2\\
                                           + ( 7.9 \pm0.6) \,&\cpW + (5.2\pm0.6) \left(\cpW\right)^2 + (0.01\pm0.1)\,\cpW \cpWt
				\end{aligned}$ & 
    $\begin{aligned} \rule[-1.0em]{0pt}{2.5em} 12 \,& + ( 1.8 \pm0.9) \,\cpWt + (6.4\pm1.0) \left(\cpWt\right)^2\\
                                           + (13.8 \pm 0.8) \,&\cpW + ( 10.2 \pm 0.8 ) \left(\cpW\right)^2+( -1.4 \pm 2.1)\,\cpW \cpWt
				\end{aligned}$\tabularnewline
				\cline{2-4}
				& $[-\frac{\pi}{2},0]$ & 
                $\begin{aligned} \rule[-1.0em]{0pt}{2.5em} 3 \,& + (0.2\pm0.4) \,\cpWt + (2.6\pm0.5) \left(\cpWt\right)^2\\
                                           + (2.9\pm0.4) \,&\cpW + (2.6\pm0.4) \left(\cpW\right)^2 + ( 0.12 \pm 0.05)\,\cpW \cpWt
				\end{aligned}$ & 
    $\begin{aligned} \rule[-1.0em]{0pt}{2.5em} 4 \,& + ( 0.4 \pm0.5  ) \,\cpWt + (3.1\pm0.6) \left(\cpWt\right)^2\\
                                           + (4.1\pm0.4) \,&\cpW + (4.1\pm0.4) \left(\cpW\right)^2+(0.3\pm0.3)\,\cpW \cpWt
				\end{aligned}$\tabularnewline
				\cline{2-4}
				& $[0,\frac{\pi}{2}]$ & 
                $\begin{aligned} \rule[-1.0em]{0pt}{2.5em} 3 \, & + (-0.5\pm0.5) \,\cpWt + (3.2\pm 0.6) \left(\cpWt\right)^2\\
                                           + ( 3.3 \pm0.4) \,&\cpW + (2.7\pm0.4) \left(\cpW\right)^2+(-0.2\pm0.1)\,\cpW \cpWt
				\end{aligned}$ & 
    $\begin{aligned} \rule[-1.0em]{0pt}{2.5em} 4 \,& + (-1.0\pm0.5) \,\cpWt + (2.4\pm0.6) \left(\cpWt\right)^2\\
                                           + (5.6\pm0.4) \,&\cpW + (4.5\pm0.4) \left(\cpW\right)^2 + (-0.2\pm0.3)\,\cpW \cpWt
				\end{aligned}$\tabularnewline
                \cline{2-4}
				& $[\frac{\pi}{2},\pi]$ & 
                $\begin{aligned} \rule[-1.0em]{0pt}{2.5em} 9 \,& + (-1.2\pm0.6) \,\cpWt + (5.3\pm0.9) \left(\cpWt\right)^2\\
                                           + ( 9.7\pm0.6) \,&\cpW + (6.7\pm0.6) \left(\cpW\right)^2+(-0.4\pm0.3)\,\cpW \cpWt
				\end{aligned}$ & 
    $\begin{aligned} \rule[-1.0em]{0pt}{2.5em} 12 \,& + (-1.7\pm0.8) \,\cpWt + (7.7\pm1.0) \left(\cpWt\right)^2\\
                                           + (11.9\pm0.7) \,&\cpW + (8.8\pm0.7) \left(\cpW\right)^2+(0.3\pm0.5)\,\cpW \cpWt
				\end{aligned}$\tabularnewline
                \hline
		\multirow{4}{*}[0pt]{$[175-250]$} & $[-\pi,-\pi/2]$ & 
    $\begin{aligned} \rule[-1.0em]{0pt}{2.5em} 175 \,& + (38 \pm 2) \,\cpWt + 128 \left(\cpWt\right)^2\\
                                           + 164 \,&\cpW + 153 \left(\cpW\right)^2+(2.5\pm0.7)\,\cpW \cpWt
				\end{aligned}$ & 
    $\begin{aligned} \rule[-1.0em]{0pt}{2.5em} 218 \,& + 40 \,\cpWt + 166 \left(\cpWt\right)^2 \\
                                           + 211 \,&\cpW + 193 \left(\cpW\right)^2 + (-4.6\pm3.3)\,\cpW \cpWt
				\end{aligned}$\tabularnewline
				\cline{2-4}
                & $[-\frac{\pi}{2},0]$ & 
                $\begin{aligned} \rule[-1.0em]{0pt}{2.5em} 50 \,& + (7.9\pm1.2) \,\cpWt + 53.2 \left(\cpWt\right)^2\\
                                           + 54 \,&\cpW + 59 \left(\cpW\right)^2+(1.6\pm0.4)\,\cpW \cpWt
				\end{aligned}$ & 
    $\begin{aligned} \rule[-1.0em]{0pt}{2.5em} 64 \,& + (8.1\pm1.4) \,\cpWt + 69.9 \left(\cpWt\right)^2\\
                                           + 72.8 \,&\cpW + 76.7 \left(\cpW\right)^2 + (8.3\pm1.0)\,\cpW \cpWt
				\end{aligned}$\tabularnewline
				\cline{2-4}
				& $[0,\frac{\pi}{2}]$ & 
                $\begin{aligned} \rule[-1.0em]{0pt}{2.5em} 50 \,& + (-9.1 \pm 1.2) \,\cpWt + 55.0 \left(\cpWt\right)^2\\
                                           + 54.1 \,&\cpW + 56.7 \left(\cpW\right)^2+(-0.9\pm0.4)\,\cpW \cpWt
				\end{aligned}$ & 
    $\begin{aligned} \rule[-1.0em]{0pt}{2.5em} 64 \,& + (-11.3 \pm  1.4) \,\cpWt + 69.6 \left(\cpWt\right)^2\\
                                           + 69.8 \,&\cpW + 75.7 \left(\cpW\right)^2 + (-5.8\pm 1.2)\,\cpW \cpWt
				\end{aligned}$\tabularnewline
                \cline{2-4}
				& $[\frac{\pi}{2},\pi]$ & 
                $\begin{aligned} \rule[-1.0em]{0pt}{2.5em} 175 \,& -41 \,\cpWt + 131 \left(\cpWt\right)^2\\
                                           +  171 \,&\cpW + 156 \left(\cpW\right)^2+(0.7\pm0.7)\,\cpW \cpWt
				\end{aligned}$ & 
    $\begin{aligned} \rule[-1.0em]{0pt}{2.5em} 217 \,& + (-38\pm2) \,\cpWt + 165 \left(\cpWt\right)^2\\
                                           + 207 \,&\cpW + 191 \left(\cpW\right)^2 + (2.5\pm2.6)\,\cpW \cpWt
				\end{aligned}$\tabularnewline
                \hline 
		\multirow{4}{*}{$[250-300]$} & $[-\pi,-\pi/2]$ & 
    $\begin{aligned} \rule[-1.0em]{0pt}{2.5em} 66 \,& + (21.5\pm1.3) \,\cpWt + 78.4 \left(\cpWt\right)^2\\
                                           + 70.2 \,&\cpW + 91.7 \left(\cpW\right)^2+(0.3\pm0.9)\,\cpW \cpWt
				\end{aligned}$ & 
    $\begin{aligned} \rule[-1.0em]{0pt}{2.5em} 82 \, & + (22.2\pm1.5) \,\cpWt + 100 \left(\cpWt\right)^2\\
                                           + 85.8 \, & \cpW + 113.2 \left(\cpW\right)^2+(1.2\pm1.8)\,\cpW \cpWt
				\end{aligned}$\tabularnewline
				\cline{2-4}
                & $[-\frac{\pi}{2},0]$ & 
                $\begin{aligned} \rule[-1.0em]{0pt}{2.5em} 18 \,& + (4.8\pm0.8) \,\cpWt + 34 \left(\cpWt\right)^2\\
                                           + 21.6 \,&\cpW + 34.1 \left(\cpW\right)^2 + (1.2 \pm 0.3)\,\cpW \cpWt
				\end{aligned}$ & 
    $\begin{aligned} \rule[-1.0em]{0pt}{2.5em} 24 \,& + (4.2\pm0.9) \,\cpWt + 41.9 \left(\cpWt\right)^2\\
                                           + 29.8 \,&\cpW + 46.7 \left(\cpW\right)^2 + (5.6\pm0.8)\,\cpW \cpWt
				\end{aligned}$\tabularnewline
				\cline{2-4}
				& $[0,\frac{\pi}{2}]$ & 
                $\begin{aligned} \rule[-1.0em]{0pt}{2.5em} 18 \,& + (-4.7\pm0.8) \,\cpWt + 33 \left(\cpWt\right)^2\\
                                           + 23.9 \,&\cpW + 36.3 \left(\cpW\right)^2+(-1.9\pm0.5)\,\cpW \cpWt
				\end{aligned}$ & 
    $\begin{aligned} \rule[-1.0em]{0pt}{2.5em} 24 \,& + ( -4.8 \pm 0.9 ) \,\cpWt + 42.5 \left(\cpWt\right)^2\\
                                           + 28.2 \,&\cpW + 45.3 \left(\cpW\right)^2 + (-4.2\pm1.1)\,\cpW \cpWt
				\end{aligned}$\tabularnewline
                \cline{2-4}
				& $[\frac{\pi}{2},\pi]$ & 
                $\begin{aligned} \rule[-1.0em]{0pt}{2.5em} 66 \,& + (-19.3\pm1.4) \,\cpWt + 79.6 \left(\cpWt\right)^2\\
                                           + 70.1 \,&\cpW + 93.1 \left(\cpW\right)^2+(-0.3\pm0.8)\,\cpW \cpWt
				\end{aligned}$ & 
    $\begin{aligned} \rule[-1.0em]{0pt}{2.5em} 82 \,& + (-20.2\pm1.6) \,\cpWt + 97 \left(\cpWt\right)^2\\
                                           + 87.0 \,&\cpW + 115.7 \left(\cpW\right)^2 + (0.1\pm1.8) \,\cpW \cpWt
				\end{aligned}$\tabularnewline
                \hline
        \multirow{4}{*}{$[300-\infty]$} & $[-\pi,-\pi/2]$ & 
    $\begin{aligned} \rule[-1.0em]{0pt}{2.5em} 25 \,& + (11.1\pm0.8) \,\cpWt + 76.6 \left(\cpWt\right)^2\\
                                           + 35.4 \,&\cpW + 83.7 \left(\cpW\right)^2+(-0.7\pm0.5)\,\cpW \cpWt
				\end{aligned}$ & 
    $\begin{aligned} \rule[-1.0em]{0pt}{2.5em} 30 \,& + (13\pm1.0) \,\cpWt + 101.7 \left(\cpWt\right)^2\\
                                           + 43.1 \,&\cpW + 108.3 \left(\cpW\right)^2+(-1.1\pm1.1)\,\cpW \cpWt
				\end{aligned}$\tabularnewline
				\cline{2-4}
                & $[-\frac{\pi}{2},0]$ & 
                $\begin{aligned} \rule[-1.0em]{0pt}{2.5em} 8 \,& + (4.7\pm0.5) \,\cpWt + 40.9 \left(\cpWt\right)^2\\
                                           + 15.8 \,&\cpW + 43.6 \left(\cpW\right)^2 + (1.1 \pm 0.2)\,\cpW \cpWt
				\end{aligned}$ & 
    $\begin{aligned} \rule[-1.0em]{0pt}{2.5em} 11 \,& + (5.7\pm0.6) \,\cpWt + 54.3 \left(\cpWt\right)^2\\
                                           + 18.6 \,&\cpW + 57.2 \left(\cpW\right)^2+(3.6\pm0.5)\,\cpW \cpWt
				\end{aligned}$\tabularnewline
				\cline{2-4}
				& $[0,\frac{\pi}{2}]$ & 
                $\begin{aligned} \rule[-1.0em]{0pt}{2.5em} 8 \,& + (-3.7 \pm 0.5) \,\cpWt + 41.9 \left(\cpWt\right)^2\\
                                           + 15.1 \,&\cpW + 44.2 \left(\cpW\right)^2 + (-0.74\pm0.18)\,\cpW \cpWt
				\end{aligned}$ & 
    $\begin{aligned} \rule[-1.0em]{0pt}{2.5em} 11 \,& + ( -4.5 \pm 0.6) \,\cpWt + 54.8 \left(\cpWt\right)^2\\
                                           + 18.2 \,&\cpW + 58.0 \left(\cpW\right)^2 + (-3.6\pm0.7)\,\cpW \cpWt
				\end{aligned}$\tabularnewline
                \cline{2-4}
				& $[\frac{\pi}{2},\pi]$ & 
                $\begin{aligned} \rule[-1.0em]{0pt}{2.5em} 24 \,& + (-10.6 \pm 0.8) \,\cpWt + 79.5 \left(\cpWt\right)^2\\
                                           + 33.6 \,&\cpW + 84.1 \left(\cpW\right)^2+(-3.9\pm2.8)\,\cpW \cpWt
				\end{aligned}$ & 
    $\begin{aligned} \rule[-1.0em]{0pt}{2.5em} 30 \,& + (-12.9\pm1.0) \,\cpWt + 102.1 \left(\cpWt\right)^2\\
                                           + 41.2 \,&\cpW + 107.6 \left(\cpW\right)^2+(-0.4\pm1.1)\,\cpW \cpWt
				\end{aligned}$\tabularnewline
                \hline
			\end{tabular}
}
				\caption{ Number of expected signal events from $pp\to WH\to \ell\nu b\bar{b}$ at the HL-LHC in the boosted category after applying the analysis described in App.~\ref{app:app_detec_analysis} and assuming $\Lambda=1$~TeV. We quote the Monte Carlo uncertainties when they are higher than $5\%$.
	}
	\label{tab:App_sigma_Wh_HLLHC_boos_LO_NLO}
	\end{table}

    \begin{table}[h!]
		\centering
		\begin{scriptsize}
   			\begin{tabular}{|@{\hspace{.35em}}c|c|c@{\hspace{.25em}}|c@{\hspace{.5em}}|c@{\hspace{.5em}}|c@{\hspace{.25em}}|c@{\hspace{.5em}}|c@{\hspace{.5em}}|}
	          \hline
				\multicolumn{8}{|c|}{Number of expected background events at the HL-LHC} \tabularnewline			
                \hline
				 \multirow{3}{*}{
				\hspace{-2.5em}
                $\phi_W$ bin
                \hspace{-2.5em}} & \multicolumn{7}{c|}{ $p_T^H$ bin $[$GeV$]$} \tabularnewline
				\cline{2-8}
				  &\multicolumn{4}{c}{Boosted category}&\multicolumn{3}{|c|}{Resolved category}\tabularnewline
				\cline{2-8} &$[0-175]$ & $[175-250]$ & $[250-300]$ & $[300-\infty]$ &$[0-175]$ & $[175-250]$ & $[250-\infty]$ \tabularnewline
				\hline
				$[-\pi,-\frac{\pi}{2}]$ & 
				$550 \pm 50 $ & $2078  $ & $390 \pm 24$ & $158\pm16$ & $58000\pm6000$ & $2300$ & $47\pm8$\tabularnewline
				$[-\frac{\pi}{2},0]$ & 
			$290 \pm 30 $ & $940 \pm 70$ & $138 \pm 17$ & $66 \pm 9 $ & $28000\pm5000 $ & $1090\pm80$ & $50\pm13$\tabularnewline
                $[0,\frac{\pi}{2}]$ & 
                $220 \pm 30  $ & $750 \pm 40$ & $156 \pm 14$ & $50 \pm 7$ & $23000\pm3000$ & $810\pm50$ & $12\pm4$\tabularnewline
                $[\frac{\pi}{2},\pi]$ &
                $ 610\pm40 $ & $ 2000 $ & $ 400\pm30 $ & $ 161\pm17 $ & $49000\pm5000$ & $2300$ & $90\pm30$\tabularnewline
                \hline
			\end{tabular}
			\end{scriptsize}
				\caption{ Number of expected background events at the HL-LHC for different angular and $p_T^H$ bins after applying the analysis described in App.~\ref{app:app_detec_analysis}.
	}
	\label{tab:App_sigma_Bkgd_HLLHC}
	\end{table}

In this appendix, we present our predictions for the expected number of signal and background events at the HL-LHC from the analysis described in section~\ref{sec:CPV_LHC_Strat} and App.~\ref{app:app_Simu_details}. We present them split into resolved and boosted categories and in bins of $p_T^H$ and $\phi_W$. In the case of the signal, we present its dependence up to quadratic order in $\cpWt$ and $\cpW$ at LO and NLO QCD with NLO EW corrections applied via $k$-factors as explained above.

Table~\ref{tab:App_sigma_Wh_HLLHC_res_LO_NLO} presents the expected number of signal events in the resolved category. Table~\ref{tab:App_sigma_Wh_HLLHC_boos_LO_NLO} contains the numbers corresponding to the boosted category. We checked that varying the coefficients of the $\cpW - \cpWt$ interference within their errors does not significantly affect our results; hence, the large errors we report are deemed acceptable. Finally, Table~\ref{tab:App_sigma_Bkgd_HLLHC} reports our predictions for the total number of background events in each of the relevant bins.

\newpage
\bibliographystyle{JHEP.bst}
\bibliography{bibCPVbosonic.bib}
\end{document}